\documentclass[10pt,aps,prc,twocolumn,superscriptaddress,showpacs,floatfix]{revtex4}

\usepackage{longtable}
\usepackage{array}
\usepackage{amsmath}
\usepackage{mathptmx}
\usepackage{mathptm}
\usepackage{amssymb}
\usepackage{bm}
\usepackage{multirow}
\usepackage{float}
\usepackage[final]{graphicx}
\usepackage{placeins}
\usepackage{xcolor}
\usepackage{wasysym}
\usepackage[caption=false]{subfig}
\usepackage{subfloat}
\usepackage{rotating}
\usepackage{booktabs}
\usepackage{tabularx}
\usepackage[sort&compress]{natbib}
\usepackage{textcomp}
\usepackage{fontenc}
\usepackage{times}
\usepackage{epsfig}
\usepackage{verbatim}
\usepackage[english]{babel}
\usepackage{chngcntr}
\usepackage{hycolor}
\usepackage[colorlinks=true,linkcolor=blue,citecolor=blue,urlcolor=blue,dvipdfm]{hyperref}

\makeatletter  
\renewcommand{\paragraph}{\@startsection{paragraph}{4}{0ex}%
   {-3.25ex plus -1ex minus -0.2ex}%
   {1.5ex plus 0.2ex}%
   {\centering\small\small}}
\counterwithin{paragraph}{subsubsection}         
\makeatother                                     
\stepcounter{secnumdepth}
\stepcounter{tocdepth}

\begin{document}


\title{Nuclear structure of $^{30}$S and its implications for nucleosynthesis in classical novae}
\author{K.~Setoodehnia}
 \email{kiana.setoodeh@gmail.com}
 \affiliation{Department of Physics \& Astronomy, McMaster University, Hamilton, ON L8S 4M1, Canada\\}
\author{A.~A.~Chen}
 \affiliation{Department of Physics \& Astronomy, McMaster University, Hamilton, ON L8S 4M1, Canada\\}
 \author{D.~Kahl}
 \affiliation{Center for Nuclear Study (CNS), the University of Tokyo, Wako Branch at RIKEN, 2-1 Hirosawa, Wako, Saitama 351-0198, Japan\\}
 \author{T.~Komatsubara}
 \affiliation{Institute of Physics, University of Tsukuba, Tennodai 1-1-1, Tsukuba, Ibaraki 305-8577, Japan\\}
 \author{J.~Jos$\acute{\mbox{e}}$}
 \affiliation{Departament de F$\acute{i}$sica i Enginyeria Nuclear, Universitat Polit$\grave{e}$cnica de Catalunya, c/Comte d'Urgell 187, 08036 Barcelona, Spain\\}
 \affiliation{Institut d'Estudis Espacials de Catalunya, c/Gran Capit$\grave{a}$ 2-4, Ed.~Nexus 201, 08034 Barcelona, Spain\\}
 \author{R.~Longland}
 \affiliation{Departament de F$\acute{i}$sica i Enginyeria Nuclear, Universitat Polit$\grave{e}$cnica de Catalunya, c/Comte d'Urgell 187, 08036 Barcelona, Spain\\}
 \affiliation{Institut d'Estudis Espacials de Catalunya, c/Gran Capit$\grave{a}$ 2-4, Ed.~Nexus 201, 08034 Barcelona, Spain\\}
\author{Y.~Abe}
 \affiliation{Institute of Physics, University of Tsukuba, Tennodai 1-1-1, Tsukuba, Ibaraki 305-8577, Japan\\}
\author{D.~N.~Binh}
 \affiliation{Center for Nuclear Study (CNS), the University of Tokyo, Wako Branch at RIKEN, 2-1 Hirosawa, Wako, Saitama 351-0198, Japan\\}
\author{J.~Chen}
 \affiliation{Department of Physics \& Astronomy, McMaster University, Hamilton, ON L8S 4M1, Canada\\}
\author{S.~Cherubini}
 \affiliation{INFN-Laboratori Nazionali del Sud and Dipartimento di Fisica ed Astronomia Universit$\grave{a}$ di Catania, Via S.~Sofia 6, 95123 Catania, Italy\\}
 \affiliation{Center for Nuclear Study (CNS), the University of Tokyo, Wako Branch at RIKEN, 2-1 Hirosawa, Wako, Saitama 351-0198, Japan\\}
\author{J.~A.~Clark}
 \affiliation{Physics Division, Argonne National Laboratory, Argonne, IL 60439, USA\\}
\author{C.~M.~Deibel}
 \altaffiliation{Present Address: Department of Physics \& Astronomy, Louisiana State University, Baton Rouge, LA 70803, USA}
 \affiliation{Joint Institute for Nuclear Astrophysics, Michigan State University, East Lansing, MI 48824, USA\\}
 \affiliation{Physics Division, Argonne National Laboratory, Argonne, IL 60439, USA\\}
\author{S.~Fukuoka}
 \affiliation{Institute of Physics, University of Tsukuba, Tennodai 1-1-1, Tsukuba, Ibaraki 305-8577, Japan\\}
\author{T.~Hashimoto}
 \affiliation{Center for Nuclear Study (CNS), the University of Tokyo, Wako Branch at RIKEN, 2-1 Hirosawa, Wako, Saitama 351-0198, Japan\\}
\author{T.~Hayakawa}
 \affiliation{Japan Atomic Energy Agency (JAEA), Tokai--mura, Ibaraki 3191195, Japan\\}
\author{J.~Hendriks}
 \affiliation{Department of Physics \& Astronomy, Western University, London, ON N6A 5B7, Canada\\}
\author{Y.~Ishibashi}
 \affiliation{Institute of Physics, University of Tsukuba, Tennodai 1-1-1, Tsukuba, Ibaraki 305-8577, Japan\\}
\author{Y.~Ito}
 \affiliation{Institute of Physics, University of Tsukuba, Tennodai 1-1-1, Tsukuba, Ibaraki 305-8577, Japan\\}
\author{S.~Kubono}
 \affiliation{Center for Nuclear Study (CNS), the University of Tokyo, Wako Branch at RIKEN, 2-1 Hirosawa, Wako, Saitama 351-0198, Japan\\}
\author{W.~N.~Lennard}
 \affiliation{Department of Physics \& Astronomy, Western University, London, ON N6A 5B7, Canada\\}
\author{T.~Moriguchi}
 \affiliation{Institute of Physics, University of Tsukuba, Tennodai 1-1-1, Tsukuba, Ibaraki 305-8577, Japan\\}
\author{D.~Nagae}
 \affiliation{Institute of Physics, University of Tsukuba, Tennodai 1-1-1, Tsukuba, Ibaraki 305-8577, Japan\\}
\author{R.~Nishikiori}
 \affiliation{Institute of Physics, University of Tsukuba, Tennodai 1-1-1, Tsukuba, Ibaraki 305-8577, Japan\\}
\author{T.~Niwa}
 \affiliation{Institute of Physics, University of Tsukuba, Tennodai 1-1-1, Tsukuba, Ibaraki 305-8577, Japan\\}
\author{A.~Ozawa}
 \affiliation{Institute of Physics, University of Tsukuba, Tennodai 1-1-1, Tsukuba, Ibaraki 305-8577, Japan\\}
\author{P.~D.~Parker}
 \affiliation{Wright Nuclear Structure Laboratory, Yale University, New Haven, CT 06520, USA\\}
\author{D.~Seiler}
 \affiliation{Physik Department E12, Technische Universit$\ddot{a}$t M$\ddot{u}$nchen, D-85748 Garching, Germany\\}
\author{T.~Shizuma}
 \affiliation{Japan Atomic Energy Agency (JAEA), Tokai--mura, Ibaraki 3191195, Japan\\}
\author{H.~Suzuki}
 \affiliation{Institute of Physics, University of Tsukuba, Tennodai 1-1-1, Tsukuba, Ibaraki 305-8577, Japan\\}
\author{C.~Wrede}
 \altaffiliation{Present Address: Department of Physics \& Astronomy, Michigan State University, East Lansing, MI 48824, USA; and National Superconducting Cyclotron Laboratory, Michigan State University, East Lansing, MI 48824, USA.}
 \affiliation{Department of Physics, University of Washington, Seattle, WA 98195, USA\\}
\author{H.~Yamaguchi}
 \affiliation{Center for Nuclear Study (CNS), the University of Tokyo, Wako Branch at RIKEN, 2-1 Hirosawa, Wako, Saitama 351-0198, Japan\\}
\author{T.~Yuasa}
 \affiliation{Institute of Physics, University of Tsukuba, Tennodai 1-1-1, Tsukuba, Ibaraki 305-8577, Japan\\}


\date{\today}


\begin{abstract}
\textbf{Background:} The uncertainty in the $^{29}$P($p, \gamma$)$^{30}$S reaction rate over the temperature range of 0.1 $\leq$ $T$ $\leq$ 1.3 GK was previously determined to span $\sim$4 orders of magnitude due to the uncertain location of two previously unobserved 3$^{+}$ and 2$^{+}$ resonances in the E$_{x}$ $=$ 4.7 -- 4.8 MeV region in $^{30}$S. Therefore, the abundances of silicon isotopes synthesized in novae, which are relevant for the identification of presolar grains of putative nova origin, were uncertain by a factor of 3. \textbf{Purpose:} (a) To investigate the level structure of $^{30}$S above the proton threshold (4394.9(7) keV) via charged-particle spectroscopy using the $^{32}$S($p, t$)$^{30}$S reaction and in-beam $\gamma$-ray spectroscopy using the $^{28}$Si($^{3}$He, $n\gamma$)$^{30}$S reaction to calculate the $^{29}$P($p, \gamma$)$^{30}$S reaction rate. (b) To explore the impact of this rate on the abundances of silicon isotopes synthesized in novae. \textbf{Method:} Differential cross sections of the $^{32}$S($p, t$)$^{30}$S reaction were measured at 34.5 MeV. Distorted Wave Born Approximation (DWBA) calculations were performed to constrain the spin-parity assignments of the observed levels, including the two astrophysically important levels. An energy level scheme was deduced from $\gamma$-$\gamma$ coincidence measurements using the $^{28}$Si($^{3}$He, $n\gamma$)$^{30}$S reaction. Spin-parity assignments based on measurements of $\gamma$-ray angular distributions and $\gamma$-$\gamma$ directional correlation from oriented nuclei were made for most of the observed levels of $^{30}$S. \textbf{Results:} The resonance energies corresponding to the states with 4.5 MeV $<$ $E_{x}$ $\lesssim$ 6 MeV, including the two astrophysically important states predicted previously, are measured with significantly better precision than before. The spin-parity assignments of both astrophysically important resonances, as well as the existence of a previously observed tentative state at $E_{x}$ $\approx$ 5.95 MeV are confirmed. The uncertainty in the rate of the $^{29}$P($p, \gamma$)$^{30}$S reaction is substantially reduced over the temperature range of interest. Finally, the influence of this rate on the abundance ratios of silicon isotopes synthesized in novae are obtained via 1D hydrodynamic nova simulations. \textbf{Conclusions:} The uncertainty in the $^{29}$P($p, \gamma$)$^{30}$S reaction rate is reduced to the point that it no longer affects the silicon isotopic abundance ratios significantly, and thus the results of our nova hydro simulation for the nucleosynthesis in the Si-Ca mass region are more reliable than before.
\end{abstract}


\maketitle


\section{\label{Astrophysics}Astrophysical Motivation}

\indent Classical nova outbursts are caused by explosive hydrogen burning as a result of a thermonuclear runaway in the envelope accreted from a main sequence star onto a white dwarf in a close semi-detached binary system. Simulations~\cite{Iliadis:2002} show that peak temperatures reached in the thermonuclear runaway are typically in the 0.1 -- 0.4 GK range, and the ejecta show significant nuclear processing. The dominant nuclear reaction flow proceeds close to the valley of stability on the proton-rich side and is dominated by a series of ($p, \gamma$) and ($p, \alpha$) reactions, as well as $\beta^{+}$-decays. Classical nova outbursts are thought to be the major source of $^{15}$N, $^{17}$O and to some extent $^{13}$C in the Galaxy~\cite{Jose:2006} and contribute to the abundances of other species with masses up to $A$ $\approx$ 40, including $^{26}$Al.\\
\indent The ejecta of classical novae are studied by systematic infrared observations~\cite{Gehrz:1998,Starrfield:2007} which reveal episodes of dust formation following a nova outburst. Several candidate presolar grains of nova origin have been found~\cite{Amari:2001,Amari:2002}, most of which are of silicon carbide (SiC) type. These grains show abundance anomalies for some isotopes (compared with the average solar system isotopic abundances), e.g., close to or slightly lower than solar $^{29}$Si/$^{28}$Si ratios and higher than solar $^{30}$Si/$^{28}$Si ratios~\cite{Jose:2004}.\\
\indent In order to reach a quantitative agreement between the isotopic abundances observed in the presolar grains~\cite{Amari:2002} and those predicted by simulations~\cite{Jose:2004}, nova nucleosynthesis models require some dilution. Thus, the mixing between the material in nova ejecta and the solar-like material must be understood to tighten the links between nova nucleosynthesis and presolar grains. Also, a better knowledge of the rates of the reactions that affect nova nucleosynthesis is required to better understand the origin of the isotopic ratios observed in the nova presolar grain candidates. Improving the reaction rates can also constrain nova models and simulations, and amend our understanding of nova nucleosynthesis~\cite{Starrfield:2007}.\\
\indent According to hydrodynamic classical nova simulations~\cite{Jose:2004}, the dominant nova nucleosynthetic path is sensitive to the chemical composition of the white dwarf, the extent to which convective mixing occurs between the material of the white dwarf's core and that of the envelope, and the thermal history of the envelope. Such questions can be partially answered via analysis of the Si isotopic abundance ratios ($^{29}$Si/$^{28}$Si and $^{30}$Si/$^{28}$Si) in SiC presolar grains of potential nova origin~\cite{Jose:2004}, and thus such ratios are of specific significance to this work's motivation.\\
\indent To explore and improve the silicon isotopic abundances in nova ejecta predicted from nova simulations, the thermonuclear reactions that most strongly affect the synthesis of silicon in novae must be determined and their rates understood. One such reaction is $^{29}$P($p, \gamma$)$^{30}$S. Over the temperature range characteristic of explosive nucleosynthesis in novae (0.1 -- 0.4 GK), the rate of the $^{29}$P($p, \gamma$)$^{30}$S reaction competes with that of $^{29}$P($\beta^{+}$) decay. If in this temperature range the $^{29}$P($p, \gamma$)$^{30}$S reaction rate is faster than the $^{29}$P($\beta^{+}$) decay rate, and if the $^{30}$P($\beta^{+}$) decay rate competes favorably with the rate of proton capture on $^{30}$P~\cite{Jose:2001}, the net effect is an increase in the production of $^{30}$Si via the $^{29}$P($p, \gamma$)$^{30}$S($\beta^{+}$)$^{30}$P($\beta^{+}$)$^{30}$Si reaction sequence, as well as a simultaneous decrease in the abundance of $^{29}$Si, which is the product of the $\beta^{+}$-decay of $^{29}$P. Therefore, an excess in $^{30}$Si together with the depletion in $^{29}$Si observed in some SiC presolar grains could indicate imprints of a nova origin. In a study on the sensitivity of nova nucleosynthesis to uncertainties in thermonuclear reaction rates~\cite{Iliadis:2002}, a change in the $^{29}$P($p, \gamma$)$^{30}$S rate by 10$^{4}$, which was consistent with the rate limits from Ref.~\cite{Iliadis:2001}, resulted in changes in $^{29,30}$Si abundances by a factor of 3.\\
\indent In the temperature range characteristic of explosive hydrogen burning (0.1 $\leq$ $T$ $\leq$ 1.3 GK), the Gamow window of the $^{29}$P($p, \gamma$)$^{30}$S reaction spans $E_{cm} \approx$ 700 -- 1770 keV, where there is a low level density. Thus, the rate depends on the properties of isolated and narrow $^{29}$P + $p$ resonances corresponding to $^{30}$S (\emph{t$_{\tiny{1/2}}$} $=$ 1175.9(17) ms~\cite{Souin:2011}) proton unbound states with 4.5 $ \lesssim E_{x} \lesssim$ 6 MeV.\\
\indent The $^{29}$P($p, \gamma$)$^{30}$S rate was evaluated by Wiescher and G$\ddot{\mbox{o}}$rres~\cite{Wiescher:1988}, and more recently by Iliadis {\em et al.}~\cite{Iliadis:2001,Iliadis:2010a} and Bardayan {\em et al.}~\cite{Bardayan:2007}. The rate calculated by Iliadis {\em et al.}~\cite{Iliadis:2001} was found to be dominated by the 3$_{1}^{+}$ and 2$_{3}^{+}$ proton unbound states in $^{30}$S. The excitation energies corresponding to these two unobserved resonances were predicted~\cite{Iliadis:2001} using the Isobaric Multiplet Mass Equation (IMME) to be 4733(40) keV and 4888(40) keV for the states with $J^{\pi}$ $=$ 3$^{+}$ and $J^{\pi}$ $=$ 2$^{+}$, respectively. Such large uncertainties in the resonance energies, $E_{r}$, resulted in an uncertainty in the rate which spanned $\sim$4 orders of magnitude~\cite{Iliadis:2001}. Prior to this prediction, several experiments had been performed to study the structure of $^{30}$S~\cite{Paddock:1972,Caraca:1972,Kuhlmann:1973,Yokota:1982,Fynbo:2000}. However, the two astrophysically important states predicted by Iliadis {\em et al.}~\cite{Iliadis:2001} were not observed in any of the previous experiments.\\
\indent A direct measurement of the $^{29}$P($p, \gamma$)$^{30}$S reaction is currently not feasible because no $^{29}$P radioactive ion beam with the required beam intensity ($>$ 10$^{8}$ pps) is available. Thus, following the prediction by Iliadis {\em et al.}~\cite{Iliadis:2001}, attempts were made to find these two states via indirect methods~\cite{Galaviz:2006,Bardayan:2007,Figueira:2008,O'Brien:2009,Galaviz:2010}. Bardayan {\em et al.}~\cite{Bardayan:2007} remeasured the excitation energies and spin-parity assignments of the states of $^{30}$S up to 7.1 MeV by means of the $^{32}$S($p, t$)$^{30}$S two-nucleon transfer reaction. As a result, a state at 4704(5) keV was discovered and was proposed to be the predicted 3$^{+}_{1}$ state. However, no trace of the other important level was found.\\
\indent Shortly thereafter, we performed two separate experiments, each with two phases, to determine the excitation energies and spin-parity assignments of several states of $^{30}$S, which were populated via the $^{32}$S($p, t$)$^{30}$S and $^{28}$Si($^{3}$He, $n$$\gamma$)$^{30}$S two nucleon transfer reactions.\\
\indent In Ref.~\cite{Setoodehnia:2010}, the resonance energies corresponding to six proton unbound states with $E_{x}$ $<$ 5.5 MeV in $^{30}$S were presented, including both astrophysically important states predicted by Iliadis {\em et al.}~\cite{Iliadis:2001} one of which was observed for the first time. Since then, we have performed a new $^{32}$S($p, t$)$^{30}$S measurement with a different target (phase II), and have improved upon the analysis of the existing data. Phase I of our $^{28}$Si($^{3}$He, $n$$\gamma$)$^{30}$S experiment was performed with the sole purpose of determining via $\gamma$-ray coincidence measurements the energies of the two important resonances predicted by Iliadis {\em et al.}~\cite{Iliadis:2001}, and phase II was carried out to measure the $\gamma$-ray angular distributions and $\gamma$-$\gamma$ angular correlations from oriented nuclei to infer the spins of the observed $^{30}$S states. The results of phase I of our $^{28}$Si($^{3}$He, $n$$\gamma$)$^{30}$S experiment are also published~\cite{Setoodehnia:2011a}.\\
\indent The present work discusses in detail the experimental setups and data analyses for the second phases of our $^{32}$S($p, t$)$^{30}$S and $^{28}$Si($^{3}$He, $n$$\gamma$)$^{30}$S experiments, and presents our unpublished data for the first phase of our $^{32}$S($p, t$)$^{30}$S experiment. This work thus presents our combined final results on the energies and spin-parity assignments of the observed $^{30}$S states, the most updated $^{29}$P($p, \gamma$)$^{30}$S reaction rate calculated via a newly developed Monte Carlo method, as well as the impact of this rate on the abundance ratios of silicon isotopes synthesized in novae. Therefore, the results in the present paper supersede those of our previous publications~\cite{Setoodehnia:2010,Setoodehnia:2011a}.

\section{\label{Experiment}Experiments}
\subsection{\label{pt_Experiment}The \boldmath$^{32}$S($p, t$)$^{30}$S experiment}
\subsubsection{\label{pt_Experiment_Setup}Experimental setup and data analysis: phase II}

\begin{figure}[ht]
\centering\vspace{-0.6cm}
  \includegraphics[width=0.46\textwidth]{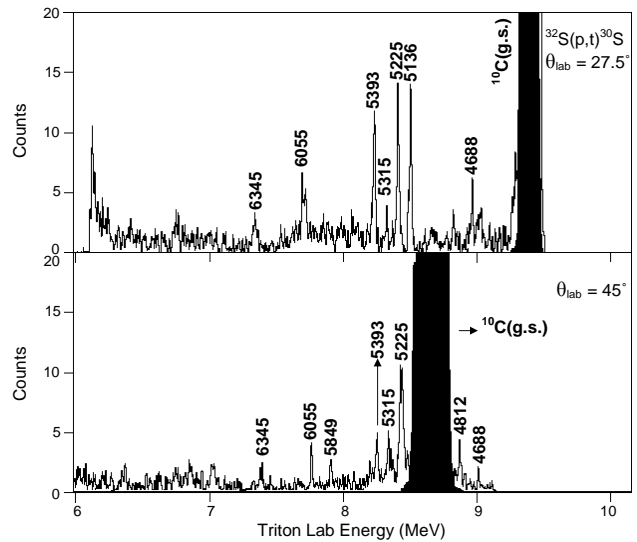}\\
  \caption{\label{figure1}Triton spectra from the $^{32}$S($p, t$)$^{30}$S reaction obtained with the implanted target. Peaks corresponding to $^{30}$S states are labeled with energies in keV. The filled histograms are background spectra measured with an isotopically enriched $^{12}$C target, normalized to the $^{32}$S($p, t$)$^{30}$S data. The main contaminant is the ground state (g.s.) of $^{10}$C. For 27.5$^{\circ}$, an aluminum plate along the focal plane blocked the region corresponding to tritons with energies higher than 9.5 MeV, where elastically scattered protons reached the focal plane. At 45$^{\circ}$ the gates cut the region to the right of the peak corresponding to the 4688-keV state.}
\end{figure}

\begin{figure}[ht]
\begin{center}
  \includegraphics[width=0.46\textwidth]{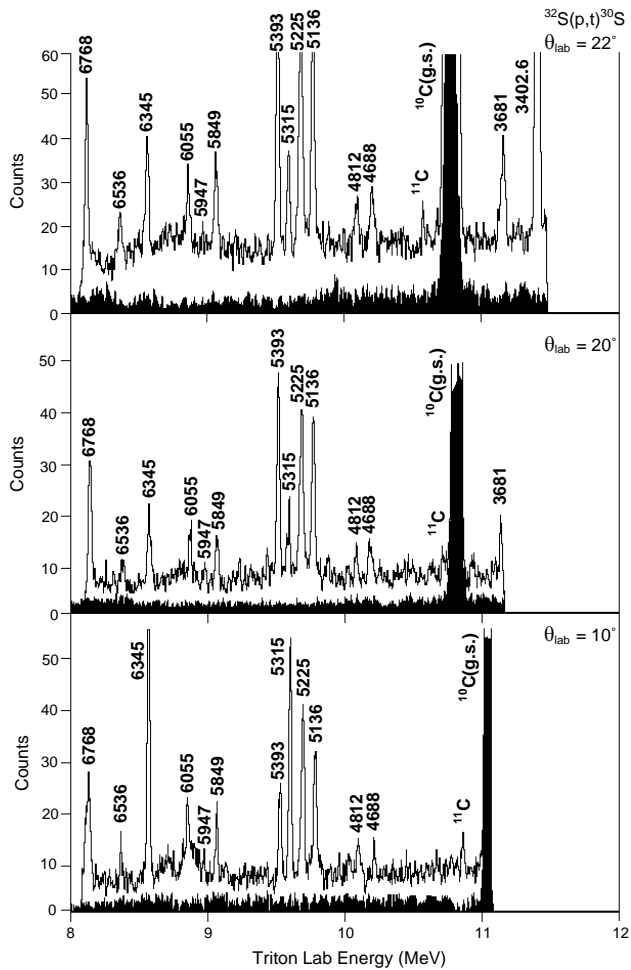}\\
\end{center}
  \caption{\label{figure2}Triton spectra measured (in phase I) from the $^{32}$S($p, t$)$^{30}$S reaction at selected angles obtained with the CdS target (for details, see Ref.~\cite{Setoodehnia:2010}). Peaks corresponding to $^{30}$S states are labeled with energies in keV. The states with $E_{x}$ $>$ 5.5 MeV were not published in Ref.~\cite{Setoodehnia:2010}. The filled histograms are background spectra measured with a $^{nat}\!$Cd target on a carbon backing, normalized to the $^{32}$S($p, t$)$^{30}$S data. A peak from the $^{13}$C($p, t$)$^{11}$C reaction is also identified and labeled by its parent nucleus. For 10$^{\circ}$ and 20$^{\circ}$, an aluminum plate along the focal plane blocked the region corresponding to tritons whose energies are higher than 11 MeV, where elastically scattered protons reached the focal plane.}
\end{figure}

\indent The experiment was performed at the Wright Nuclear Structure Laboratory (WNSL) at Yale University. A proton beam was accelerated, using the ESTU tandem Van de Graaff accelerator, to 34.5 MeV ($\Delta$$E$/$E$ $\sim$ 6 $\times$ 10$^{-4}$)~\cite{Setoodehnia:2010,Setoodehnia:2011b}.\\
\indent The beam impinged on a 55.9 $\pm$ 5.6 $\mu$g/cm$^{2}$ isotopically pure (99.9\% enriched) $^{12}$C foil implanted with 10.4 $\pm$ 0.4 $\mu$g/cm$^{2}$ of $^{32}$S. This target was fabricated specifically to reduce the relatively flat background produced by the $^{nat}\!$Cd, where $nat$ refers to natural, component of the CdS target used in phase I of our $^{32}$S($p, t$)$^{30}$S experiment~\cite{Setoodehnia:2010}. The production procedure for the implanted target is described elsewhere~\cite{Setoodehnia:2011b,Lennard:2011}. The thicknesses of the $^{32}$S and $^{12}$C layers in the implanted target were obtained through a Rutherford backscattering measurement~\cite{Setoodehnia:2011b,Lennard:2011}.\\
\indent In addition to the aforementioned target, a free-standing 311-$\mu$g/cm$^{2}$ natural Si foil was used for calibration purposes. Also, a stand alone 40-$\mu$g/cm$^{2}$-thick 99.9\% isotopically enriched $^{12}$C foil was used to measure the background from ($p, t$) reactions on the carbon substrate in the implanted target. The method of measuring the thicknesses of these targets is described in Ref.~\cite{Setoodehnia:2011b}.\\
\indent The reaction ejectiles were dispersed according to their momenta with an Enge split-pole magnetic spectrograph, with vertical and horizontal aperture settings of $\Delta\phi = \pm$40 mrad, and $\Delta\theta = \pm$30 mrad, respectively. The study was carried out at multiple angles with magnetic field strengths of 10 kG for $\theta\,=$ 22${^\circ}$; 9.5 kG for $\theta\,=$ 27.5${^\circ}$; and 9.2 kG for $\theta\,=$ 45${^\circ}$, where $\theta$ is the scattering angle in the laboratory system.\\
\indent The tritons were focused at the spectrograph's focal plane, where they were detected with an isobutane-filled position sensitive ionization drift chamber~\cite{Setoodehnia:2011b}, together with a plastic scintillator. The ionization chamber measured the positions along the focal plane and energy losses ($\Delta E$) of the tritons. Those that passed through this detector deposited their residual energy ($E_{res}$) in the plastic scintillator.\\
\indent $\Delta E$, $E_{res}$ and position (proportional to momentum) were measured to identify tritons and determine their momenta. The tritons were selected according to $\Delta E$ and $E_{res}$, which were plotted vs.~focal plane position gates. The spectra of the tritons' momenta were then plotted for each spectrograph angle (see Fig.~\ref{figure1}). Triton peaks corresponding to $^{30}$S states in these spectra were clearly identified through kinematic analysis.\\
\indent The only contaminant peak observed was the ground state of $^{10}$C (see Fig.~\ref{figure1}). The $^{32}$S implanted target produced a background that was decreased by about a factor of 2 compared with the relatively flat background produced by the $^{nat}\!$Cd component of the CdS target used in phase I of the $^{32}$S($p, t$)$^{30}$S experiment.\\
\indent The triton peaks observed in the presented spectra were fitted using a least-squares multi-Gaussian fit function to determine the peak centroids, widths and areas. The energy cali-

\noindent bration was determined from a combination of known levels of $^{26}$Si (measured with the $^{28}$Si($p, t$) reaction using the Si-target) and of $^{30}$S, whose adopted energies are weighted averages of previous work on $^{26}$Si levels~\cite{Bell:1969,Paddock:1972,Bohne:1982,Caggiano:2002,Bardayan:2002,Parpottas:2004,Seweryniak:2007,Matic:2010} and on $^{30}$S~\cite{Paddock:1972,Caraca:1972,Kuhlmann:1973,Yokota:1982}. Since the earlier publication~\cite{Setoodehnia:2010}, the previous calibration fits were improved through reanalysis of the previous data (corresponding to phase I) by accounting for the angle of the target with respect to the beam (details are provided in Ref.~\cite{Setoodehnia:2011b}). Fig.~\ref{figure2} presents the $E_{x}$ $>$ 5.5 MeV excited states in $^{30}$S observed in phase I of the $^{32}$S($p, t$)$^{30}$S experiment that were not published in Ref.~\cite{Setoodehnia:2010}.\\
\indent The final excitation energy uncertainties for the data of both phases of this experiment arise from: (1) statistical uncertainties ($\leq$ 2 keV), (2) uncertainties in the thicknesses of the implanted target (1 keV) and the CdS target (2 keV) taking into account the uncertainty in the thickness of the $^{nat}\!$Si target used for calibration, and (3) uncertainty in the $Q$-values of the $^{28}$Si($p, t$)$^{26}$Si and $^{32}$S($p, t$)$^{30}$S reactions (0.3 keV~\cite{Kwiatkowski:2010} and 0.4 keV~\cite{Souin:2011}, respectively). Therefore, the $^{30}$S excitation energy uncertainties, when added in quadrature, were 3 keV and 2 keV for the CdS and implanted targets, respectively. Lastly, to obtain the final $^{30}$S excitation energies, a weighted average was calculated for each state over all the angles, and thus over both targets.\\
\indent With respect to the previous publication~\cite{Setoodehnia:2010}, all the measured excitation energies from the present work have smaller uncertainties by at least 40\% as a result of a reduction in the uncertainty of the $Q$-value of the $^{32}$S($p, t$)$^{30}$S reaction due to a recent improved measurement~\cite{Souin:2011} on the $^{30}$S mass, and our

\noindent improved calibration fits for the previous data obtained by the CdS target.\\
\indent The energy resolution was approximately 28 keV and 22 keV (FWHM) for the spectra obtained with the CdS and implanted targets, respectively. Therefore, our achieved energy resolution is a factor of 3 -- 5 smaller than those of previous $^{32}$S($p, t$)$^{30}$S measurements~\cite{Paddock:1972,Bardayan:2007}.

\subsubsection{\label{pt_Experiment_Results}Results: both phases combined}

\indent Over both phases of the $^{32}$S($p, t$)$^{30}$S experiment, 12 proton unbound states of $^{30}$S with $E_{\tiny{x}} <$ 6.8 MeV were observed, and
\begin{table}[ht]
\vspace{-0.7cm}
\caption{Weighted average excitation energies of $^{30}$S from both phases of our $^{32}$S($p, t$)$^{30}$S experiment. States used for energy calibration are marked by an asterisk.}
\begin{minipage}{\linewidth}
\renewcommand\thefootnote{\thempfootnote}
\centering
\begin{ruledtabular}
\begin{tabular}{clcl}
$E_{x}$ (keV) & $J^{\pi}$ & $E_{x}$ (keV) & $J^{\pi}$ \\
\midrule[0.05em]\addlinespace[0.2mm]
2208(3) &  & 5393(2) & 3$^{+}$ \\
3402.6$^{\ast}$ &  & 5849(2) & (1$^{-}$, 2$^{+}$, 4$^{+}$) \\
3681(3) & (1$^{+}$, 0$^{+}$) & 5947(2) &  \\
4688(2) & 3$^{+}$ & 6055(3) & (1$^{-}$) \\
4812(2) & 2$^{+}$ & 6345(3) & (0$^{+}$) \\
\hspace{-2mm}5136$^{\ast}$ & (4$^{+}$) & 6536(3) & (2, 3) \\
5225(2) & (0$^{+}$) & 6768(3) & 2$^{(-)}$ \\
5315(2) & (3$^{-}$, 2$^{+}$) & & \\[0.2ex]
\end{tabular}
\end{ruledtabular}
\end{minipage}
\label{tab:1}
\end{table}

\noindent their weighted average energies (over all angles) are listed in Table~\ref{tab:1}.\\
\indent Most of the measured energies in the present work are in agreement within 1 -- 2$\sigma$ with those measured in the previous $^{32}$S($p, t$)$^{30}$S measurements~\cite{Paddock:1972,Bardayan:2007}. The 5947-keV state observed in the present work (see Fig.~\ref{figure2}) is in good agreement with the 5945-keV tentative level observed in Ref.~\cite{Fynbo:2000}. For the state with an expected excitation energy of $E_{x}$ $\approx$ 4.7 MeV~\cite{Iliadis:2001}, our measured energy of 4688(2) keV does not agree with the 4704(5) keV measured in Ref.~\cite{Bardayan:2007}. Most of the levels observed in our $^{32}$S($p, t$)$^{30}$S experiments whose $E_{x}$ $>$ 5 MeV have been measured previously but have spin-parity assignments that are either unknown or tentative.\\
\indent To obtain the spin-parity assignments of $^{30}$S states observed in both phases of the $^{32}$S($p, t$)$^{30}$S experiment, the equivalent thickness of the sulfur content of the CdS target was required. This thickness was determined to be 53 $\pm$ 5 $\mu$g/cm$^{2}$ through the reanalysis of the data of a previous scattering experiment~\cite{Wrede:2010}, where an 8-MeV $^{4}$He$^{+}$ beam along with the Enge spectrograph at WNSL and a silicon surface barrier detector were used to determine the composition and thickness of the CdS target. The theoretical angular distributions of the cross sections were then computed, following the work of Ref.~\cite{Bardayan:2007}, using (i) the one-step finite range transfer formalism via {\fontfamily{pcr}\selectfont\small DWUCK5}~\cite{DWUCK} for the natural-parity states, and (ii) the coupled reaction channels formalism via {\fontfamily{pcr}\selectfont\small FRESCO}~\cite{Thompson:1988} (under the assumption of finite-range interaction potential) for the unnatural-parity states. {\fontfamily{pcr}\selectfont\small DWUCK5} is unable to handle unnatural-parity states. {\fontfamily{pcr}\selectfont\small FRESCO}, on the other hand, can be used for all states. The former code was used for the cases of natural-parity states due to its simplicity. Its output for a few natural-parity states was examined against that of {\fontfamily{pcr}\selectfont\small FRESCO}, and the results were identical in terms of the shapes of the theoretical DWBA curves.\\
\indent To determine the orbital angular momenta that were transferred in the $^{32}$S($p, t$)$^{30}$S reaction to populate the observed $^{30}$S excited states, the same shell-model assumptions made in Ref.~\cite{Bardayan:2007} are made here. The angular distributions for the sequence $^{32}$S($p, d$)$^{31}$S$_{g.s.}$($d, t$)$^{30}$S were obtained for each of the unnatural-parity final states in $^{30}$S.\\
\indent The distorted waves in the entrance and exit channels were calculated for optical interaction potentials, the parameters of which were taken from Ref.~\cite{Bardayan:2007} (and references therein), and

\begin{table*}[ht]\small
\caption{Optical model parameters used in conjunction with {\fontfamily{pcr}\selectfont\small DWUCK5} and {\fontfamily{pcr}\selectfont\small FRESCO} for the analysis of the angular distributions.}
\centering
\begin{minipage}{\linewidth}
\renewcommand\thefootnote{\thempfootnote}
\setlength{\tabcolsep}{6pt} 
\begin{tabular}{cccccccccccccc}
\toprule[1.0pt] \\ [0.2ex] Reaction & $V_{0}$ & $W_{0}$ & $W_{D}$ & $V_{s}$\footnote{This parameter, as well as $r''_{0}$ and $a''$, is taken from Ref.~\cite{Paddock:1972}.} & $r_{0}$ & $a$ & $r\sp{\prime}_{0}$ & $a\sp{\prime}$ & $r''_{0}$ & $a''$ & $r_{0c}$ & $\lambda$ & PNLOC \\[0.1ex]
Channel & (MeV) & (MeV) & (MeV) & (MeV) & (fm) & (fm) & (fm) & (fm) & (fm) & (fm) & (fm) & (fm) & \\ [1ex] \hline\hline \\ [0.2ex]
$p$ $+$ $^{32}$S & \hspace{1.5mm}37.1 & 0 & 6.875 & 7.5 & 1.18 & 0.66 & 1.18 & 0.66 & 1.18 & 0.7 & 1.25 &  &  \\
$t$ $+$ $^{30}$S & 144 & 30 & 0 & 0 & 1.24 & 0.68 & 1.45 & 0.84 & 0 & 0 & 1.25 &  &  \\
$d$ $+$ $^{31}$S & \hspace{-1mm}90 & 0 & 25 &  & 1.30 & 0.62 & 1.18 & 0.58 &  &  & 1.25 &  &  \\
\hspace{0.1cm}$n$ $+$ $^{31}$S\footnote{The input parameters corresponding to this channel are taken from Ref.~\cite{Tribble:1977}.} &  & 0 & 0 &  & 1.20 & 0.65 &  &  &  &  & 1.30 & 25 & 0.85 \\
$2n$ $+$ $^{30}$S &  & 0 & 0 &  & 1.25 & 0.65 &  &  &  &  &  & 25 & \\[1ex]
\bottomrule[1.0pt]
\end{tabular}
\end{minipage}
\label{tab:2}
\end{table*}

\begin{figure*}[ht]
  \begin{center}\vspace{0.4cm}
  \includegraphics[width=\textwidth]{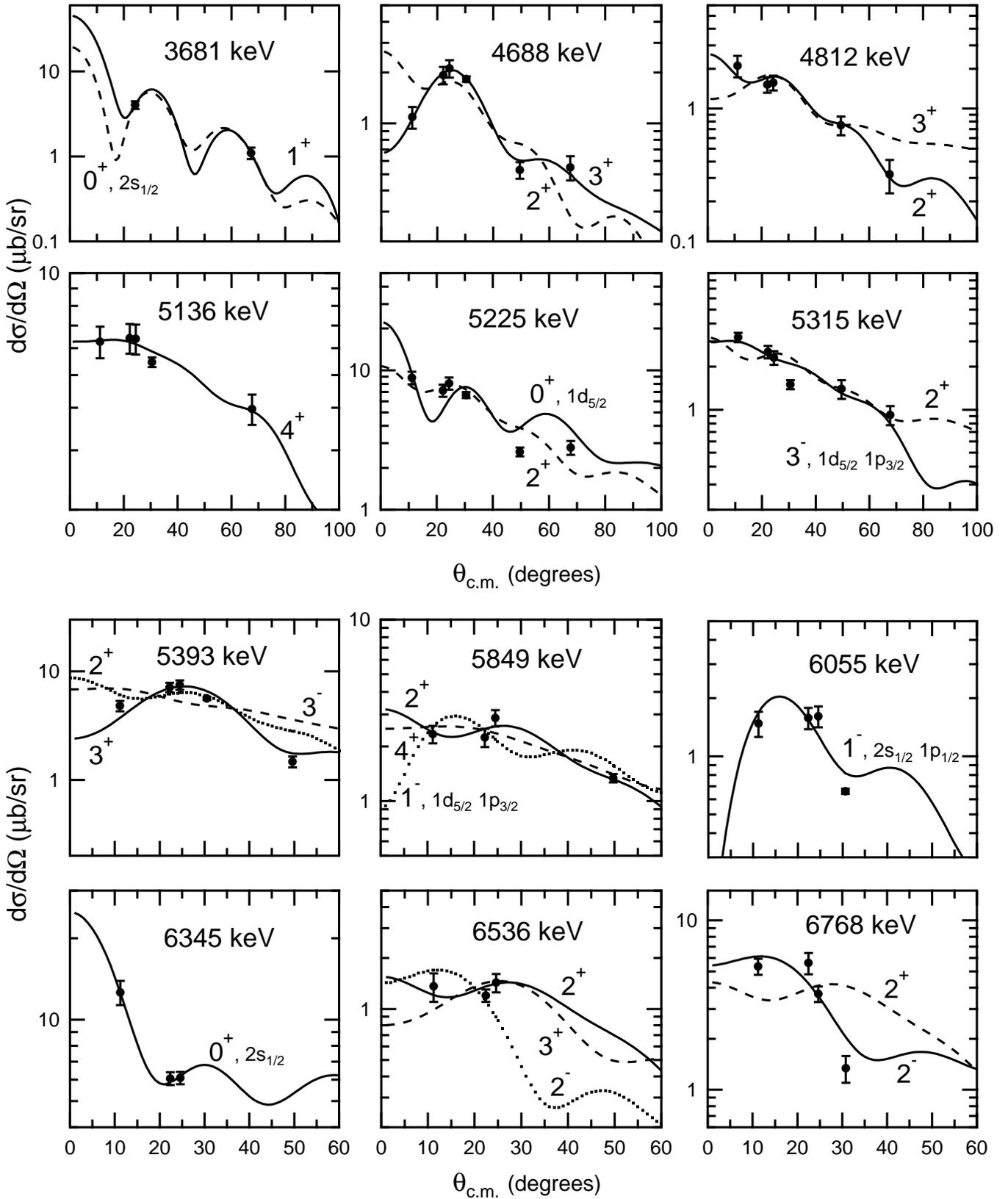}\\
  \end{center}
  \caption{\label{figure3}Triton angular distributions populating states of $^{30}$S compared with the DWBA curves obtained by {\fontfamily{pcr}\selectfont\small DWUCK5} and {\fontfamily{pcr}\selectfont\small FRESCO} for natural- and unnatural-parity levels, respectively. The filled circles with error bars are the measured differential cross sections in the center-of-mass system, and the solid, dashed or dotted curves are the theoretical angular distributions. If not shown, the error bar is smaller than the point size. The excitation energies are given on the top middle of each plot. The shell-model orbital from which the di-neutron is stripped is given for the $0^{+}$ and the negative natural-parity levels.}
\end{figure*}

\noindent are given in Table~\ref{tab:2}.\\
\indent Furthermore, the widely used Reid soft core potential~\cite{Reid:1968} was used to derive the deuteron and triton wave functions, as well as the p-n and d-n interactions for {\fontfamily{pcr}\selectfont\small FRESCO} calculations.\\
\indent Triton angular distribution plots are shown in Fig.~\ref{figure3}. Angular distributions of the states with $E_{x}$ $\leq$ 5.136 MeV are discussed in \S~\ref{spin-Parity Assignments}, since those states were also observed in our $\gamma$-ray measurements. In the following, we will only discuss the angular distributions for $^{30}$S levels with $E_{x}$ $\geq$ 5.225 MeV.\\
\indent $\bullet$ \textit{The 5225-keV level:} This state is a prominent peak that was observed at every angle measured in the $^{32}$S($p, t$)$^{30}$S experiments. There is no conclusive information regarding the $J^{\pi}$ assignment of this state in the literature. Our only guide comes from a shell-model calculation~\cite{Wiescher:1988}, which suggested that there should be a 0$^{+}$ level around 5.2 MeV. Although reasonable fits are obtained with $J^{\pi} =$ 0$^{+}$ and 2$^{+}$, our fit using $J^{\pi} =$ 0$^{+}$ is more consistent with the data. So we suggest that this state is the mirror to the 0$^{+}_{2}$ state in $^{30}$Si at 5372.2 keV~\cite{Basunia:2010}.\\
\indent $\bullet$ \textit{The 5315-keV level:} This state is also a prominent peak observed at all angles. It is known to be a 3$^{-}$ state~\cite{Yokota:1982}. Our angular distribution is better fitted by an \textit{l} $=$ 3 angular momentum transfer, but \textit{l} $=$ 2 would also be reasonably consistent (see Fig.~\ref{figure3}). We suggest that this state is most likely the mirror to the 5487.5-keV state in $^{30}$Si with $J^{\pi}$ $ =$ 3$^{-}$~\cite{Basunia:2010}.\\
\indent $\bullet$ \textit{The 5393-keV level:} This state was observed at all angles measured in the $^{32}$S($p, t$)$^{30}$S experiments. Its spin was tentatively assigned to be $J =$ 1 or 2 in previous work~\cite{Yokota:1982}. In Ref.~\cite{Bardayan:2007}, tentative $J^{\pi}$ $=$ 3$^{-}$ and 2$^{+}$ assignments were given to this state. Our triton angular distribution is more consistent with $J^{\pi}$ $=$ 3$^{+}$ assignment, and thus we assign this state to be 3$^{+}$, making it the mirror to the 3$^{+}_{2}$ state in $^{30}$Si at 5231.38 keV~\cite{Basunia:2010}.\\
\indent $\bullet$ \textit{The 5849-keV level:} This state was tentatively assigned to be a 1$^{-}$ state in Ref.~\cite{Bardayan:2007}. However, \textit{l} $=$ 2, 3 and 4 transfers could not be excluded. In our data, this level was observed at 10${^\circ}$, 20${^\circ}$, 22${^\circ}$ and 45${^\circ}$. We can rule out $J^{\pi}$ $=$ 4$^{-}$ and 2$^{-}$ assignments but 1$^{-}$, 2$^{+}$ and 4$^{+}$ are all in reasonable agreement with our data (see Fig.~\ref{figure3}).\\
\indent $\bullet$ \textit{The 5947-keV level:} This level was too weakly populated to extract a significant angular distribution.\\
\indent $\bullet$ \textit{The \boldmath$E_{x}$ $>$ 6 MeV states:} With the exception of the 6055-keV and 6768-keV states, which are observed at four angles (see Fig.~\ref{figure3}), all other states of $^{30}$S observed in the present experiment whose excitation energies are above 6 MeV are only observed at most at three angles, 10${^\circ}$, 20${^\circ}$ and 22${^\circ}$ (see Fig.~\ref{figure2}). Nonetheless, we propose a tentative assignment of 1$^{-}$ to the 6055-keV state, which is consistent with the results of Ref.~\cite{Yokota:1982}, but the energy of this state from our data differs by 62 keV. We also make a tentative assignment of 0$^{+}$ to the 6345-keV state consistent with a definite $J^{\pi}$ $=$ 0$^{+}$ assignment made in Ref.~\cite{Yokota:1982}. Furthermore, we tentatively assign $J =$ 2 or 3 to the 6536-keV state, which is consistent with what was suggested in Ref.~\cite{Yokota:1982}. Lastly, for the 6768-keV state, we confirm $J =$ 2 suggested in Ref.~\cite{Bardayan:2007}, and likely rule out \textit{l} $=$ 3 and 4 transfers. Though, our angular distribution data are best fitted with a negative parity assignment. Therefore, we propose a $J^{\pi}$ $=$ 2$^{(-)}$ assignment to this state.

\subsection{\label{ng_Experiment}The \boldmath$^{28}$Si($^{3}$He, $n\gamma$)$^{30}$S experiment}
\subsubsection{\label{ng_Experiment_Setup}Experimental setup and data analysis: phase II}

\begin{figure}[ht]
\vspace{0.4cm}
\begin{center}
  \includegraphics[width=0.47\textwidth]{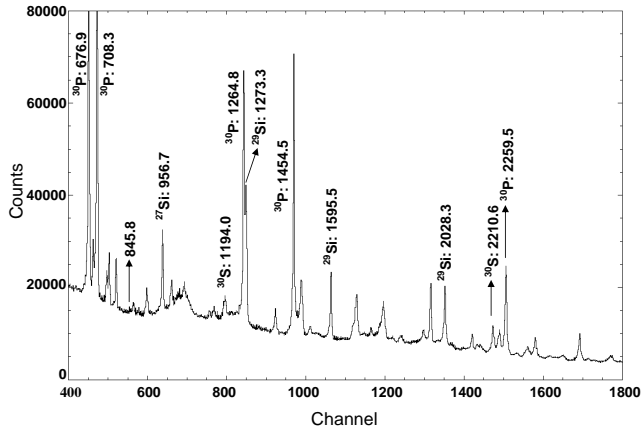}\\
  \end{center}
  \caption{\label{figure4}Singles $\gamma$-ray spectrum measured during phase II of the experiment at 90$^{\circ}$ using detector 1. Selected strong transitions are labeled by their parent nuclei and with energies (in keV) that are weighted averages between both phases of the experiment. The 2210.6-keV (2$_{1}^{+}$ $\rightarrow$ 0$_{1}^{+}$) and 1194-keV (2$_{2}^{+}$ $\rightarrow$ 2$_{1}^{+}$) peaks originate from levels in $^{30}$S.}
\end{figure}

\indent An in-beam $\gamma$-ray spectroscopy experiment using the $^{28}$Si($^{3}$He, $n\gamma$)$^{30}$S reaction was carried out to assign spins to the populated $^{30}$S levels based on measurements of $\gamma$-ray angular distributions and $\gamma$-$\gamma$ angular correlations from oriented nuclei. This experiment was performed at the University of Tsukuba Tandem Accelerator Complex (UTTAC) in Japan. A $^{3}$He$^{2+}$ beam was accelerated to 9 MeV via the $12$UD Pelletron tandem accelerator at UTTAC. The details of this beam are described in Ref.~\cite{Setoodehnia:2011a}. The beam impinged on a self-standing 25 $\mu$m-thick foil of $^{nat}\!$Si, of which the $^{28}$Si abundance is 92.23\%.\\
\indent High-purity germanium detectors with 50\% and 70\% relative efficiency were placed at 90${^\circ}$ and 135${^\circ}$ with respect to the beam axis, respectively. We hereafter refer to these detectors as 1 and 2, respectively. These detectors were located on opposite sides with respect to the beam line. The energy resolution of detectors 1 and 2 was determined to be 4.4 keV and 3.2 keV (FWHM) at $E_{\gamma} =$ 1333 keV, respectively. $\gamma$-$\gamma$ coincidence data were accumulated during a total of 4 days, and was corrected hourly for detector gain shifts. A sample $\gamma$-ray spectra can be seen in Fig.~\ref{figure4}.\\
\indent To extract the centroid and area of each peak, the peaks

\begin{figure}[ht]
  \centering\vspace{0.5cm}
  \subfloat[$\gamma$-$\gamma$ coincidence spectrum measured at 90${^\circ}$]
  {\label{figure5a}%
    \includegraphics[width=8cm]{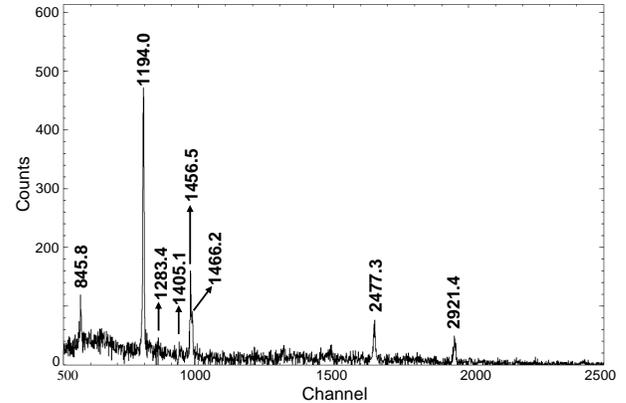}}%
    \vspace{0.1cm}
  \quad%
  \subfloat[$\gamma$-$\gamma$ coincidence spectrum measured at 135${^\circ}$]
  {\label{figure5b}%
    \includegraphics[width=8cm]{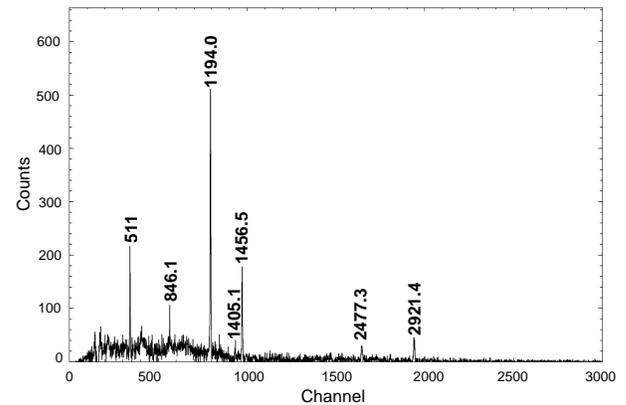}}
  \quad%
\caption{\label{figure5}The $\gamma$-$\gamma$ coincidence spectrum measured during phase II at (top) 90${^\circ}$ and (bottom) 135${^\circ}$ obtained from gating on the 2210.6-keV (2$_{1}^{+}$ $\rightarrow$ 0$_{1}^{+}$) transition of $^{30}$S. Peaks corresponding to the transitions from known $^{30}$S states are labeled with energies (in keV). At 135${^\circ}$, the labeled energies are corrected for Doppler shift except that of the 846.1-keV $\gamma$-ray (see text). The 2477.1-keV and 2599.5-keV peaks are from the decays of proton-unbound states at 4688.0 keV and 4810.4 keV, respectively.}
\end{figure}

\indent were fitted using a single-Gaussian function whenever they were reasonably isolated from each other, and with a multi-Gaussian function for the partially resolved or unresolved doublets. Those peaks that were affected by Doppler shift at higher angles were fitted using Gaussian-plus-exponential functions to account for the exponential tail. Background subtraction was performed by assuming a linear function under each peak.\\
\indent The Ge-detectors' initial energy calibration and energy-dependent efficiencies were determined with a standard $^{152}$Eu calibration source. The initial energy calibration fit was improved via internal calibration by using strong $^{30}$P $\gamma$-rays emitted from the $^{28}$Si($^{3}$He, $p\gamma$)$^{30}$P reaction, whose cross section is higher than that of the $^{28}$Si($^{3}$He, $n\gamma$)$^{30}$S reaction at this beam energy~\cite{Groenevbld:1970,Bass:1972}. The resulting uncertainties in the detection efficiencies were estimated to be 5\%.\\
\indent The coincidence analysis was performed via construction of a $\gamma$-$\gamma$ coincidence matrix. Fig.~\ref{figure5} presents the coincidence spectra. The $\gamma$-ray angular distribution and $\gamma$-$\gamma$ angular correlation measurements and their results will be discussed in \S~\ref{angular_distribution} and \S~\ref{angular_correlation}, respectively.

\subsubsection{\label{ng_Experiment_Results}Results: both phases combined}
\paragraph{\label{decay_scheme}\textit{Decay scheme of $^{30}$S}}

\indent\indent In the singles $\gamma$-ray spectra of both Ge-detectors during each phase of the experiment, two $\gamma$-rays were clearly observed at 2210.6(3) keV and 1194.0(1) keV, which correspond to the 2$_{1}^{+}$ $\rightarrow$ 0$_{1}^{+}$ and 2$_{2}^{+}$ $\rightarrow$ 2$_{1}^{+}$ transitions in $^{30}$S, respectively (see Fig.~\ref{figure4}).\\
\indent A few $\gamma$-rays with energies in the range of 3 MeV were expected to be observed in the singles spectra according to the measured branching ratios~\cite{Caraca:1972,Kuhlmann:1973} of the $\gamma$-rays from decays of the bound states and the lowest-lying resonances of $^{30}$S. However, these $\gamma$-rays did not appear as separate observable peaks in the singles $\gamma$-ray spectra obtained during either phase of the experiment. This was most likely because they were obscured by the Compton scattered $\gamma$-rays from $^{30}$P transitions.\\
\indent After placing software gates on the 2210.6- and 1194.0-keV peaks, $\gamma$-decay cascades from higher-lying states were observed in the $\gamma$-$\gamma$ coincidence spectra (see Fig.~\ref{figure5}). In particular, we observed transitions with energies of 2477.3(3) keV, 2599.0(4) keV (see Fig.~2 in Ref.~\cite{Setoodehnia:2011a}) and 2921.4(4) keV from $^{30}$S proton-unbound states at 4688.1(4) keV, 4809.8(5) keV and 5132.3(5) keV, respectively.\\
\indent Recoil energies were taken into account when constructing the final excitation energies of $^{30}$S from its $\gamma$-ray decay scheme. The results are given in Table~\ref{tab:3}. The final uncertainties in the energies are due to the statistical uncertainties

\begin{table*}[ht]\small
\caption{Weighted average energies (between both phases of the experiment) and relative intensities of the observed transitions in $^{30}$S. The latter are calculated with respect to the strongest $\gamma$-ray measured at the same angle. The uncertainties in the recoil energies ($E_{recoil}$) were negligible, and thus are not presented. The energies of initial and final states ($E_{i}$ and $E_{f}$, respectively) are corrected for the corresponding recoil energies. The results obtained in the $\gamma$-ray measurement of Ref.~\cite{Kuhlmann:1973} are also shown for comparison.}
\begin{minipage}{\linewidth}
\renewcommand\thefootnote{\thempfootnote}
\centering
\setlength{\tabcolsep}{6pt} 
\begin{tabular}{ccccclllllll}
\toprule[1.0pt] \\ [0.2ex]
\multicolumn{7}{c}{Present Work} & \phantom{ab} & \multicolumn{3}{c}{Ref.~\cite{Kuhlmann:1973}} \\
\cmidrule[0.05em]{1-7} \cmidrule[0.05em]{9-11}\addlinespace[0.3mm]
Gate\footnote{The transition on which the coincidence gate is placed.} & $E_{\gamma}$ & $E_{recoil}$ & $E_{i}$ & $E_{f}$ & $I^{90{^\circ}}_{\gamma}$ & $I^{135{^\circ}}_{\gamma}$ && $E_{\gamma}$ & $E_{i}$ & $E_{f}$ \\
 & (keV) & (keV) & (keV) & (keV) & (\%) & (\%) && (keV) & (keV) & (keV) \\ [1ex]
\midrule[0.05em]
2210.6(3) & 846.0(4) & 0.01 & unplaced & unplaced & 3.9(6) & 2.8(5) &&  &  &  \\
 & 1194.0(1) & 0.03 & 3404.7(3) & 2210.7(3) & 33.5(5) & 43.3(10) && 1192.0(5) & 3402.6(13) & 2210.7(5) \\
1194.0(1) & \hspace{1.6mm}1283.4(3)\footnote{This transition is not observed at 135$^{\circ}$.} & 0.03 & 4688.1(4) & 3404.7(3) & 1.2(2) &  &&  &  &  \\
1194.0(1) & 1405.1(4) & 0.04 & 4809.8(5) & 3404.7(3) & 3.1(4) & 1.9(4) &&  &  &  \\
2210.6(3) & 1456.5(3) & 0.04 & 3667.2(4) & 2210.7(3) & 11(3) & 13.9(9) && 1456.8(9) & 3667.5(10) & 2210.7(5) \\
2210.6(3) & 1466.2(3) & 0.04 & 3676.9(4) & 2210.7(3) & 3.1(1) & 3.6(6) && 1465(3) & 3676(3) & 2210.7(5) \\
 & 2210.6(3) & 0.10 & 2210.7(3) & g.s. & 100(1) & 100(1) && 2210.7(5) & 2210.7(5) & g.s. \\
 &  &  &  &  &  &  && 3402.6(13) & 3402.6(13) & g.s. \\
 &  &  &  &  &  &  && 3676(3) & 3676(3) & g.s. \\
2210.6(3) & 2477.3(3) & 0.10 & 4688.1(4) & 2210.7(3) & 6.0(4) & 9.3(9) &&  &  &  \\
2210.6(3) & \hspace{1.6mm}2599.0(4)\footnote{This transition is too weak at 135$^{\circ}$ to obtain a reasonable yield.} & 0.10 & 4809.8(5) & 2210.7(3) & 1.6(3) & &&   &  &  \\
2210.6(3) & 2921.4(4) & 0.20 & 5132.3(5) & 2210.7(3) & 9.7(4) & 18.3(10) && 2925(2) & 5136(2) & 2210.7(5) \\
\bottomrule[1.0pt]
\end{tabular}
\end{minipage}
\label{tab:3}
\end{table*}

\begin{figure*}[ht]
\begin{center}
\begin{tabular}{cc}
\epsfig{file=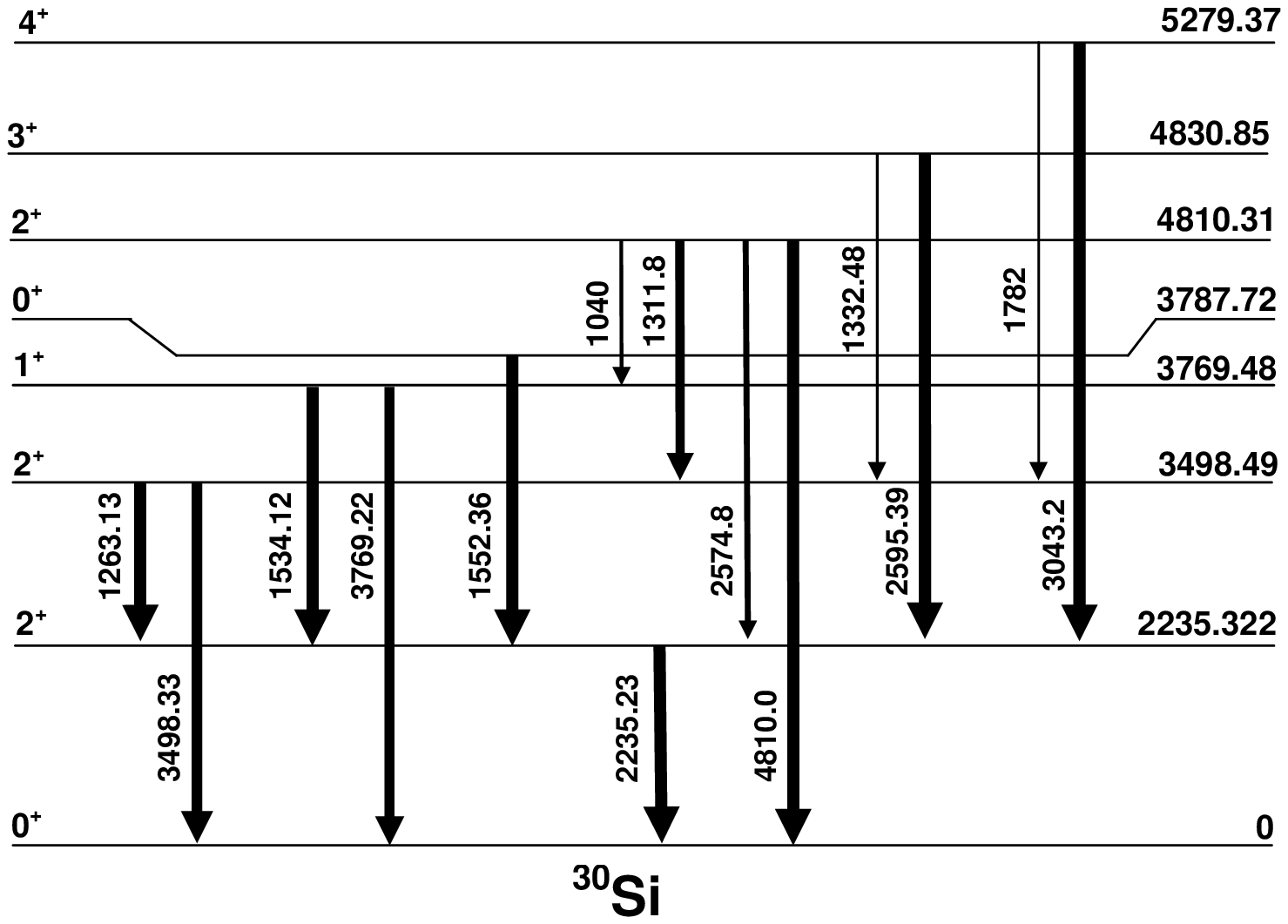,width=0.46\linewidth,clip=} &
\epsfig{file=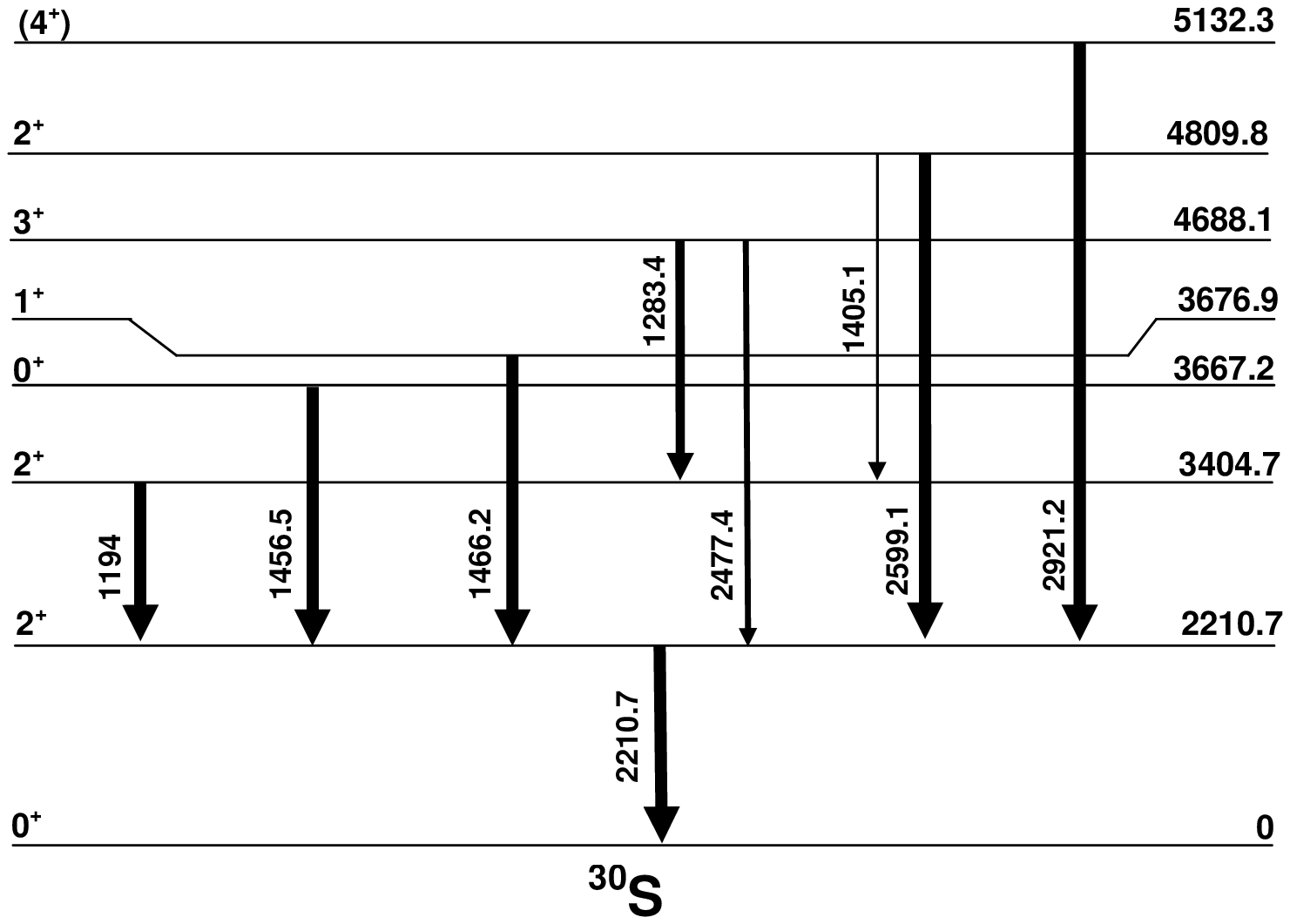,width=0.46\linewidth,clip=}
\end{tabular}
\end{center}
\caption{\label{figure6}(left) A portion of $^{30}$Si decay scheme~\cite{Basunia:2010} in comparison with (right) that of its mirror nucleus, $^{30}$S, based on the results of the present work. The $\gamma$-ray branches are not to scale; however, the thicker the arrow, the stronger the branch. All the observed $\gamma$-rays are shown with their energies (in keV) corrected for the recoil energies of the corresponding $^{30}$S excited states, which are also shown (in keV). The $\gamma$-ray transitions with energies above 3 MeV in $^{30}$S could not be resolved in our experiments due to the presence of strong $^{30}$P transitions in that region.}
\end{figure*}
\begin{figure}[ht]
\vspace{-0.6cm}
\begin{center}
  \includegraphics[width=0.47\textwidth]{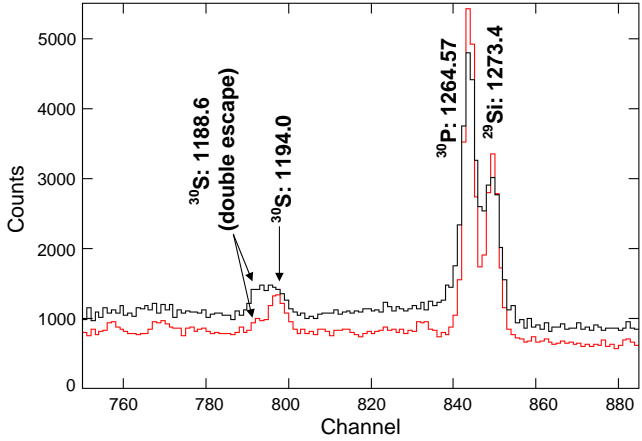}\\
  \end{center}
  \caption{\label{figure7}(Color online) The 1194-keV peak together with the double escape peak of the 2210.6-keV line of $^{30}$S. The other two peaks are identified by their parent nucleus and their energy (in keV). The black and red spectra are measured by the 50\% and 70\% relative efficiency detectors, respectively. For a short time during Phase II of the experiment, these detectors were placed at $\pm$90${^\circ}$ with respect to the beam axis. The 1188.6-keV transition is the double escape peak of the 2210.6-keV $\gamma$-ray of $^{30}$S, and its yield has decreased significantly when measured by the larger detector at -90${^\circ}$. The peak corresponding to the 1194-keV $\gamma$-ray is one of the two that stands out in the spectrum measured by the larger detector at -90${^\circ}$.}
\end{figure}


\noindent in the corresponding centroids only, because all the calibration energies have negligible uncertainties.\\
\indent From the recoil energies and the $\gamma$-ray energies, the excitation energies of the first few states were reconstructed to obtain the level scheme of $^{30}$S (see Table~\ref{tab:3} and Fig.~\ref{figure6}). The measured energies of most of the levels are in agreement with the results of the $^{32}$S($p, t$)$^{30}$S measurements discussed earlier, as well as those of previous measurements on the $\gamma$-rays of $^{30}$S~\cite{Caraca:1972,Kuhlmann:1973}. In particular, the measured energies of the two astrophysically important excited states at 4688.1(4) keV and 4809.8(5) keV from our $\gamma$-ray measurements are in excellent agreement with the 4688(2) keV and 4812(2) keV energies from the $^{32}$S($p, t$)$^{30}$S experiments presented in Table~\ref{tab:1}.\\
\indent However, there are discrepancies in the energies of two $^{30}$S levels: the energies of the 3404.7- and 5132.3-keV states deduced from our $\gamma$-ray energies are $\sim$2 keV higher and $\sim$4 keV lower, respectively, than those measured in the $\gamma$-ray study of Ref.~\cite{Kuhlmann:1973}. The reason for the discrepancy in the energy of the 5-MeV state is unclear; however, we suggest that the inconsistency between the measured energies of the 3-MeV state originates from the presence of a double escape peak at 1188.6 keV (see Fig.~\ref{figure7}) just beside the peak at 1194 keV, correspond-

\indent ing to the 2$_{2}^{+}$ $\rightarrow$ 2$_{1}^{+}$ transition in $^{30}$S, observed in our singles spectra.\\
\indent The energy of the 1194-keV $\gamma$-ray results in the level energy of the 2$^{+}_{2}$ state of $^{30}$S to be $\sim$2 keV higher than that measured by Kuhlmann {\em et al.}~\cite{Kuhlmann:1973}. The latter measurement was performed in the early 1970's when the Ge-detectors were smaller. Thus, it may be possible that the 1188.6-keV double escape peak was also present in their spectra but because of the lower detector efficiency the two peaks were assumed to be one.\\
\indent We expected to observe the $\gamma$-rays emitted from de-excitations of the 3407.7- and 3676.9-keV states directly to the ground state in the singles spectra. In addition, if the 4809.8-keV state is the 2$^{+}_{3}$ state in $^{30}$S, then according to the decay scheme of its mirror level we expect that the 2$^{+}_{3}$ $\rightarrow$ 0$^{+}_{1}$ transition in $^{30}$S is a strong branch (with respect to the strength of the other decay branches of the same level). Therefore, we also expected to observe the 4809.8-keV $\gamma$-rays of $^{30}$S in the singles spectra. However, the detection efficiency for detecting such high energy $\gamma$-rays is relatively low, and the high energy regions of the spectra obtained in the $^{28}$Si($^{3}$He, $n\gamma$)$^{30}$S experiments are obscured mostly by wide peaks originating from transitions in $^{30}$P. Therefore, the 3407.7-, 3676.9-, and 4809.8-keV transitions are not resolved. Hence, the fact that the 2$^{+}_{3}$ $\rightarrow$ 0$^{+}_{1}$ transition in $^{30}$S is not observed in these experiments does not imply that this transition is weak. Based on Ref.~\cite{Alburger:1974} where the intensities of the $\gamma$-rays of the mirror nucleus $^{30}$Si were measured at 90$^{\circ}$, we estimate that the branching ratio of the 2$^{+}_{3}$ $\rightarrow$ 0$^{+}_{1}$ transition in $^{30}$Si is 36 $\pm$ 3~\% which should be similar to that of the transition from the 4809.8-keV state to the ground state in $^{30}$S.\\
\indent We have observed a weak line at 846 keV in the singles $\gamma$-ray spectrum measured at 90$^{\circ}$ (see Fig.~\ref{figure4}), which also appears in the coincidence spectra at 90$^{\circ}$ and 135$^{\circ}$ as a more noticeable peak (see Fig.~\ref{figure5}). The energy of this peak does not seem to be Doppler shifted at 135$^{\circ}$, which suggests that this $\gamma$-ray may originate from a state whose half-life is more than 2 ps~\cite{Singh}. This $\gamma$-ray transition is also in coincidence with the 1194-keV transition in $^{30}$S. A weighted average between independent measured energies at 90$^{\circ}$ and 135$^{\circ}$ for this $\gamma$-ray results in $E_{\gamma}$ $=$ 846.0(4) keV (see Table~\ref{tab:3}). The energy of this transition does not add up to any of the known levels of $^{30}$S; however, the fact that it is a fairly prominent peak and is in coincidence with two transitions of $^{30}$S suggests that this $\gamma$-ray may also belong to the decay scheme of this nucleus. The higher lying resonances ($E_{x}$ $>$ 6 MeV) of $^{30}$S may emit $\gamma$-rays in this energy range, e.g., the decay transition from the state with $E_{x}$ $=$ 7123(10) keV to that with $E_{x}$ $=$ 6280.1(12) keV~\cite{Basunia:2010}. In particular, if one of these resonances has a high spin, its proton decay might be suppressed by the centrifugal barrier, and thus it can decay via $\gamma$-ray emission. The 846-keV $\gamma$-ray transition has so far remained unplaced in the level scheme obtained from the present experiment.\\
\indent The relative intensities from full-energy-peaks of all the observed transitions were calculated at 90${^\circ}$ and 135${^\circ}$. For the coincidence spectra, first the yield of the 1194-keV transition observed in the singles spectrum was normalized to that of the 2210.6-keV $\gamma$-ray transition also obtained from the same spectrum. Then the relative intensity of the 1194-keV $\gamma$-ray transition was used to convert the yields of all the other $\gamma$-ray transitions in the coincidence spectra into relative intensities. These results are tabulated in Table~\ref{tab:3}.

\paragraph{\label{angular_distribution}\textit{Singles measurements: $\gamma$-ray angular distributions}}

\indent\indent For a transition $J_{i}$ $\rightarrow$ $J_{f}$, where $J$ represents the spin of the state, the theoretical $\gamma$-ray angular distribution function is defined as~\cite{Litherland:1961}:
\begin{equation}
W(\theta)_{theo}\,=\,\displaystyle\sum_{\substack{k\,=\,0\\k\,=\,even}}^{4} A_{k}P_{k}(\cos\theta) , \label{eq:2}
\end{equation}

\noindent where the coefficients $P_{k}(\cos\theta)$ are the Legendre Polynomials, and the $A_{k}$ coefficients are defined as~\cite{Morinaga:1976} (p.~55):
\begin{eqnarray}
A_{k}(j_{i}\lambda\lambda'j_{f}) &=& \dfrac{\alpha_{k}B_{k}}{1\,+\,\delta^{2}}[F_{k}(j_{f}\lambda\lambda\!j_{i})\,+\,2\delta\,F_{k}(j_{f}\lambda\lambda'j_{i}) \nonumber \\
&+& \delta^{2}F_{k}(j_{f}\lambda'\lambda'j_{i})] , \label{eq:3}
\end{eqnarray}

\noindent where $j_{i}$ and $j_{f}$ are the spins of the initial and final states involved in the transition, respectively; $\lambda$ and ${\lambda\sp{\prime}}$ are transition multipolarities; $\alpha_{k}$ are the alignment factors (see Eq.~(\ref{eq:5})); $B_{k}$ and $F_{k}$ coefficients are tabulated~\cite{Yamazaki:1967} for different $j_{i}$ $\rightarrow$ $j_{f}$ transitions; and $\delta$ is the mixing ratio of a $\gamma$-ray transition defined as~\cite{Iliadis:2007} (p.~54):
\begin{equation}
\delta^{2}_{j}\,=\,\dfrac{\Gamma_{j}(\overline{\omega}L\,+\,1)}{\Gamma_{j}(\overline{\omega\sp{\prime}}L)}, \label{eq:4}
\end{equation}

\begin{figure*}[ht]
\begin{center}
\begin{tabular}{cc}
\epsfig{file=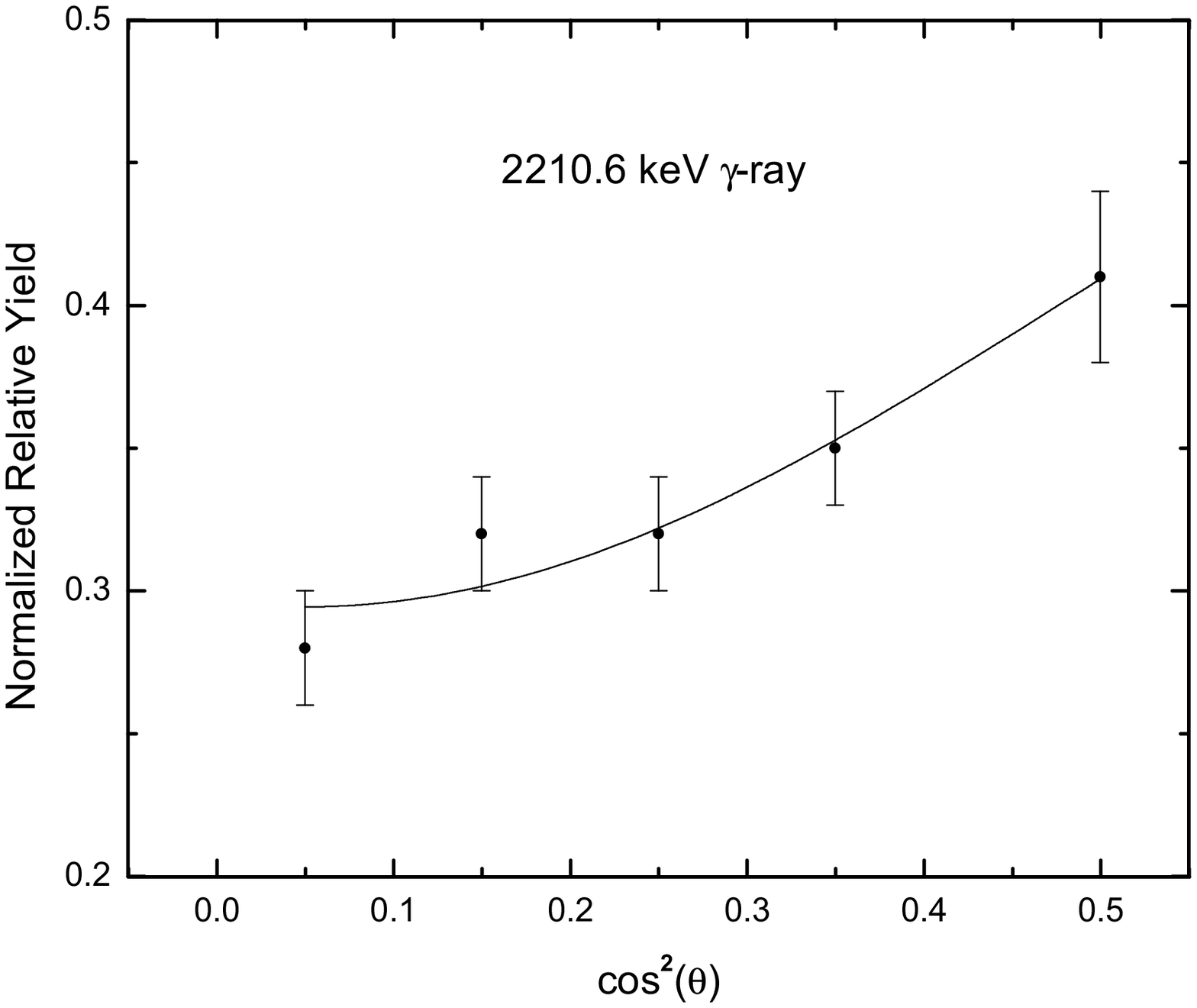,width=0.46\linewidth,clip=} &
\epsfig{file=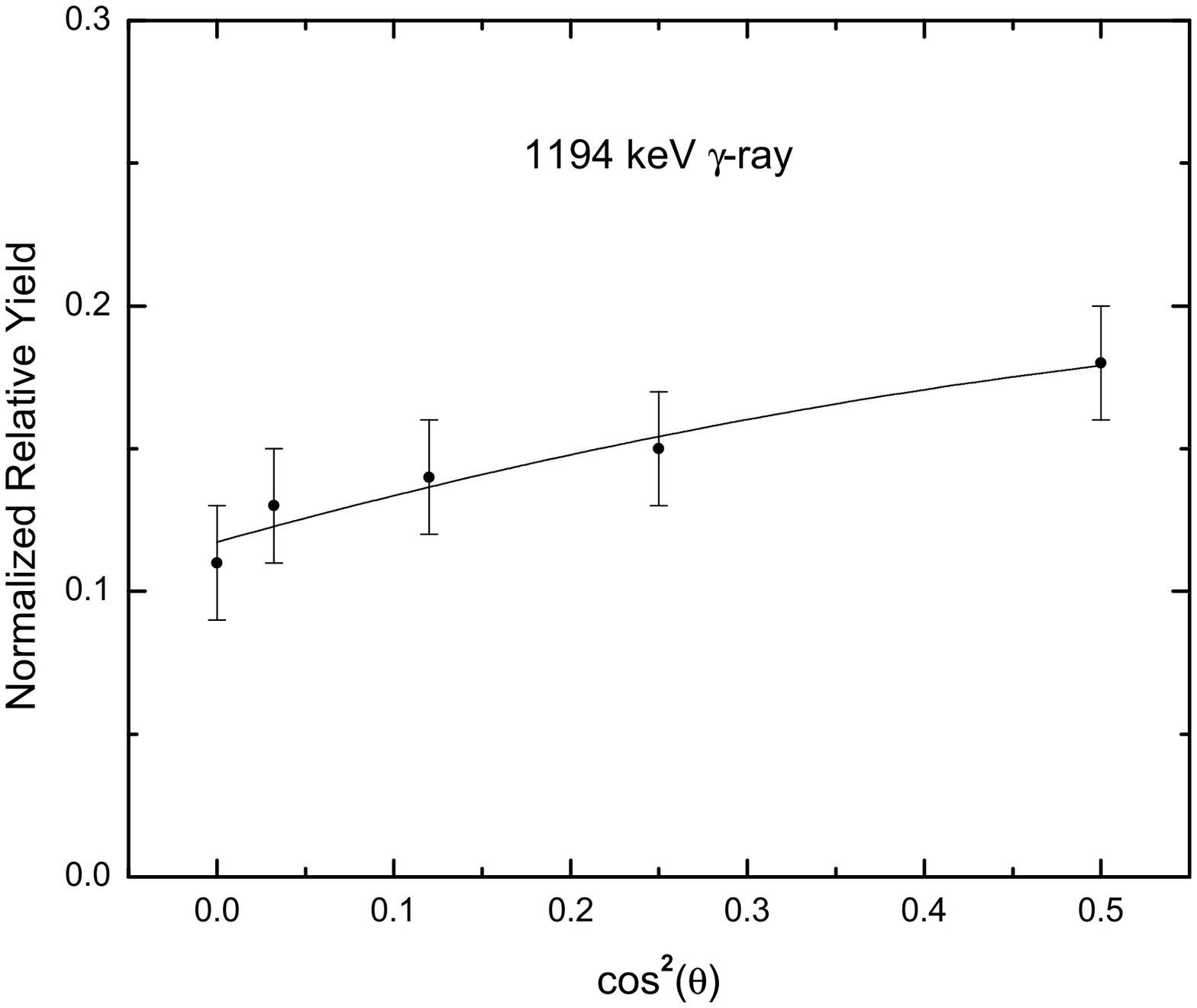,width=0.46\linewidth,clip=}
\end{tabular}
\end{center}
  \caption{\label{figure8}Experimental $\gamma$-ray angular distributions of the 2210.6-keV (left) and the 1194-keV (right) transitions. Both $\gamma$-rays are observed in the singles spectra obtained during phase II of the experiment. They correspond to the 2$^{+}_{1}$ $\rightarrow$ 0$^{+}_{1}$ and 2$^{+}_{2}$ $\rightarrow$ 2$^{+}_{1}$ transitions in $^{30}$S, respectively. The solid lines are best fits to Legendre polynomials.}
\end{figure*}

\noindent where $\overline{\omega\sp{\prime}}L$ and $\overline{\omega}L\,+\,1$ are the magnetic and electric transitions of multipolarity $L$, and $L\,+\,1$, respectively; and $\Gamma_{j}$ is the partial $\gamma$-ray width corresponding to a state with spin $j$.\\
\indent The alignment factors are defined as~\cite{Morinaga:1976} (p.~55):
\begin{equation}
\alpha_{k}\,=\,\displaystyle\sum\limits_{m\,=\,-j}^{j} \alpha_{k}^{(m)}\,P(m) , \label{eq:5}
\end{equation}

\noindent where $k$ is even and $k$ $\geq$ 6 are ignored due to a rapid decrease of transition probabilities of higher order multipoles. An individual aligned state with spin $j$ can be represented as a Gaussian probability distribution $P(m)$ of $2j\,+\,1$ magnetic substates $m_{j}$, where $m_{j}$ $=$ $-j$, $\cdots$, $j$, with the FWHM of $\sigma$ along the beam axis. $P(m)$ is the population parameter and is defined as~\cite{Morinaga:1976} (p.~56):
\begin{equation}
P(m)\,=\,\dfrac{\exp\left(\dfrac{-m^{2}}{2\sigma^{2}}\right)}{\displaystyle\sum\limits_{m{\sp{\prime}}\,=\,-j}^{j} \exp\left(\dfrac{-m{\sp{\prime}}^{2}}{2\sigma^{2}}\right)} , \label{eq:6}
\end{equation}

\noindent where $\sigma$ can be defined experimentally.\\
\indent An incomplete alignment of a state relative to the beam axis results in an attenuation of the population parameter. The alignment factors, $\alpha_{k}$, describe the degree of the attenuation of the population parameter. While $\alpha_{0}$ is considered to be unity, $\alpha_{2}$ and $\alpha_{4}$ coefficients are determined experimentally.\\
\indent For the $\gamma$-ray angular distribution measurement during phase II of the experiment, the total charge deposited by the beam could not be determined due to a faulty beam current integrator. Therefore, to take into account the fluctuations in the beam intensity and possible target degradations or changes in the target profile that could affect the areas under the peaks of interest, detector 1 was used as a monitor detector. It was kept fixed at 90${^\circ}$ with respect to the beam axis 10 cm away from the target. Detector 2, on the other hand, was positioned 7 cm away from the target and on the opposite side of detector 1. Detector 2 was moved between 90${^\circ}$ to 120${^\circ}$ in intervals of 10${^\circ}$ and was lastly positioned at 135${^\circ}$ with respect to the beam axis. It could not be place at angles higher than 135${^\circ}$ due to the presence of the beam line.\\
\indent The singles $\gamma$-ray spectra were then obtained for 1 hour from both detectors at five different angular pairs and were calibrated as explained before.\\
\indent For every ($\theta_{1}$, $\theta_{2}$) angular pair corresponding to detectors 1 and 2, the intensities of the 2210.6-keV and 1194-keV $\gamma$-ray transitions from $^{30}$S were normalized to the intense 1$^{+}$ $\rightarrow$ 1$^{+}$ transition at 708.7 keV in $^{30}$P.\\
\indent The normalized relative yields for each peak of interest were plotted against $\cos^{2}(\theta)$, where $\theta$ is the detection angle, and these data were fitted (see Fig.~\ref{figure8}) using the function:
\begin{equation}
W(\theta)_{exp}\,=\,A_{0}\,+\,A_{2}P_{2}(\cos\theta)\,+\,A_{4}P_{4}(\cos\theta) , \label{eq:1}
\end{equation}

\noindent where the coefficients $A_{i}$ are extracted from the fit, and $P_{2}(\cos\theta)$ and $P_{4}(\cos\theta)$ are Legendre polynomials. $W(\theta)_{exp}$ represents the experimental $\gamma$-ray angular distribution function, which can be used to normalize $W(\theta)_{theo}$. From the latter, one can infer the alignment probability of an excited state involved in a $\gamma$-ray transition.\\
\indent The angular distributions of the two observed $^{30}$S peaks in the singles spectra are discussed below.

\begin{itemize}
  \item \textit{The 2210.6-keV transition of \boldmath$^{30}$S:}
\end{itemize}

\indent The 2210.6-keV $\gamma$-ray corresponds to the 2$^{+}_{1}$ $\rightarrow$ 0$^{+}_{1}$ transition in $^{30}$S. This transition is a pure $E2$ ($\delta$ $=$ 0) and is a so-called stretched quadrupole transition~\cite{Rasmussen:1966}.\\
\indent The experimental intensities of the 2$^{+}_{1}$ $\rightarrow$ 0$^{+}_{1}$ transition (see Fig.~\ref{figure8}) was used to normalize the $W(\theta)_{theo}$ of this transition
\begin{table*}[ht]\small
\caption{Results of the $\gamma$-ray angular distribution studies for $^{30}$S transitions observed in the present work. Energies are in keV.}
\begin{minipage}{\linewidth}
\renewcommand\thefootnote{\thempfootnote}
\centering
\setlength{\tabcolsep}{6pt} 
\begin{tabular}{c c l l c l l}
\toprule[1.0pt] \\ [0.2ex] $E_{\gamma}$ & $J^{\pi}_{i}\,\rightarrow\,J^{\pi}_{f}$ & $A_{2}$/$A_{0}$\footnote{This value is normalized such that Eq.~(\ref{eq:1}) becomes $W_{exp}(\theta)\,=\,1\,+\,\left(A_{2}/A_{0}\right)P_{2}(\cos\theta)\,+\,\left(A_{4}/A_{0}\right)P_{4}(\cos\theta)$, which resembles Eq.~(\ref{eq:2}), where $A_{0}P_{0}(\cos\theta)$ $=$ 1.} & $A_{4}$/$A_{0}$~$^{a}$ & $\sigma$/j\footnote{The attenuation factors are~\cite{Mateosian:1974a} ($\alpha_{2}$,$\alpha_{4}$) $=$ (0.41482,0.048393) for $\sigma$/j $=$ 0.6, and ($\alpha_{2}$,$\alpha_{4}$) $=$ (0.53784,0.095181) for $\sigma$/j $=$ 0.5.} & Mult.\footnote{Transition multipolarity} & $\delta$ \\[1ex] \hline\hline \\ [0.2ex]
2210.6(3)\footnote{$E_{i}$ $\rightarrow$ $E_{f}$: 2210.7(3) keV $\rightarrow$ g.s.} & 2$^{+}_{1}$ $\rightarrow$ 0$^{+}_{1}$ & 0.4(2) & -0.0091(1800) & 0.6 & $E2$ & 0 \\[0.4ex]
1194.0(1)\footnote{$E_{i}$ $\rightarrow$ $E_{f}$: 3404.7(3) keV $\rightarrow$ 2210.7(3) keV} & 2$^{+}_{2}$ $\rightarrow$ 2$^{+}_{1}$ & 0.38(25) & -0.14(22) & 0.5 & $M\!1$, $E2$ & 0.16 \\[1ex]
\bottomrule[1.0pt]
\end{tabular}
\end{minipage}
\label{tab:4}
\end{table*}

\begin{figure*}[ht]
\begin{center}
\includegraphics[width=0.8\textwidth]{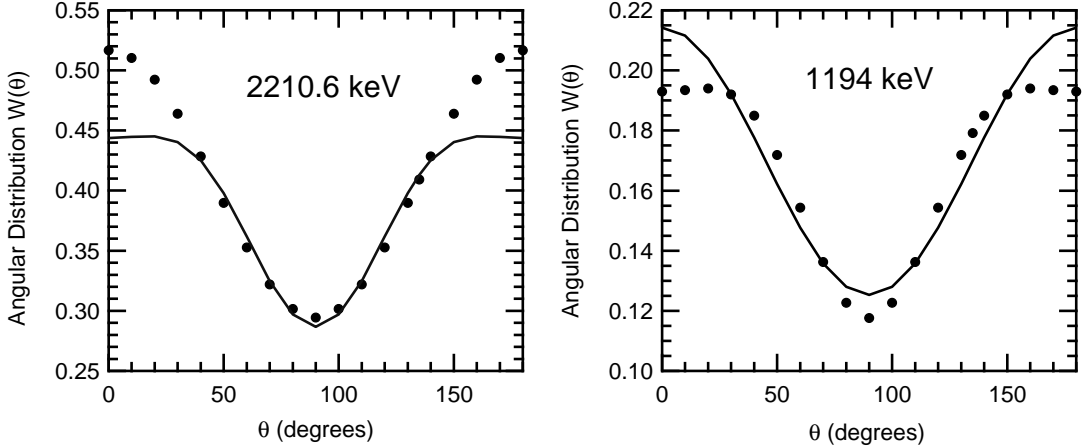}\\
\end{center}
\caption{\label{figure9}Experimental $\gamma$-ray angular distributions shown in circles in comparison with the theoretical angular distributions normalized to the data shown with solid lines. The former were obtained from fitting Eq.~(\ref{eq:1}) to relative intensities shown in Fig.~\ref{figure8}, and the latter were calculated using Eqs.~(\ref{eq:2}) and (\ref{eq:3}). The comparison is made for the (left) 2210.6-keV and (right) 1194-keV $\gamma$-rays. The normalization of the theoretical angular distributions is best performed with $\sigma$/$j$ $=$ 0.6 and $\delta$ $=$ 0 for the 2210.6-keV $\gamma$-ray, and with $\sigma$/$j$ $=$ 0.5 and $\delta$ $=$ 0.16 for the 1194-keV $\gamma$-ray. The agreement between the theoretical and experimental curves over most of the angles is good for the 2210.6-keV $\gamma$-ray and less satisfactory for the 1194-keV $\gamma$-ray (see text). For those angles lower than 40${^\circ}$ and higher than 150${^\circ}$, the $W(\theta)^{norm}_{theo}$ diverges significantly from $W(\theta)_{exp}$ due to the lack of data points for normalization at those angles.}
\end{figure*}

\noindent obtained via Eqs.~(\ref{eq:2}) and (\ref{eq:3}) using $\delta$ $=$ 0.\\
\indent To normalize $W(\theta)_{theo}$ to $W(\theta)_{exp}$, the coefficients $B_{2}F_{2}$ and $B_{4}F_{4}$ for the 2$^{+}$ $\rightarrow$ 0$^{+}$ transition were taken to be 0.7143 and -1.7143, respectively, from Ref.~\cite{Morinaga:1976} (p.~82). Therefore, the only parameters that were free to vary were the alignment factors $\alpha_{2}$ and $\alpha_{4}$. These coefficients are given in the literature~\cite{Mateosian:1974a} for 0.1 $\leq$ $\sigma$/$j$ $\leq$ 2.\\
\indent Thus, for each ($\alpha_{2}$, $\alpha_{4}$) pair corresponding to a specific $\sigma$/$j$ value, the theoretical angular distribution was calculated at the same angles at which a relative yield was measured in phase II of the $^{28}$Si($^{3}$He, $n\gamma$)$^{30}$S experiment. An average normalization factor was thus obtained and was used to normalize $W(\theta)_{theo}$ to the intensity at each angle. Then $|W(\theta)^{norm}_{theor}\,-\,I_{exp}|$/$\delta$$I_{exp}$, where $I_{exp}$ and $\delta$$I_{exp}$ are respectively the intensity and its uncertainty obtained from the data at the angle $\theta$, was plotted against $\cos^{2}\theta$. Hence, the specific ($\alpha_{2}$, $\alpha_{4}$) pair, which yielded the minimum difference between $W(\theta)_{theo}$ and $I_{exp}$, was found. Finding the ($\alpha_{2}$, $\alpha_{4}$) pair uniquely determines the parameter $\sigma$/$j$, where $\sigma$ is the FWHM of the population parameter. The results for the 2$^{+}_{1}$ $\rightarrow$ 0$^{+}_{1}$ transition in $^{30}$S are presented in Table~\ref{tab:4} and Fig.~\ref{figure9}.

\begin{itemize}
  \item \textit{The 1194-keV transition of \boldmath$^{30}$S:}
\end{itemize}
\indent\indent From a comparison of the 1194-keV $\gamma$-ray, corresponding to the 2$^{+}_{2}$ $\rightarrow$ 2$^{+}_{1}$ transition in $^{30}$S, with the mirror transition in $^{30}$Si, it was assumed that this transition is a mixed $M\!1$/$E2$. For this transition, the mixing ratio $\delta$ is an additional free parameter that is required for normalization of $W(\theta)_{theo}$ to $W(\theta)_{exp}$.\\
\indent For the 1194-keV $\gamma$-ray, the coefficients $A_{2}/A_{0}$ and $A_{4}/A_{0}$ were first extracted from the experimental fit (see Eq.~(\ref{eq:1})). $W(\theta)_{theo}$ was calculated for all ($\alpha_{2}$, $\alpha_{4}$) pairs corresponding to 0.1 $\leq$ $\sigma/j$ $\leq$ 2 for a 2$^{+}$ $\rightarrow$ 2$^{+}$ transition~\cite{Mateosian:1974a}. The mixing ratio was set to a constant free parameter from a prechosen set of values. The parameters $B_{2}F_{2}$ and $B_{4}F_{4}$ are constants given in the literature~\cite{Morinaga:1976} (p.~82). With these, $W(\theta)_{theo}$ was calculated for each value of $\delta$. Then, a $\chi^{2}$ statistical test was
\begin{figure*}[ht]
\vspace{0.2cm}
\begin{center}
\hspace{-0.4cm}\includegraphics[width=0.25\textwidth,angle=270]{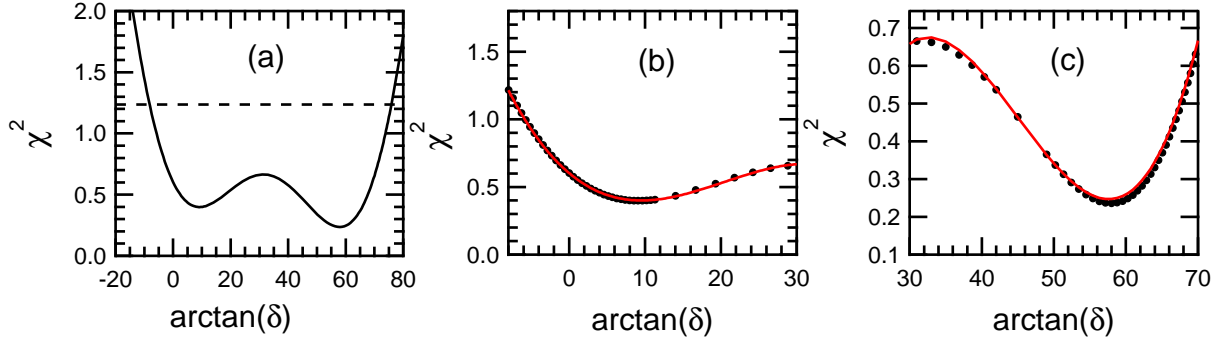}\\
\end{center}
\caption{\label{figure10}(Color online) (a) $\chi^{2}$ (solid line) vs.~$\arctan\delta$ for the 1194-keV $\gamma$-ray de-exciting the 3404.7-keV state of $^{30}$S. The dashed line shows $\chi^{2}_{min}$ $+$ 1, and is therefore our 1$\sigma$ confidence level (see text). (b) and (c) Polynomial fits of the 3$^{rd}$ degree are shown with solid red lines passing through a selected portion of $\chi^{2}$ (circles) vs.~$\arctan\delta$.}
\end{figure*}

\noindent performed with $\chi^{2}$ defined by:
\begin{equation}
\chi^{2}\,=\,\left(\dfrac{A^{exp}_{2}\,-\,A^{theo}_{2}}{\delta\!A^{exp}_{2}}\right)^{2}\,+\,\left(\dfrac{A^{exp}_{4}\,-\,A^{theo}_{4}}{\delta\!A^{exp}_{4}}\right)^{2} , \label{eq:7}
\end{equation}

\noindent where the $A^{exp}_{i}$ parameters are the yields of $^{30}$S $\gamma$-rays observed in the singles spectra and normalized to that of a $^{30}$P $\gamma$-ray peak as discussed earlier; $\delta\!A_{i}^{exp}$ are the experimental uncertainties in $A_{i}$ normalized to $\delta\!A_{0}$ extracted from the fit given by Eq.~(\ref{eq:1}); and $A^{theo}_{i}$ is calculated using Eq.~(\ref{eq:3}).\\
\indent The $\chi^{2}$ was plotted against $\arctan\delta$ (see panel (a) in Fig.~\ref{figure10}) and had local minima at $\arctan\delta$ $\simeq$ 10 and $\arctan\delta$ $\simeq$ 60. $\chi^{2}$ was again separately plotted for two regions around these minima, and each region was fitted (see panels (b) and (c) in Fig.~\ref{figure10}) with a polynomial of the third degree to obtain the functional forms of $\chi^{2}$ with respect to $\delta$ for these regions. Then a $\chi^{2}$ minimization procedure was used to find the best possible local solutions, which were $\delta$ $=$ 0.16 and $\delta$ $=$ 1.5.\\
\indent Those $\delta$'s that are within 1.0 of the best $\chi^{2}$ (see the dashed line in panel (a) in Fig.~\ref{figure10}) are located at approximately $\pm$1$\sigma$. Therefore, all $\delta$'s within -0.13 $\lesssim$ $\delta$ $\lesssim$ 3.73 are valid, which means our uncertainty in delta is very large. However, our choice of $\delta$ from the aforementioned range is determined by the consistency with the mixing ratio of the mirror transition ($\delta$ $=$ 0.18(5)~\cite{Basunia:2010}) and the agreement between the theoretical and experimental angular distributions for the 1194-keV $\gamma$-ray. The latter is best for $\sigma$/$j$ $=$ 0.4; however, $\delta$ in that case is calculated to be 0.04, which is not consistent (within 2$\sigma$) with the mixing ratio of the mirror transition. Therefore, the next best value is $\sigma$/$j$ $=$ 0.5, for which $\delta$ $=$ 0.16 consistent with that of the 2$^{+}_{2}$ $\rightarrow$ 2$^{+}_{1}$ mirror transition in $^{30}$Si. $\delta$ $=$ 0.16, as mentioned before, also represents a local minimum in the $\chi^{2}$ vs.~$\arctan\delta$ plot. We thus adopted $\delta$ $=$ 0.16 and held it fixed. For the sign of $\delta$, we have followed the convention adopted by Krane and Steffen~\cite{Krane:1970} as opposed to that of Rose and Brink~\cite{Rose:1967}.\\
\indent To confirm that we can reject $\delta$ $=$ 1.5, the single particle $E2$ transition strength $B(E2;2^{+}\,\rightarrow\,0^{+}$) in Weisskopf units was determined as follows:
\begin{equation}
B(E2) (\mbox{in W.u.})\,=\,\dfrac{9.527\,\times\,10^{6}\,BR}{E_{\gamma}^{5}\,(1\,+\,\alpha)\,A^{4/3}\,t_{\tiny 1/2}}\,\,, \label{eq:8}
\end{equation}
\noindent where $t_{\tiny 1/2}$ is the half-life of the state under consideration; $A$ is the mass number; $E_{\gamma}$ is in keV; $\alpha$ is the internal conversion coefficient, which is ignored for our case as this coefficient decreases with increasing $E_{\gamma}$; and $BR$ is the branching ratio of the transition of interest.\\
\indent We obtained $B(E2;2^{+}\rightarrow0^{+})$ $=$ 0.41 for the 1194-keV $\gamma$-ray transition. As a rule of thumb~\cite{Singh}, if the $B(E2;2^{+}\rightarrow0^{+})$ of a transition is larger than one, the corresponding state which emits the $\gamma$-ray of interest is most likely a collective state, for which the mixing ratio should be large. On the other hand, when a transition has $B(E2;2^{+}\rightarrow0^{+})$ $<$ 1, the state which initiates the transition is to a good approximation estimated as a single particle state with a small mixing ratio. Since our estimated $B(E2;2^{+}\rightarrow0^{+})$ value for the 1194-keV transition falls into the latter category, we concluded that the 3404.7-keV state is a single particle state with a small mixing ratio. Therefore, we adopted $\delta =$ 0.16.\\
\indent Finally, the procedure which was described for the 2210.6-keV $\gamma$-ray transition was repeated for the 1194-keV $\gamma$-ray transition to determine the FWHM of its population parameter. The results are given in Table~\ref{tab:4} and Fig.~\ref{figure9}.\\
\indent The previously described $\chi^{2}$ method was also performed as a check for the 2210.6-keV transition, and a sharp minimum at $\delta$ $=$ 0 confirmed the stretched $E2$ profile for this $\gamma$-ray.\\
\indent The large uncertainties in the experimentally determined $A_{2}/A_{0}$ and $A_{4}/A_{0}$ (see Table~\ref{tab:4}) are mostly due to the low statistics in each peak in the singles $\gamma$-ray spectra. However, they are still consistent with the typical values~\cite{Singh,Mateosian:1974b,NuclearDataSheets} expected for a stretched quadrupole with $\Delta$$J$ $=$ 2 (for the transition from the 2210.7-keV state to the ground state) and a mixed dipole-plus-quadrupole with $\Delta$$J$ $=$ 0 (for the transition from the 3404.7-keV state to the 2210.7-keV state).

\paragraph{\label{angular_correlation}\textit{Coincidence measurements: $\gamma$-$\gamma$ angular correlations}}

\indent\indent Measurements of the Directional Correlations of $\gamma$-rays de-exciting Oriented states (DCO ratios) allow to deduce the angular correlation
information from the $\gamma$-$\gamma$ coincidence data. The method of measuring DCO ratios is applied to determine the multipolarities of the $\gamma$-rays involved in a cascade, and thus

\begin{table*}[ht]
\caption{The experimental DCO ratios for $^{30}$S $\gamma$-rays observed in the present experiment. $j_{i}$, $j_{m}$ and $j_{f}$ are the spins of the initial, intermediate and final states, respectively. Theoretical DCO ratios are from Refs.~\cite{Singh,NuclearDataSheets}, and are obtained from known transitions for which $\sigma/j$ $=$ 0.3. See text for further explanations of the theoretical ratios.}
\begin{minipage}{\linewidth}
\renewcommand\thefootnote{\thempfootnote}
\centering
\setlength{\tabcolsep}{4pt} 
\begin{tabular}[b]{ccclllcccccc}
\toprule[1pt]\addlinespace[0.4mm]
\multicolumn{6}{c}{The $^{28}$Si($^{3}$He, $n\gamma$)$^{30}$S Experiment -- Phase II} & \phantom{ab} & \multicolumn{5}{c}{Theory} \\
\cmidrule[0.05em]{1-6} \cmidrule[0.05em]{8-12}\addlinespace[0.4mm]
$E_{\gamma}$ (keV) & $j_{i}$$\rightarrow{j_{m}}$$\rightarrow{j_{f}}$ & $\sigma/j$ & Mult.\footnote{Transition multipolarity, $D$ and $Q$ refer to dipole and quadrupole, respectively.} & $\delta$ & $R_{DCO}$ && $\Delta$$J$ & Mult.$^{a}$ & $\Delta$$J$ & Mult.$^{a}$ & $R_{DCO}$\\
& & & & & && ($j_{m}$$\rightarrow{j_{f}}$) & ($j_{m}$$\rightarrow{j_{f}}$) & ($j_{i}$$\rightarrow{j_{m}}$) & ($j_{i}$$\rightarrow{j_{m}}$) & \\ \addlinespace[0.2mm]\hline\addlinespace[0.2mm]
1194.0(1) & 2\,$\rightarrow{2}$\,$\rightarrow{0}$ & 0.5\footnote{Determined experimentally from angular distribution measurements.} & $M\!1/E2$ & 0.16$^{b}$ & 0.92(4) && 2 & $E2$ & 0 & $D$ & 1.0 \\
1456.5(3) & 0\,$\rightarrow{2}$\,$\rightarrow{0}$ & 0.3\footnote{The alignment factor of $\sigma/j$ $=$ 0.3 is usually adopted when no experimental information is available for this parameter. Since $\gamma$-ray angular distribution measurements were only obtained for the 2210.6-keV and 1194-keV $\gamma$-rays, we have assigned $\sigma/j$ $=$ 0.3 for all other $\gamma$-rays of $^{30}$S.} & $E2$ & 0\footnote{From selection rules.} & 0.94(9) && 2 & $E2$ & 2 & $Q$ & 1.0 \\
1466.2(3) & 1\,$\rightarrow{2}$\,$\rightarrow{0}$ & 0.3$^{c}$ & $M\!1$, $E2$ & -0.09(3)\footnote{This mixing ratio was adopted from the mirror transition (see Ref.~\cite{Basunia:2010}).} & 0.40(8) && 2 & $E2$ & 1 & $D$ & 0.5 \\
2477.3(3) & 3\,$\rightarrow{2}$\,$\rightarrow{0}$ & 0.3$^{c}$ & $M\!1$, $E2$ & 0.73(9)$^{e}$ & 0.37(4) && 2 & $E2$ & 1 & $D$ & 0.5 \\
2921.4(4) & 4\,$\rightarrow{2}$\,$\rightarrow{0}$ & 0.3$^{c}$ & $E2$ & 0$^{d}$ & 0.99(11) && 2 & $E2$ & 2 & $Q$ & 1.0 \\[1ex]
\bottomrule[1pt]
\end{tabular}
\end{minipage}
\label{tab:5}
\end{table*}

\noindent it can be used as a guide for determination of the spins of the associated states involved in the transitions.\\
\indent For a $j_{f}$ $\rightarrow$ $j_{m}$ $\rightarrow$ $j_{i}$ cascade, where $j_{f}$, $j_{m}$ and $j_{i}$ are the spins of the final, intermediate and initial states, respectively, the DCO ratio is generally defined as~\cite{Kramer-Flecken:1989}:
\begin{equation}
R_{DCO}\,=\,\dfrac{I^{\gamma_{2}}_{\theta_{1}}\,(\mbox{Gate}^{\gamma_{1}}_{\theta_{2}})}{I^{\gamma_{2}}_{\theta_{2}}\,(\mbox{Gate}^{\gamma_{1}}_{\theta_{1}})}\,\,, \label{eq:9}
\end{equation}

\noindent where $\theta_{1}$ and $\theta_{2}$ are the angles with respect to the beam axis at which detectors 1 and 2 are placed, respectively; $I$ is the intensity; and $\gamma_{1}$ and $\gamma_{2}$ are transitions observed in coincidence, which originate from the $j_{m}$ $\rightarrow$ $j_{i}$ and $j_{f}$ $\rightarrow$ $j_{m}$ decays, respectively. $\gamma_{1}$ transition is the one on which the coincidence gate is placed.\\
\indent The theoretical DCO ratios are given in Table~\ref{tab:5} for the cases where the $\gamma_{2}$ transition from the $j_{f}$ $\rightarrow$ $j_{m}$ decay is a pure transition. If on the other hand, $\gamma_{2}$ is a mixed transition, the theoretical DCO ratio is expected~\cite{NuclearDataSheets} to differ from what is listed in Table~\ref{tab:5}. The significance of such a difference depends upon the severity of the dipole-plus-quadrupole admixture of the $\gamma_{2}$ transition~\cite{Singh}.\\
\indent For our $\gamma$-$\gamma$ angular correlation measurement, detectors 1 and 2 were positioned, with respect to the beam axis, at 90${^\circ}$ and 135${^\circ}$, respectively, and on opposite sides with respect to the beam line. The 2$^{+}_{1}$ $\rightarrow$ 0$^{+}_{1}$ transition in $^{30}$S was measured with detector 1 and after gating on this transition, the higher lying transitions were observed in the coincidence spectrum measured by detector 2. Both detectors were placed as close to the target as possible, i.e., 3 cm and 7 cm away from the target, respectively.\\
\indent The $\gamma$-$\gamma$ angular correlations of $^{30}$S $\gamma$-rays were determined by measuring the DCO ratios for each $^{30}$S $\gamma$-ray that was observed at both angles. Since the statistics under the 1283.4-keV, 1405.1-keV, and 2599-keV $\gamma$-ray transitions corresponding to the 3$^{+}_{1}$ $\rightarrow$ 2$^{+}_{2}$, 2$^{+}_{3}$ $\rightarrow$ 2$^{+}_{2}$ and 2$^{+}_{3}$ $\rightarrow$ 2$^{+}_{1}$ decays in $^{30}$S, respectively, are too poor, the DCO ratio could not be determined for these transitions.\\
\indent The experimental DCO ratios of all other transitions of $^{30}$S were determined after a gate was set around the stretched quadrupole transition with 2210.6-keV energy (2$^{+}_{1}$ $\rightarrow$ 0$^{+}_{1}$), observed in the singles $\gamma$-ray spectra at both angles, to obtain the corresponding coincidence spectra. The peaks of interest in the coincidence spectra were then fitted, and their yields, corrected for detector efficiencies, were obtained and used to calculate the DCO ratios via Eq.~(\ref{eq:9}). The results are given in Table~\ref{tab:5}.\\
\indent The $\gamma$-$\gamma$ directional correlations of $\gamma$-ray transitions in $^{30}$S, and the spin-parity assignments of $^{30}$S states with $E_{x}$ $\leq$ 5.136 MeV from our ($p, t$) measurements are discussed below.

\paragraph{\label{spin-Parity Assignments}\textit{Spin-parity assignments}}

\indent\indent Prior to discussing the spin-parity assignments, it should be noted that in the following discussion, the energies of the adopted $\gamma$-ray transitions are corrected for $^{30}$S recoil energies (see Table~\ref{tab:3} and Fig.~\ref{figure6}).\\
\indent $\bullet$ \textit{The 2210.7-keV $\gamma$-ray Transition:} corresponds to the 2210.7-keV $\rightarrow$ ground state decay transition. The 2210.7-keV state was observed only at 62${^\circ}$ during phase I of the $^{32}$S($p, t$)$^{30}$S experiment, thus no $J^{\pi}$ assignment is available from that experiment. However, our present $\gamma$-ray angular distribution parameters for this transition (see Table~\ref{tab:4}) confirm $J^{\pi}$ $=$ 2$^{+}$.\\
\indent $\bullet$ \textit{The 1194-keV $\gamma$-ray Transition:} corresponds to the 3404.7-keV $\rightarrow$ 2210.7-keV decay transition. Due to the lack of triton angular distribution data from our $^{32}$S($p, t$)$^{30}$S experiments for the 3404.7-keV state, no conclusive spin-parity assignment was obtained for this state from those experiments. Nevertheless, the $J^{\pi}$ assignment for the 3404.7-keV state is already established as 2$^{+}$ from various previous measurements, e.g., Ref.~\cite{Bardayan:2007}, and the results of our $\gamma$-ray angular distribution measurements for the 1194-keV $\gamma$-ray transition agree with a $\Delta$$J$ $=$ 0 transition from a $J^{\pi}$ $=$ 2$^{+}$ state (see Table~\ref{tab:4}). Moreover, our experimental $R_{DCO}$ ratio for the 2$^{+}_{2}$ $\rightarrow$ $2^{+}_{1}$ $\rightarrow$ $0^{+}_{1}$ cascade agrees with the theoretical ratio within 2$\sigma$, and is consistent with an $M\!1$ transition with a small $E2$ admixture for the 1194-keV $\gamma$-ray transition for the 2$^{+}_{2}$ $\rightarrow$ 2$^{+}_{1}$ decay.\\
\indent $\bullet$ \textit{The 1456.5-keV $\gamma$-ray Transition:} corresponds to the 3667.2-keV $\rightarrow$ 2210.7-keV decay transition. The 3667.2-keV state could not be resolved in our $^{32}$S($p, t$)$^{30}$S experiments, and thus no information on its energy or spin-parity is available from those experiments. The present experimental and theoretical $R_{DCO}$ ratios for the 0$^{+}_{2}$ $\rightarrow$ 2$^{+}_{1}$ $\rightarrow$ 0$^{+}_{1}$ cascade are consistent with unity, suggesting that the transition from the 3667.2-keV state to the 2210.7-keV state has the same multipolarity as that of the decay of the 2210.7-keV state to the ground state (see Ref.~\cite{Kramer-Flecken:1989} and Table~\ref{tab:5}). This implies that the 1456.5-keV $\gamma$-ray is a pure quadrupole transition. Therefore, we confirm the assignment of $J^{\pi}$ $=$ 0$^{+}$ for the 3667.2-keV state, because from the mirror nucleus no other possibilities are expected in this energy range for a $\Delta$$J$ $=$ 2 transition corresponding to the 3667.2-keV $\rightarrow$ 2210.7-keV decay transition.\\
\indent $\bullet$ \textit{The 1466.2-keV $\gamma$-ray Transition:} corresponds to the 3676.9-keV $\rightarrow$ 2210.7-keV decay transition. From our $^{32}$S($p, t$)$^{30}$S experiment, we obtained an energy of 3681(3) keV, consistent with the 3676.9(4) keV obtained from our in-beam $\gamma$-ray spectroscopy experiment within 2$\sigma$. The present triton angular distribution data for the 3681-keV state agree with both $J^{\pi}$ $=$ 0$^{+}$ and $J^{\pi}$ $=$ 1$^{+}$ (see Fig.~\ref{figure3}). Previous measurements~\cite{Kuhlmann:1973,Bardayan:2007} have assigned a $J^{\pi}$ $=$ 1$^{+}$ to this state. According to Table~\ref{tab:5}, the theoretical $R_{DCO}$ is expected to be 0.5 if the 1466.2-keV $\gamma$-ray transition is a stretched dipole ($E1$ or $M\!1$ transition with $\delta$ $=$ 0) $\Delta$$J$ $=$ 1 transition from a state with $J^{\pi}$ $=$ 1$^{+}$ or $J^{\pi}$ $=$ 3$^{+}$. If, on the other hand, the aforementioned transition is a mixed dipole-plus-quadrupole instead of a stretched dipole, the theoretical $R_{DCO}$ should differ from 0.5~\cite{NuclearDataSheets}. Considering the $J^{\pi}$ $\rightarrow$ 2$^{+}$ $\rightarrow$ 0$^{+}$ cascade as the 3676.9-keV $\rightarrow$ 2210.7-keV $\rightarrow$ ground state decay transitions, our previous discussion implies that the 3676.9-keV state could either be the 1$^{+}_{1}$ or 3$^{+}_{1}$ state in $^{30}$S. A $J^{\pi}$ $=$ 0$^{+}$,1$^{+}$ doublet is thought~\cite{Caraca:1972,Kuhlmann:1973} to exist in $E_{x}$ $=$ 3.6 -- 3.8 MeV region in $^{30}$S. Being very close in energy to the 3667.2-keV state, the 3676.9-keV state must be the 1$^{+}$ member of the aforementioned doublet, now that we have confirmed the former as the 0$^{+}$ member. Our experimental $R_{DCO}$ ratio for the 3676.9 keV $\rightarrow$ 2210.7 keV $\rightarrow$ ground state cascade is slightly lower than 0.5 (see Table~\ref{tab:5}), which implies that the 3676.9-keV state is most likely the 1$^{+}_{1}$ state of $^{30}$S and the 3676.9 keV $\rightarrow$ 2210.7 keV decay transition is a likely an $M\!1$ transition with a small $E2$ admixture. We could not determine the mixing ratios of any of the transitions observed via the present $\gamma$-$\gamma$ directional correlation measurements. Therefore, we have adopted the mixing ratio of -0.09(3)~\cite{Basunia:2010} (from the mirror transition) for the 1466.2-keV $\gamma$-ray transition of $^{30}$S. In conclusion, we suggest a $J^{\pi}$ $=$ 1$^{+}$ for the 3676.9-keV state of $^{30}$S.\\
\indent $\bullet$ \textit{The 2477.3-keV $\gamma$-ray Transition:} corresponds to the 4688.1-keV $\rightarrow$ 2210.7-keV decay transition. The present triton angular distribution for the 4688.1-keV state is consistent with a $J^{\pi} =$ 3$^{+}$ assignment (see Fig.~\ref{figure3}). Moreover, the decay branches of the 4688-keV state, observed in our in-beam $\gamma$-ray spectroscopy experiments, also agree with those of the mirror state~\cite{Setoodehnia:2011a} assuming that the 4688-keV state is the 3$^{+}_{1}$ state of $^{30}$S. Our experimental DCO ratio for the 3$^{+}_{1}$ $\rightarrow$ $2^{+}_{1}$ $\rightarrow$ $0^{+}_{1}$ cascade is significantly different from the theoretical $R_{DCO}$ $=$ 0.5 (see Table~\ref{tab:5}). Therefore, based on the previous discussion, we expect the 2477.3-keV $\gamma$-ray to be a $\Delta$$J$ $=$ 1 mixed $M\!1$/$E2$ transition from a $J^{\pi}$ $=$ 3$^{+}$ or a $J^{\pi}$ $=$ 1$^{+}$ state. According to the mirror states in $^{30}$Si~\cite{Basunia:2010}, only one $J^{\pi}$ $=$ 1$^{+}$ state is expected in this energy range, and that is most likely the 3676.9-keV state. These arguments suggest that the 4688.1-keV state is the 3$^{+}_{1}$ state of $^{30}$S. Therefore, our experimental $R_{DCO}$ ratio also supplements the other present results with regards to the $J^{\pi}$ value of the 4688-keV state. We have adopted the mixing ratio of the mirror transition ($\delta$ $=$ 0.73(9)~\cite{Basunia:2010}) for the 2477.3-keV $\gamma$-ray due to the lack of knowledge of its own mixing ratio. We conclude that the 4688-keV state is the mirror to the 3$^{+}_{1}$ state in $^{30}$Si at 4831 keV~\cite{Basunia:2010}. Thus, the 4688-keV level in $^{30}$S is the 3$^{+}_{1}$ astrophysically important state predicted by Iliadis {\em et al.}~\cite{Iliadis:2001}.\\
\indent $\bullet$ \textit{The 2599.1-keV $\gamma$-ray Transition:} corresponds to the 4809.8-keV $\rightarrow$ 2210.7-keV decay transition, which is a very weak transition observed at 135$^{\circ}$ in the present $^{28}$Si($^{3}$He,$n\gamma$)$^{30}$S experiment. Therefore, no experimental $R_{DCO}$ ratio could be obtained for this transition. The present triton angular distribution data agree with both $J^{\pi}$ $=$ 2$^{+}$ and 3$^{+}$ (see Fig.~\ref{figure3}) but the former is a better fit. The $\gamma$-ray branching ratios for the $\gamma$-decay of the 4809.8-keV state to the 2$^{+}_{1}$ and 2$^{+}_{2}$ states in $^{30}$S were measured at 90${^\circ}$~\cite{Setoodehnia:2011a}, and were in good agreement within their uncertainties with those of the decay of the 4810-keV state in $^{30}$Si to its 2$^{+}_{1}$ and 2$^{+}_{2}$ lower-lying states (also see \S~\ref{ng_Experiment_Results}). Also, in a recent shell model calculation for the sd-shell in $A$ $=$ 30 nuclei using the USD Hamiltonian with inclusion of a charged-dependent term~\cite{Brown,Richter}, the energy of the 2$^{+}_{3}$ state in $^{30}$S was derived to be near 4800 keV, while that of the 3$^{+}_{1}$ state was calculated to be near 4700 keV. These results altogether strongly support a $J^{\pi} =$ 2$^{+}$ assignment for the 4809.8-keV state (mirror to the 2$^{+}_{3}$ state at 4810-keV in $^{30}$Si~\cite{Basunia:2010}), making it the next astrophysically important state predicted by Iliadis {\em et al.}~\cite{Iliadis:2001}.\\
\indent $\bullet$ \textit{The 2921.4-keV $\gamma$-ray Transition:} corresponds to the 5132.3-keV $\rightarrow$ 2210.7-keV decay transition. In the shell-model analysis by Wiescher and G$\ddot{\mbox{o}}$rres~\cite{Wiescher:1988}, they concluded that there are most likely at least two levels with energy near 5 MeV: a 4$^{+}$ near 5.1 MeV and a 0$^{+}$ near 5.2 MeV. Kuhlmann {\em et al.}~\cite{Kuhlmann:1973} observed a state at 5136(2) keV, and concluded that this level is most likely a 4$^{+}$ state. In Ref.~\cite{Bardayan:2007}, a state was observed at 5168(6) keV. The triton angular distribution data in that work could not be fitted with a single angular momentum transfer, which suggested that the latter state was an unresolved doublet consisting of a 4$^{+}$ and a 0$^{+}$ state. Our triton angular distribution data are best fitted with \textit{l} $=$ 4 transfer. Our experimental $R_{DCO}$ ratio for the 4$^{+}_{1}$ $\rightarrow$ $2^{+}_{1}$ $\rightarrow$ $0^{+}_{1}$ cascade is consistent with the theoretical ratio given in Table~\ref{tab:5} under the assumption that the 5132.3-keV $\rightarrow$ 2210.7-keV transition is a stretched quadrupole with $\Delta$$J$ $=$ 2. This indicates that the 5132.3-keV state, observed in our in-beam $\gamma$-ray spectroscopy experiments, is either the 4$^{+}_{1}$ or the 0$^{+}_{3}$ state of $^{30}$S. The former is much more probable because a comparison with the mirror transitions in $^{30}$Si reveals that the 4$_{1}^{+}$ level at 5279.37 keV in $^{30}$Si decays with a 100\% branch to the first excited 2$_{1}^{+}$ state~\cite{Basunia:2010}. This is consistent with what we observe for the
\begin{table*}[ht]\small
\caption{Energy levels of $^{30}$S from this work with $E_{x}$ $<$ 6 MeV . The energies of the states used as internal calibration energies in our $^{32}$S($p, t$)$^{30}$S measurements are not shown here.}\vspace{0.2cm}
\begin{minipage}{\textwidth}
\renewcommand\thefootnote{\thempfootnote}
\centering
\setlength{\tabcolsep}{4pt} 
\begin{tabular}{cccccccccl}
\toprule[1.0pt] \\
\multicolumn{2}{c}{Present Work} & \phantom{ab} & \multicolumn{2}{c}{Present Work} & \phantom{ab} & \multicolumn{2}{c}{Adopted Energy} &  & $E_{r}$\footnote{$E_{r}$ $=$ $E_{x}\,-\,Q$, where $E_{r}$ is the resonance energy, $E_{x}$ is the weighted average excitation energy and $Q$ is the proton threshold of the $^{29}$P($p, \gamma$)$^{30}$S reaction (4394.9 keV). Those excitation energies for which no resonance energy is reported correspond to the bound states of $^{30}$S.}\\
\multicolumn{2}{c}{$^{32}$S($p, t$)$^{30}$S} & \phantom{ab} & \multicolumn{2}{c}{$^{28}$Si($^{3}$He, $n\gamma$)$^{30}$S} & \phantom{ab} & \multicolumn{2}{c}{(keV)} &  & (keV) \\
\cmidrule[0.05em]{1-2} \cmidrule[0.05em]{4-5} \cmidrule[0.05em]{7-8}
$E_{x}$ (keV) & $J^{\pi}$ && $E_{x}$ (keV) & $J^{\pi}$ && $E_{x}$ (keV) & $J^{\pi}$ &  & \\
\midrule[0.05em]
 &  && g.s. &  && g.s. & 0$^{+}$ &  & \\
2208(3) &  && 2210.7(3) & 2$^{+}$ && 2210.6(3) & 2$^{+}$ &  & \\
 &  && 3404.7(3) & 2$^{+}$ && 3403.6(6) & 2$^{+}$ &  & \\
 &  && 3667.2(4) & 0$^{+}$ && 3667.0(5) & 0$^{+}$ &  & \\
3681(3) & (1$^{+}$, 0$^{+}$) && 3676.9(4) & 1$^{+}$ && 3677.0(4) & 1$^{+}$ &  & \\
4688(2) & 3$^{+}$ && 4688.1(4) & 3$^{+}$ && 4688.1(4) & 3$^{+}$ &  & 293.2(8) \\
4812(2) & 2$^{+}$ && 4809.8(5) &  && 4809.8(6) & 2$^{+}$ &  & 414.9(9) \\
 & \hspace{1.8mm}(4$^{+}$)\footnote{The corresponding energy (5136(2) keV) was used as internal calibration energy, and is thus not reported here.} && 5132.3(5) & (4$^{+}$) && 5132.6(8) & (4$^{+}$) &  & 737.7(11)\\
5225(2) & (0$^{+}$) &&  &  && \hspace{-2mm}5221(2) & (0$^{+}$) &  & 826(2) \\
5315(2) & (3$^{-}$, 2$^{+}$) &&  &  && \hspace{-.1cm}5314(4)\footnote{This state is most likely the 5288-keV state observed by Yokota {\em et al.}~\cite{Yokota:1982}, which was assigned to be the 3$^{-}_{1}$ state in $^{30}$S.} & (3$^{-}$) &  & 919(4) \\
5393(2) & 3$^{+}$ &&  &  && \hspace{-2mm}5391(2) & 3$^{+}$ &  & 996(2) \\
5849(2) & (1$^{-}$, 2$^{+}$, 4$^{+}$) &&  &  && \hspace{-2mm}5847(2) & (2$^{+}$) &  & 1452(2) \\
5947(2) &  &&  &  && \hspace{-2mm}5946(3) & (4$^{+}$) &  & 1551(3) \\[1ex]
\bottomrule[1.0pt]
\end{tabular}
\end{minipage}
\label{tab:6}
\end{table*}

\noindent 5132.3-keV state in $^{30}$S, as well as what was observed for the same state in Ref.~\cite{Kuhlmann:1973}. If the 5132.3-keV state were the 0$^{+}_{3}$ state, based on its decay scheme in the mirror nucleus, we would have expected to observe other decay branches from this state with comparable strengths, in addition to the 2921.4-keV $\gamma$-ray transition~\cite{Basunia:2010}. From these arguments, we tentatively assign $J^{\pi}$ = 4$^{+}$ to the 5132.7-keV level of $^{30}$S.\\
\indent In the following subsection, the spin-parity assignments for a few other $^{30}$S states with $E_{x}$ $\lesssim$ 6 MeV are discussed.

\paragraph{\label{Adopted_levels}\textit{Adopted energy levels in $^{30}$S}}

\indent\indent Table~\ref{tab:6} presents the combined results of both phases of both our experiments on $^{30}$S excitation energies below 6 MeV, and the corresponding recommended spin-parity assignments. The adopted energies in Table~\ref{tab:6} are the $^{30}$S weighted average excitation energies over all independent measurements in the literature, including the present work. States used as internal calibration energies were excluded in the calculations of the adopted energies. In a few cases where the uncertainty in the weighted average was smaller than the smallest uncertainty in the measured excitation energies, the latter was adopted as the final uncertainty only if the energy was measured in fewer than 4 independent measurements~\cite{Singh}.\\
\indent From our ($p, t$) measurements, a unique spin-parity assignment could not be determined for the 1452-keV and 1551-keV resonances. To calculate the $^{29}$P($p, \gamma$)$^{30}$S reaction rate, a $J^{\pi}$ value had to be assumed for each of these resonances.\\
\indent A tentative $J^{\pi}$ value of 1$^{-}$ was assigned~\cite{Bardayan:2007} to the 1452-keV resonance (see Table~\ref{tab:6}), but due to poor statistics for this particular resonance, \textit{l} $=$ 2 or 3 transfers were not excluded. In our ($p, t$) measurements, the $J^{\pi}$ value for the 996-keV resonance fits best with a 3$^{+}$ assignment, and we have assigned the 919-keV resonance to be the 3$^{-}_{1}$ state (see \S~\ref{pt_Experiment_Results}). Hence, we have tentatively assigned the 1452-keV resonance to be the 2$^{+}_{4}$ state in $^{30}$S.\\
\indent The 1551-keV resonance has only been observed in the measurement of Ref.~\cite{Fynbo:2000} and in our $^{32}$S($p, t$)$^{30}$S measurements. However, the data obtained in these measurements were not enough to assign a conclusive $J^{\pi}$ value to this resonance. From the results of a recent shell model calculation~\cite{Brown,Richter}, the energies of the 4$^{+}_{2}$ states in $^{30}$S and its mirror nucleus -- $^{30}$Si -- are almost identical to each other. The excitation energy of the 4$^{+}_{2}$ state in $^{30}$Si is 5950.73(15) keV~\cite{Basunia:2010}. On the other hand, the weighted average energy between the results of Ref.~\cite{Fynbo:2000} and our $^{32}$S($p, t$)$^{30}$S measurements for the corresponding state in $^{30}$S is 5946(3) keV. Therefore, we concluded that this latter state is most likely the mirror to the 4$^{+}_{2}$ state in $^{30}$Si at 5950.73(15) keV.\\
\indent The states presented in Table~\ref{tab:6} are the only ones that could play a crucial role in determination of the $^{29}$P($p, \gamma$)$^{30}$S reaction rate in the temperature range characteristic of explosive hydrogen burning (0.1 GK $\leq$ $T$ $\leq$ 1.3 GK). The excited states whose energies are below 4.5 MeV become important in determining the non-resonant contributions to the $^{29}$P($p, \gamma$)$^{30}$S reaction rate.

\section{\label{Rate}The \boldmath$^{29}$P($p, \gamma$)$^{30}$S reaction rate}

\begin{figure}[ht]
\begin{center}
\includegraphics[width=0.5\textwidth]{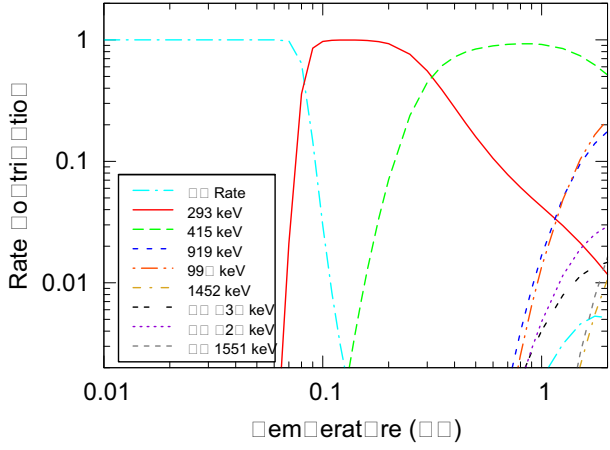}\\
\end{center}
\caption{\label{figure11}(Color online) Resonant and DC contributions to the $^{29}$P($p, \gamma$)$^{30}$S Monte Carlo rate as a function of temperature. Abbreviations are as follows: DC: Direct Capture; UL: Upper Limit. The latter is used for those resonances for which the proton partial width is estimated to be an upper limit.}
\end{figure}

\begin{figure}[ht]
  \vspace{0.4cm}
  \centering
  \hspace{-0.5cm}
  {\label{figure12a}%
  \includegraphics[width=0.3\textwidth,angle=270]{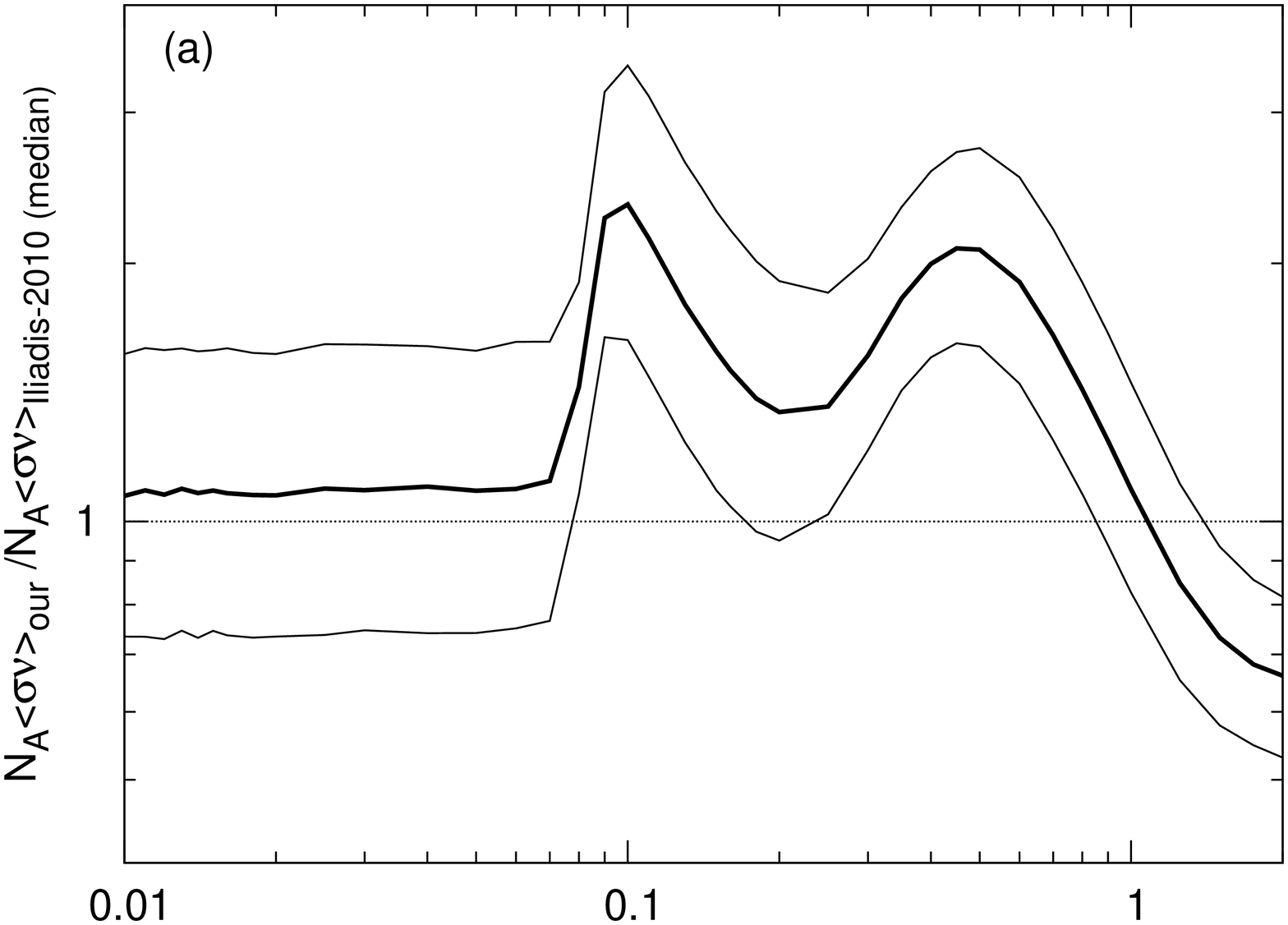}}%
  \vspace{0.3cm}\quad%
  {\label{figure12b}%
  \includegraphics[width=0.3\textwidth,angle=270]{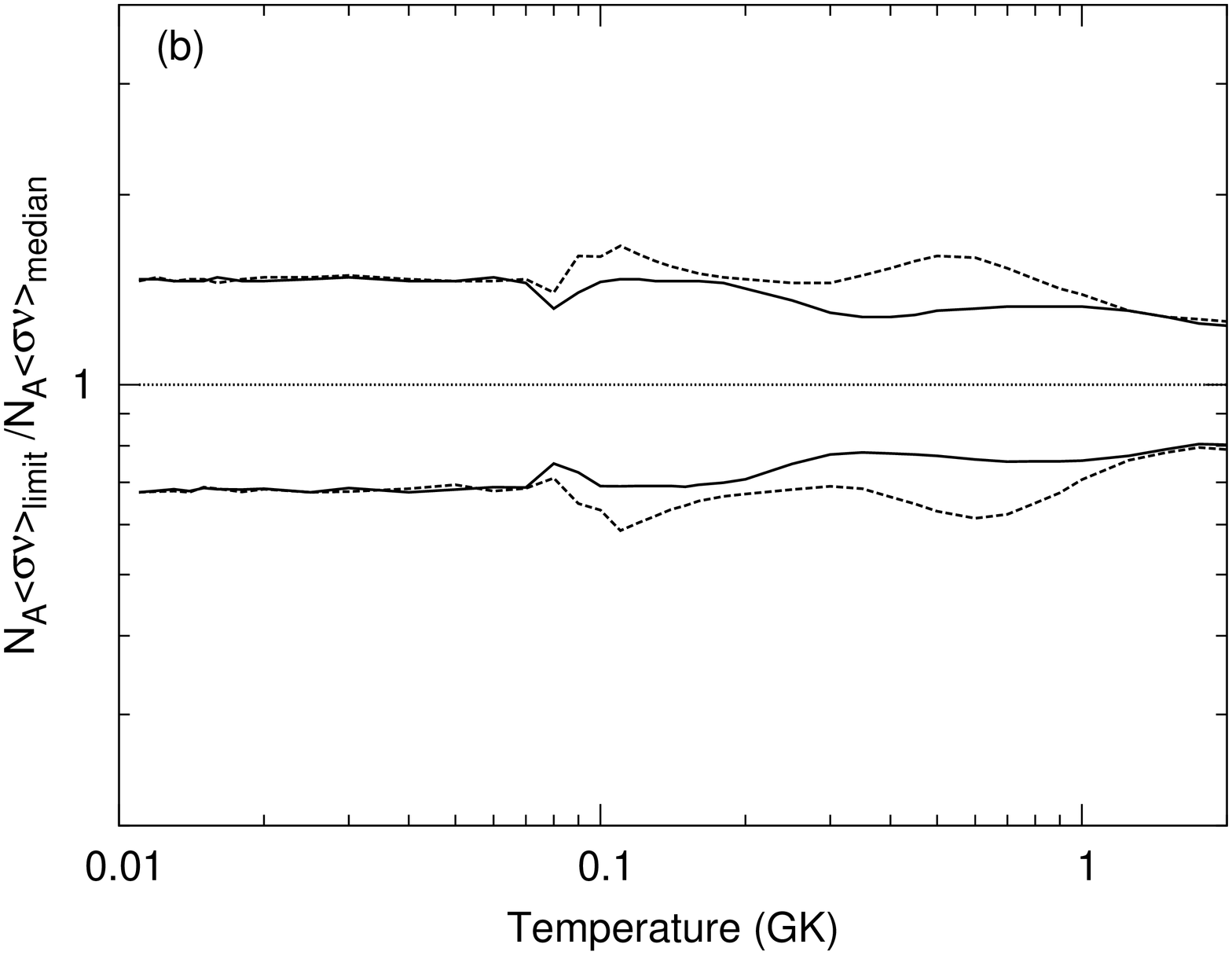}}
  \quad%
  \vspace{0.6cm}\caption{\label{figure12}(top) The ratio of our Monte Carlo low (lower thin line), median (middle thick line) and high (upper thin line) rates to the Monte Carlo median rate of Ref.~\cite{Iliadis:2010a}. Our median rate is 2.3 times larger than that of Ref.~\cite{Iliadis:2010a} at $T$ $=$ 0.1 GK. (bottom) The uncertainty bands corresponding to $N_{A}\!<\sigma\upsilon\!>_{\tiny \mbox{high}}$/$N_{A}\!<\sigma\upsilon\!>_{\tiny \mbox{median}}$ and $N_{A}\!<\sigma\upsilon\!>_{\tiny \mbox{low}}$/$N_{A}\!<\sigma\upsilon\!>_{\tiny \mbox{median}}$ from our Monte Carlo rate (solid lines) compared to those of Ref.~\cite{Iliadis:2010a} (dashed lines). At $T$ $=$ 0.1 GK, the ratio of the $N_{A}\!<\sigma\upsilon\!>_{\tiny \mbox{high}}$/$N_{A}\!<\sigma\upsilon\!>_{\tiny \mbox{low}}$ from our Monte Carlo rate is 72\% smaller than that of the Monte Carlo rate reported in Ref.~\cite{Iliadis:2010a}.}
\end{figure}

\indent To obtain the non-resonant contribution to the $^{29}$P($p, \gamma$)$^{30}$S reaction rate, one has to determine the astrophysical S-factor, $S(E)$, from:
\begin{equation}
S(E)\,\approx\,S(0)\,+\,S\sp{\prime}(0)E\,+\,\dfrac{1}{2}S''(0)E^{2} , \label{eq:10}
\end{equation}
\noindent where the primes indicate derivatives with respect to $E$. $S(E)$ can be integrated to give the non-resonant reaction rate:
\begin{equation}
N_{A}<\!\sigma\upsilon\!>\,=\,\sqrt{\dfrac{8}{\pi\mu}}\,\dfrac{N_{A}}{(kT)^{3/2}} \int_0^\infty S(E) \exp\left(\dfrac{-E}{kT}\,-\,\sqrt{\dfrac{E_{G}}{E}}\right)dE , \label{eq:11}
\end{equation}

\noindent where $N_{A}<\!\sigma\upsilon\!>$ is the reaction rate, $N_{A}$ is Avogadro's number, $\mu$ is the reduced mass, $k$ is Boltzmann's constant, $T$ is the temperature, $E$ is the center-of-mass energy in keV, and $E_{G}$ is the Gamow energy (in MeV).\\
\indent The $^{29}$P($p, \gamma$)$^{30}$S direct capture (DC) reaction rate to all bound states, including the ground state, was calculated assuming proton transfer into 2$\textit{s}$ and 1$\textit{d}$ final orbitals. For each final state, the S-factor was calculated by taking into account the $E1$ and $M\!1$ nature of the transitions, which were then weighted by the corresponding spectroscopic factors determined from those of the mirror states~\cite{Mackh:1973}. The weighted S-factor contributions from each state of $^{30}$S were then summed to derive the total S-factor as a function of proton bombarding energy for each transition multipolarity.\\
\indent The S-factor was then fitted with a polynomial of the form given in Eq.~(\ref{eq:10}) to determine the fit parameters, i.e., $S(0)$, $S\sp{\prime}(0)$ and $S''(0)$. An uncertainty of 40\% for the direct capture S-factor is adopted following the approach of Ref.~\cite{Iliadis:2010b}.\\
\indent With increasing center-of-mass energy, resonances become important, and therefore the non-resonant S-factor in Eq.~(\ref{eq:10}) is truncated at the so-called cutoff energy, after which the direct capture S-factor deviates from the total astrophysical S-factor. The cutoff energy was chosen~\cite{Iliadis:2010b} at $\sim$1000 keV for the $^{29}$P($p, \gamma$)$^{30}$S reaction rate.\\
\indent To calculate the resonant contributions to the rate, the proton widths were determined using the expression:
\begin{equation}
\Gamma_{\tiny{p}} = 2\dfrac{\hbar^{2}}{\mu a^{2}}P_{\tiny{l}}C^{2}S\theta_{\tiny{sp}}^{2}\,\,, \label{eq:12}
\end{equation}

\noindent where $\mu$ is the reduced mass, $P_{\tiny{l}}$ is the barrier penetrability (calculated using $r_{\tiny{0}}$ $=$ 1.25 fm) for orbital angular momentum $l$, $a\,=\,r_{\tiny{0}}(A_{\tiny{t}}^{\tiny{1/3}}+A_{\tiny{p}}^{\tiny{1/3}})$ is the interaction radius in terms of target and projectile mass numbers ($A_{\tiny{t}}$ and $A_{\tiny{p}}$, respectively), $C$ and $S$ are the isospin Clebsch-Gordan coefficient and spectroscopic factor, respectively, and $\theta_{\tiny{sp}}^{2}$ is the observed dimensionless single-particle reduced width.\\
\indent The $\theta^{2}_{sp}$ factors were estimated using Eq.~(11) together with Table~1 of Ref.~\cite{Iliadis:1997}. The only exceptions were the 4$^{+}$ states corresponding to the resonances at 737.7 keV and 1551 keV. The reduced widths of these resonances could not be determined from the approach of Ref.~\cite{Iliadis:1997}, which is limited to single-particle states in the $sd$ -- $fp$ shells. Consequently, $\theta_{\tiny{sp}}^{2} \leq$ 1 is assumed for these states.\\
\indent Spectroscopic factors were determined from neutron spectroscopic factors of the mirror states measured with $^{29}$Si($d, p$)$^{30}$Si~\cite{Mackh:1973}. The mirror levels in $^{30}$Si corresponding to the resonances of $^{30}$S at 737.7 keV, 826 keV and 1551 keV were populated very weakly in the measurement of Ref.~\cite{Mackh:1973}, and thus no $C^{2}S$ values could be determined experimentally for these levels. Hence, an upper limit of $C^{2}S$ $\leq$ 0.01 is adopted for these states, based on the sensitivity for the extraction of small spectroscopic factors. Following the procedure of Ref.~\cite{Iliadis:2010b}, the uncertainties in the proton widths were estimated to be 40\%.\\
\indent To determine the $\gamma$-ray partial widths ($\Gamma_{\tiny{\gamma}}$), the corresponding widths of the mirror states in $^{30}$Si were calculated from measured half-lives, branching ratios, multipolarities, and mixing ratios~\cite{Basunia:2010}. For the cases where mixing ratios of the transitions of interest in $^{30}$Si have not been determined experimentally or theoretically, we have assumed that such transitions are pure, with multipolarities assumed to be the dominant multipolarity of the actual mixed transition. These widths were then scaled to account for the energy difference between each mirror pair, assuming similar decay branches and reduced transition probabilities.\\
\indent Only an upper limit is known for the half-life of the 2$^{+}_{4}$ state in $^{30}$Si. Hence, the aforementioned method of calculation of the $\gamma$-ray partial width is not possible for the corresponding mirror state in $^{30}$S. Therefore, we considered the 0.012 eV value from Ref.~\cite{Iliadis:2010b} for the total $\gamma$-ray width of the 2$^{+}_{4}$ resonance in $^{30}$S, and scaled it to account for the differences in the measured energies. Following the procedure discussed in Ref.~\cite{Iliadis:2010b}, the uncertainties in $\gamma$-ray widths are assumed to be 50\%.\\
\indent The $^{29}$P($p, \gamma$)$^{30}$S reaction rate was calculated using the Monte Carlo method presented in Refs.~\cite{Longland:2010,Iliadis:2010b,Iliadis:2010a} using 10$^{4}$ random samples. The input file required for calculation of the $^{29}$P($p, \gamma$)$^{30}$S Monte Carlo reaction rate, including the resonance parameters, is given in Ref.~\cite{Setoodehnia:2011b}. The numerical values of the $^{29}$P($p, \gamma$)$^{30}$S rate are given in Table~\ref{tab:7}.\\
\indent Since the experimental input is truncated at 6 MeV (see Table~\ref{tab:6}), there exists a temperature denoted by $T_{match}$~\cite{Iliadis:2010a}, above which the nuclear physics input is insufficient to calculate a reliable reaction rate. For our case, $T_{match}$ was found to be 3 GK. The reaction rate above this temperature was therefore obtained as follows: the theoretical $^{29}$P($p, \gamma$)$^{30}$S rate as a function of temperature was calculated based on the Hauser-Feshbach statistical model~\cite{Newton:2008} using the {\fontfamily{pcr}\selectfont\small NON-SMOKER$^{\mbox{WEB}}$} code~\cite{NONSMOKER}. Then, these rates were scaled and normalized to the experimental Monte Carlo rate at $T_{match}$ $=$ 3 GK.\\
\indent Fig.~\ref{figure11} compares the contributions of the direct capture rate (DC Rate) and those of the resonances listed in Table~\ref{tab:6} to the total $^{29}$P($p, \gamma$)$^{30}$S thermonuclear reaction rate. The direct capture rate dominates the total rate for T $\leq$ 0.08 GK, whereas at higher temperatures characteristic of explosive nucleosynthesis in novae, the total reaction rate is dominated by a single 3$^{+}_{1}$ resonance at 293.2 keV in the range of 0.09 -- 0.3 GK. The 414.9-keV resonance with $J^{\pi}$ $=$ 2$^{+}_{3}$ is the main contributor to the total rate from 0.35 GK to 2 GK. The 996-keV resonance becomes important at temperatures higher than 2 GK, which are beyond the temperature range of interest to this work. The other resonances, including those for which only an upper limit proton partial width is known, do not contribute significantly to the $^{29}$P($p, \gamma$)$^{30}$S total rate in the temperature range of interest.\\
\indent Fig.~\ref{figure12} compares our Monte Carlo rate for the $^{29}$P($p, \gamma$)$^{30}$S reaction with that of Ref.~\cite{Iliadis:2010a}, where the energies of the 3$^{+}_{1}$ and 2$^{+}_{3}$ states of $^{30}$S were assumed to be 4704(5) keV~\cite{Bardayan:2007} and

\begin{table*}[ht]\small
\caption{Total Monte Carlo rate for the $^{29}$P($p, \gamma$)$^{30}$S thermonuclear reaction. See text for details.}
\centering
\begin{tabular}{c c c c c c c c}
\toprule[1.0pt] \\[0.1ex] T (GK) & Low Rate & Median Rate & High Rate & T (GK) & Low Rate & Median Rate & High Rate \\[0.1ex]\hline\hline\\[0.1ex]
0.010 & 5.06$\times$10$^{-42}$ & 7.38$\times$10$^{-42}$ & 1.08$\times$10$^{-41}$ & 0.130 & 2.77$\times$10$^{-10}$ & 4.01$\times$10$^{-10}$ & 5.87$\times$10$^{-10}$ \\
0.011 & 1.93$\times$10$^{-40}$ & 2.86$\times$10$^{-40}$ & 4.19$\times$10$^{-40}$ & 0.140 & 1.61$\times$10$^{-09}$ & 2.33$\times$10$^{-09}$ & 3.41$\times$10$^{-09}$ \\
0.012 & 4.88$\times$10$^{-39}$ & 7.19$\times$10$^{-39}$ & 1.06$\times$10$^{-38}$ & 0.150 & 7.37$\times$10$^{-09}$ & 1.07$\times$10$^{-08}$ & 1.56$\times$10$^{-08}$ \\
0.013 & 8.95$\times$10$^{-38}$ & 1.31$\times$10$^{-37}$ & 1.91$\times$10$^{-37}$ & 0.160 & 2.79$\times$10$^{-08}$ & 4.02$\times$10$^{-08}$ & 5.87$\times$10$^{-08}$ \\
0.014 & 1.20$\times$10$^{-36}$ & 1.77$\times$10$^{-36}$ & 2.59$\times$10$^{-36}$ & 0.180 & 2.56$\times$10$^{-07}$ & 3.66$\times$10$^{-07}$ & 5.29$\times$10$^{-07}$ \\
0.015 & 1.29$\times$10$^{-35}$ & 1.88$\times$10$^{-35}$ & 2.74$\times$10$^{-35}$ & 0.200 & 1.53$\times$10$^{-06}$ & 2.16$\times$10$^{-06}$ & 3.07$\times$10$^{-06}$ \\
0.016 & 1.12$\times$10$^{-34}$ & 1.64$\times$10$^{-34}$ & 2.42$\times$10$^{-34}$ & 0.250 & 4.23$\times$10$^{-05}$ & 5.65$\times$10$^{-05}$ & 7.67$\times$10$^{-05}$ \\
0.018 & 5.17$\times$10$^{-33}$ & 7.58$\times$10$^{-33}$ & 1.11$\times$10$^{-32}$ & 0.300 & 4.42$\times$10$^{-04}$ & 5.70$\times$10$^{-04}$ & 7.39$\times$10$^{-04}$ \\
0.020 & 1.41$\times$10$^{-31}$ & 2.06$\times$10$^{-31}$ & 3.01$\times$10$^{-31}$ & 0.350 & 2.53$\times$10$^{-03}$ & 3.23$\times$10$^{-03}$ & 4.14$\times$10$^{-03}$ \\
0.025 & 1.04$\times$10$^{-28}$ & 1.54$\times$10$^{-28}$ & 2.27$\times$10$^{-28}$ & 0.400 & 9.65$\times$10$^{-03}$ & 1.24$\times$10$^{-02}$ & 1.59$\times$10$^{-02}$ \\
0.030 & 1.62$\times$10$^{-26}$ & 2.36$\times$10$^{-26}$ & 3.49$\times$10$^{-26}$ & 0.450 & 2.76$\times$10$^{-02}$ & 3.56$\times$10$^{-02}$ & 4.61$\times$10$^{-02}$ \\
0.040 & 2.49$\times$10$^{-23}$ & 3.69$\times$10$^{-23}$ & 5.38$\times$10$^{-23}$ & 0.500 & 6.40$\times$10$^{-02}$ & 8.30$\times$10$^{-02}$ & 1.09$\times$10$^{-01}$ \\
0.050 & 4.64$\times$10$^{-21}$ & 6.80$\times$10$^{-21}$ & 9.90$\times$10$^{-21}$ & 0.600 & 2.23$\times$10$^{-01}$ & 2.93$\times$10$^{-01}$ & 3.88$\times$10$^{-01}$ \\
0.060 & 2.47$\times$10$^{-19}$ & 3.58$\times$10$^{-19}$ & 5.33$\times$10$^{-19}$ & 0.700 & 5.35$\times$10$^{-01}$ & 7.09$\times$10$^{-01}$ & 9.43$\times$10$^{-01}$ \\
0.070 & 6.05$\times$10$^{-18}$ & 8.81$\times$10$^{-18}$ & 1.28$\times$10$^{-17}$ & 0.800 & 1.02$\times$10$^{+00}$ & 1.35$\times$10$^{+00}$ & 1.80$\times$10$^{+00}$ \\
0.080 & 1.41$\times$10$^{-16}$ & 1.88$\times$10$^{-16}$ & 2.49$\times$10$^{-16}$ & 0.900 & 1.67$\times$10$^{+00}$ & 2.21$\times$10$^{+00}$ & 2.95$\times$10$^{+00}$ \\
0.090 & 5.33$\times$10$^{-15}$ & 7.34$\times$10$^{-15}$ & 1.03$\times$10$^{-14}$ & 1.000 & 2.47$\times$10$^{+00}$ & 3.26$\times$10$^{+00}$ & 4.34$\times$10$^{+00}$ \\
0.100 & 1.66$\times$10$^{-13}$ & 2.39$\times$10$^{-13}$ & 3.47$\times$10$^{-13}$ & 1.250 & 5.14$\times$10$^{+00}$ & 6.67$\times$10$^{+00}$ & 8.71$\times$10$^{+00}$ \\
0.110 & 3.07$\times$10$^{-12}$ & 4.45$\times$10$^{-12}$ & 6.52$\times$10$^{-12}$ & 1.500 & 8.85$\times$10$^{+00}$ & 1.12$\times$10$^{+01}$ & 1.43$\times$10$^{+01}$ \\
0.120 & 3.53$\times$10$^{-11}$ & 5.11$\times$10$^{-11}$ & 7.49$\times$10$^{-11}$ \\[1ex]
\bottomrule[1.0pt]
\end{tabular}
\label{tab:7}
\end{table*}

\noindent 4888(40) keV~\cite{Iliadis:2001}, respectively.\\
\indent Both rates shown in Fig.~\ref{figure12} are calculated using the Monte Carlo technique. Our lower resonance energies, relative to those adopted in Ref.~\cite{Iliadis:2010a}, for the two astrophysically important resonances, cause our median rate to be up to 2.3 times larger (at $T_{9}$ $=$ 0.1) (see top panel in Fig.~\ref{figure12}) and up to 11.5 times larger (at $T_{9}$ $=$ 0.1) than the recommended rate of Ref.~\cite{Iliadis:2001}. For this last rate, the energies of both astrophysically important resonances (corresponding to the 3$^{+}_{1}$ and 2$^{+}_{3}$ states in $^{30}$S) were determined theoretically based on the IMME, since none of these resonances were observed at the time.\\
\indent For our present rate, the energy of the resonance corresponding to the 2$^{+}_{3}$ state of $^{30}$S is determined experimentally. Thus, its uncertainty of 0.9 keV is reduced by a factor of $\sim$44 with respect to the theoretical estimate of 40 keV adopted in Ref.~\cite{Iliadis:2010a}. Furthermore, the 2-keV uncertainty in the energy of the resonance corresponding to the 3$^{+}_{1}$ state in $^{30}$S, measured in this work, is also reduced by 40\% with respect to the 5 keV measured in Ref.~\cite{Bardayan:2007} that is used to derive the rate in Ref.~\cite{Iliadis:2010a}.\\
\indent Therefore, since these two resonances together dominate the total rate over 0.08 $<$ $T$ $\leq$ 2 GK, the reductions in their associated uncertainties also reduce the uncertainty in the total reaction rate (see the bottom panel in Fig.~\ref{figure12}). For example, at $T$ $=$ 0.1 GK, where the uncertainty in both our rate and that of Ref.~\cite{Iliadis:2010a} is maximum, the $N_{A}\!<\sigma\upsilon\!>_{\tiny \mbox{high}}$/$N_{A}\!<\sigma\upsilon\!>_{\tiny \mbox{low}}$ ratio from our Monte Carlo rate is 72\% smaller than that of the Monte Carlo rate reported in Ref.~\cite{Iliadis:2010a}.

\section{\label{Abundances}Nova isotopic abundances}

\indent In \S~\ref{Astrophysics}, it was emphasized that $^{29}$P($p, \gamma$)$^{30}$S is one of the two reactions that are thought to affect the silicon isotopic ratios in nova ejecta.\\
\indent To investigate the impact of the updated $^{29}$P($p, \gamma$)$^{30}$S rate on the isotopic abundances of silicon synthesized in classical novae, we have computed three different models of nova outbursts, with identical input physics except for the adopted

\noindent $^{29}$P($p, \gamma$)$^{30}$S rate. Results from our nova nucleosynthesis simulations are presented next.

\subsection{\label{Abundances_Simulations}Nova simulations}

\indent Three nova nucleosynthesis models were computed with the Lagrangian one-dimensional (spherically symmetric) full hydrodynamic and implicit code called {\fontfamily{pcr}\selectfont\small SHIVA}. Detailed information about this code is provided in Refs.~\cite{Jose:1998,Jose:1999}.\\
\indent {\fontfamily{pcr}\selectfont\small SHIVA} simulates the evolution of nova outbursts from the onset of accretion to the explosion and ejection of the nova ejecta. The hydrodynamic code is coupled directly to the nuclear reaction network. Thus, in the present work, the explosion simulations at each stage are complemented with detailed nova nucleosynthesis calculations using the most updated reaction rate libraries.\\
\indent As pointed out in Ref.~\cite{Jose:1998}, the material is dredged up on short timescales from the outermost shells of the CO- or ONe-rich core to the surface of the white dwarf by convective mixing processes. Nuclear reactions in stellar environments are sensitive to the temperature, and thus the ejected abundances of fragile nuclei that would have been destroyed if they had not been carried to higher and cooler layers, are increased by considering the convection process during the evolution of the nova outburst. This, in turn, makes the present simulations more realistic and suitable for defining absolute isotopic abundances resulting from nova nucleosynthesis than the previous post-processing nucleosynthesis simulations used in Refs.~\cite{Iliadis:2002,Bardayan:2007}, where the nucleosynthesis is decoupled from the hydrodynamics of the outburst.\\
\indent The absolute abundances observed in nova ejecta or in presolar grains of potential nova origin provide strong constraints for improvement of nova simulations. Thus, a more precise set of constraints can be obtained if predictions on specific isotopic abundances are available.\\
\indent For the present full hydrodynamic simulations, the thermodynamic profiles are identical to those of hydrodynamical simulations, given in Ref.~\cite{Jose:1998}, for a massive ONe nova with a 1.35 $M_{\astrosun}$ underlying white dwarf. Such an extreme white dwarf is adopted because a CO white dwarf shows limited activity in the Si-Ca mass region. This, in turn, is due to very little, if any, Ne, Mg and Si seed nuclei available in the outer core of a CO white dwarf, and the lower temperature achieved in a CO nova outburst~\cite{Jose:2004}. Thus the nucleosynthesis of silicon isotopes in CO novae, with even the most massive underlying white dwarf, is negligible.\\
\indent An accretion at a rate of $\dot{M}_{\tiny \mbox{acc}}\,=\,2\,\times\,10^{-10}\,M_{\astrosun}/\mbox{yr}$ of solar-like matter onto a 1.35 $M_{\astrosun}$ ONe white dwarf is assumed in all three present models. 50\% enrichment by the white dwarf's core material is adopted for the accreted matter to mimic the unknown mechanism responsible for the enhancement in metals, which ultimately powers the explosion through hydrogen burning~\cite{Jose:2004}. The initial abundances of the seed isotopes used in the present simulations are given in Ref.~\cite{Jose:2004}. The impact of the new solar metallicity~\cite{Grevesse:2007} (decreased by about a factor of 2) on the overall results presented here has been tested and is insignificant.\\
\indent In addition to hydrodynamics, a reaction rate network including 370 nuclear reactions involving 117 isotopes ranging from $^{1}$H to $^{48}$Ti is used. Monte Carlo reaction rates are adopted from the most updated compilation of Ref.~\cite{Iliadis:2010a} with additional reactions selected from the reaction rate library of Iliadis (2005). The only exception is the $^{29}$P($p, \gamma$)$^{30}$S reaction, whose rate is chosen (one at a time for each of the three models) from the present work, as well as from Refs.~\cite{Iliadis:2001,Iliadis:2010a} for comparison.\\
\indent These rates are corrected for the stellar enhancement factors to allow for the increase in reaction rates associated with participation of excited states of nuclei in the reactions. Lastly, the impact of the $^{29}$P($p, \gamma$)$^{30}$S stellar reaction rate on nova nucleosynthesis was compared for the three different reported rates: the recommended classical rate from Ref.~\cite{Iliadis:2001}, hereafter model A; the median Monte Carlo rate from Ref.~\cite{Iliadis:2010a}, hereafter model B; and the high Monte Carlo rate from this work, henceforth model C. The main distinctions in the three $^{29}$P($p, \gamma$)$^{30}$S rates used in the present nova simulations arise from different input energies and uncertainties for two resonances corresponding to the 3$^{+}_{1}$ and 2$^{+}_{3}$ states of $^{30}$S (see \S~\ref{Rate} for discussion).\\
\indent The selection of the high Monte Carlo rate from this work instead of the median rate is to account for the highest possible effect of the new rate on the abundances of elements synthesized in novae. While our median rate is 2.3 times larger (at 0.1 GK) than that of Ref.~\cite{Iliadis:2010a} (see top panel in Fig.~\ref{figure12}), the present high rate is a factor of 3.5 and 17 larger (at 0.1 GK) than the median rate of Ref.~\cite{Iliadis:2010a} and the recommended rate of Ref.~\cite{Iliadis:2001}, respectively.

\subsection{\label{Abundances_Results}Results}

\indent To assign different weights to individual shells of the underlying white dwarf, the isotopic abundances obtained from the three aforementioned hydrodynamic nova simulations were averaged over mass within each shell. The total ejected envelope mass for each of the three models is 4.55 $\times$ 10$^{-6}$ $M_{\astrosun}$. The resulting mean abundances (in mass fractions) in the ejected envelope shells for models A to C are given in Table~\ref{tab:8} for a selection of the stable isotopes in the Si-Ca mass region, whose abundances (in mass fractions) are greater than or equal to 10$^{-5}$. Those stable isotopes not included in Tables~\ref{tab:8} did not change significantly between models.\\
\indent For the stable isotopes with 14 $\leq$ $Z$ $\leq$ 20 which are products of the decays of the short-lived radioactive species, a comparison was made between the mean abundances obtained from model C and those obtained from models A and B. With respect to models A and B, the largest abundance change observed from the results of model C is a 6\% decrease in the abundance of $^{29}$Si. This percentage difference is defined to be: [($\mbox{new value}\,-\,\mbox{old value})\,\div\,\mbox{old value}]$, where the ``new'' value is an isotopic abundance or ratio resulting from model C, and the ``old'' values are those resulting from models A or B, whichever gives a higher percentage difference. A negative (positive) percentage difference indicates that the isotopic abundance or ratio resulting from model C is decreased (increased) with respect to that obtained from model A or B.\\
\indent Therefore, changing the $^{29}$P($p, \gamma$)$^{30}$S rate seems to have only a small effect on the abundances of isotopes with $A$ $\approx$ 30 produced in a nova outburst. However, because of the reduced uncertainty in the updated $^{29}$P($p, \gamma$)$^{30}$S rate, we are now more confident in the reliability of the isotopic abundances obtained using model C.\\
\indent The abundance of each stable isotope alone does not provide much useful information. Instead, to compare the isotopic abundances obtained from nova simulations with those observed in presolar grains, one has to investigate an isotopic abundance ratio. For example, the silicon isotopic ratios measured in presolar grains are usually expressed as~\cite{Jose:2004}:
\begin{equation}
\delta\left(\dfrac{^{29,30}\mbox{Si}}{^{28}\mbox{Si}}\right)\,=\,\left[\left(\dfrac{^{29,30}\mbox{Si}}{^{28}\mbox{Si}}\right)_{\mbox{ejecta}}/\left(\dfrac{^{29,30}\mbox{Si}}{^{28}\mbox{Si}}\right)_{\astrosun}\,-\,1\right]\,\times\,1000 , \label{eq:13}
\end{equation}

\begin{table}[ht]
\vspace{-0.2cm}
\caption{\label{tab:8}Selected mean composition of nova ejecta (in mass fractions, for the Si-Ca isotopes) from models of nova explosions on 1.35 $M_{\astrosun}$ ONe white dwarfs. The only difference between models A, B and C is the $^{29}$P($p, \gamma$)$^{30}$S rate used.}
\centering
\begin{ruledtabular}
\begin{tabular}{lccc}
Isotope & \multicolumn{3}{c}{Hydrodynamic Model}\\
\cmidrule[0.05em]{2-4}\addlinespace[0.2mm]
 & A & B & C \\
\cmidrule[0.05em]{2-2}\cmidrule[0.05em]{3-3}\cmidrule[0.05em]{4-4}\addlinespace[0.2mm]
 & Ref.~\cite{Iliadis:2001} & Ref.~\cite{Iliadis:2010a} & present work \\
 & (recommended) & (median) & (high)\\
\midrule[0.05em]\addlinespace[0.5mm]
$^{28}$Si & 3.08$\times$10$^{-02}$ & 3.08$\times$10$^{-02}$ & 3.08$\times$10$^{-02}$ \\
$^{29}$Si & 2.38$\times$10$^{-03}$ & 2.39$\times$10$^{-03}$ & 2.24$\times$10$^{-03}$ \\
$^{30}$Si & 1.54$\times$10$^{-02}$ & 1.54$\times$10$^{-02}$ & 1.51$\times$10$^{-02}$ \\
$^{31}$P & 8.71$\times$10$^{-03}$ & 8.73$\times$10$^{-03}$ & 8.61$\times$10$^{-03}$ \\
$^{32}$S & 5.27$\times$10$^{-02}$ & 5.27$\times$10$^{-02}$ & 5.30$\times$10$^{-02}$ \\
$^{33}$S & 8.02$\times$10$^{-04}$ & 8.01$\times$10$^{-04}$ & 8.17$\times$10$^{-04}$ \\
$^{34}$S & 3.63$\times$10$^{-04}$ & 3.63$\times$10$^{-04}$ & 3.71$\times$10$^{-04}$ \\
$^{35}$Cl & 3.85$\times$10$^{-04}$ & 3.85$\times$10$^{-04}$ & 3.95$\times$10$^{-04}$ \\
$^{36}$Ar & 5.14$\times$10$^{-05}$ & 5.14$\times$10$^{-05}$ & 5.29$\times$10$^{-05}$ \\
$^{38}$Ar & 2.19$\times$10$^{-05}$ & 2.19$\times$10$^{-05}$ & 2.21$\times$10$^{-05}$ \\
\end{tabular}
\end{ruledtabular}
\end{table}

\begin{table}[ht]
\caption{Deviations (in permil) from solar abundances in simulated and measured nova silicon isotopic abundances. Models A to C are explained in the text, and are obtained from hydrodynamic simulations of classical nova outbursts. The measured values (the first four rows) are for SiC presolar grains reported in Refs.~\cite{Jose:2004,Amari:2002}.}
\centering\vspace{0.2cm}
\setlength{\tabcolsep}{4pt}
\begin{tabular}{lllc}
\hline\hline\addlinespace[0.4mm] \hspace{0.6cm}Grain & $\delta(^{29}$Si/$^{28}$Si) & $\delta(^{30}$Si/$^{28}$Si) & Hydrodynamic\\
& \hspace{0.5cm}(\permil) & \hspace{0.5cm}(\permil) & Model \\ [1ex] \hline\addlinespace[0.2mm]
AF15bB-429-3 & 28 $\pm$ 30 & 1118 $\pm$ 44 & \\
AF15bC-126-3 & -105 $\pm$ 17 & 237 $\pm$ 20 & \\
KJGM4C-100-3 & 55 $\pm$ 5 & 119 $\pm$ 6 & \\
KJGM4C-311-6 & -4 $\pm$ 5 & 149 $\pm$ 6 & \\
 & 527.1 & 13970 & A \\
 & 533.5 & 13970 & B \\
 & 437.3 & 13678 & C \\[1ex]
\hline\hline
\end{tabular}
\label{tab:10}
\end{table}

\noindent where $\delta$ represents deviations from solar abundances in permil, and the adopted numerical values for the solar silicon isotopic ratios are~\cite{Evans:2008} (p.~130): $\left(^{29}\mbox{Si}/^{28}\mbox{Si}\right)_{\astrosun}\,=\,0.0506$ and $\left(^{30}\mbox{Si}/^{28}\mbox{Si}\right)_{\astrosun}\,=\,0.0334$. The deviations from solar abundances are computed for silicon isotopic abundance ratios obtained from models A, B and C, and the results are shown in Table~\ref{tab:10}, along with $^{29,30}$Si/$^{28}$Si ratios measured~\cite{Amari:2002} from some SiC presolar grains with proposed classical nova paternity.\\
\indent As seen in Table~\ref{tab:10}, the theoretically predicted $\delta$-values are much larger than the measured counterparts. Overall, however, regardless of the $^{29}$P($p, \gamma$)$^{30}$S reaction rate used, the $^{29}$Si/$^{28}$Si ratio in the ejecta resulting from the simulations is only slightly higher ($\sim$1.5 times larger, see Table~\ref{tab:8}) than the solar value. Using the measured $\delta(^{29}$Si/$^{28}$Si) values, given in Table~\ref{tab:10}, as inputs to Eq.~(\ref{eq:13}), we extract a measured $^{29}$Si/$^{28}$Si ratio that varies between a factor of 0.9 -- 1.1 times the solar ratio, and thus is again only slightly lower or higher than the solar value. Therefore, even though the new $^{29}$P($p, \gamma$)$^{30}$S rate does not significantly improve the theoretical $\delta$-values, the simulated signatures are qualitatively consistent with the $^{29}$Si/$^{28}$Si ratios measured in presolar grains identified to have a nova origin. In other words, the simulated and measured $\delta$-values both show enhancements in the same direction.\\
\indent On the other hand, the $^{30}$Si/$^{28}$Si ratio in the ejecta resulting from the simulations is much higher ($\sim$15 times larger) than the solar value (see Table~\ref{tab:8}), such that the classical nova ejecta resulting from the hydrodynamic models is significantly enriched in $^{30}$Si. The simulated and measured values again are in qualitative agreement with each other, i.e., enhanced in the same direction, but the magnitudes of the enhancements are not in agreement.\\
\indent Our results support the indication that in order for the models to predict the $^{30}$Si/$^{28}$Si ejecta ratio that quantitatively matches the grain data, one has to assume a mixing process between material newly synthesized in the nova outburst and more than 10 times as much unprocessed, isotopically close to solar, material before the process of grain formation~\cite{Jose:2004,Amari:2001}. The details of the ejecta dilution and the grain formation processes are still unknown.\\
\indent In addition to invoking the mixing with solar composition material, an increase in the $^{30}$P($p, \gamma$)$^{31}$S reaction rate also helps reduce the $^{30}$Si/$^{28}$Si ratio by moving the nucleosynthesis flow away from $^{30}$P toward the heavier isotopes. A decrease in the abundance of $^{30}$P consequently reduces that of $^{30}$Si produced from $^{30}$P($\beta^{+}$)$^{30}$Si. The rate of the $^{30}$P($p, \gamma$)$^{31}$S reaction has been evaluated in Refs.~\cite{Jose:2001,Wrede:2009,Parikh:2011} and more recently in Ref.~\cite{Doherty:2012}. This last rate is found to be $\sim$10 times greater, at $T$ $\sim$ 0.25 GK, than the lower limit set in Ref.~\cite{Parikh:2011}. A factor of $\sim$10 increase in the $^{30}$P($p, \gamma$)$^{31}$S reaction rate results in a typical factor of $\sim$10 reduction in the expected abundance of $^{30}$Si~\cite{Doherty:2012}. This new information may now help to better constrain the dilution process in new nova model predictions.\\
\indent In comparison with the high Monte Carlo rate from the present work, the present median and low Monte Carlo rates show smaller deviations with respect to the median rates of Refs.~\cite{Iliadis:2001,Iliadis:2010a}. Therefore, we did not extend our investigation to study the effects of these rates on the nova yields.

\section{Conclusions}

\indent The $^{29}$P($p, \gamma$)$^{30}$S reaction rate at the temperature range of 0.1 $\leq$ $T$ $\leq$ 1.3 GK is dominated by two low energy resonances just above the proton threshold (4394.9(7) keV) corresponding to two excited states in $^{30}$S in the $E_{x}$ $\approx$ 4.7 -- 4.8 MeV range, whose $J^{\pi}$ values were previously estimated~\cite{Iliadis:2001} to be 3$^{+}$ and 2$^{+}$, respectively. We have observed these excited states in $^{30}$S at 4688.1(4) keV and 4809.8(6) keV, respectively, via two separate experiments: the $^{32}$S($p, t$)$^{30}$S two-nucleon transfer reaction and an in-beam $\gamma$-ray spectroscopy experiment via the $^{28}$Si($^{3}$He, $n\gamma$)$^{30}$S reaction.\\
\indent Both of our experiments result in measured resonance energies, corresponding to the aforementioned excited states, which are in excellent agreement with each other. Moreover, we have been able to reduce the uncertainty in the energies of these resonances with respect to what was previously observed~\cite{Bardayan:2007} for the 3$^{+}$ resonance and predicted~\cite{Iliadis:2001} for the 2$^{+}$ resonance. Furthermore, we have confirmed the spin-parity assignments of both of these resonances. As a result, our new $^{29}$P($p, \gamma$)$^{30}$S reaction rate is increased by a factor of 2 over the temperature range of 0.1 $\leq$ $T$ $\leq$ 1.3 GK. Also, the uncertainty in our new rate in this temperature range has been reduced by 72\% relative to that previously determined~\cite{Iliadis:2010a}.\\
\indent This updated rate have been used to compute a full hydrodynamic nova simulation which is more realistic than the post-processing nucleosynthesis simulation performed in Ref.~\cite{Iliadis:2002}.\\
\indent Our new $^{29}$P($p, \gamma$)$^{30}$S rate has only marginally improved the agreement between the abundances observed in presolar grains of potential nova origin and those obtained from simulations. Although, our updated rate does not affect the silicon isotopic abundance ratios significantly, due to a reduction in its uncertainty, the present nova hydro simulations can be compared with more reliability to the isotopic ratios measured in presolar grains of potential nova paternity.\\
\indent As for the nuclear structure of $^{30}$S, improvements in spin-parity assignments may be made by theoretical estimates via the IMME for those $^{30}$S states whose spin-parity assignments are still tentative. However, this method is currently unreliable for $A$ $=$ 30 because many of the relevant analog states in $^{30}$P also have unknown or tentative spin-parity assignments~\cite{Basunia:2010}. Thus, if such properties of the levels of $^{30}$P are constrained better in the future, this in turn will help with the determination of those of $^{30}$S.

\section*{\label{Acknowledgment}ACKNOWLEDGEMENTS}

\indent K.S.~wishes to thank C.~Iliadis, D.~W.~Bardayan, B.~Singh and A.~M.~Moro for their assistance with the data analysis, as well as for providing some crucial software programs without which the data analysis could not be completed. Also, we would like to thank the staff of WNSL and the UTTAC for their contributions. This work was supported by the Natural Sciences and Engineering Research Council of Canada; the U.S.~Department of Energy under Grant Nos.~DE-FG02-91ER40609, DE-AC02-06CH11357 and DE-FG02-97ER41020; the Grant-in-Aid for Science Research KAKENHI 21540295 of Japan; the JSPS KAKENHI and JSPS Bilateral Joint Project of Japan; Japan Society for the Promotion of Science Core-to-Core Program on International Research Network for Exotic Femto Systems under Grant No.~18002; the DFG cluster of excellence ``Origin and Structure of the Universe''; ESF EUROCORES Program EuroGENESIS through the MICINN Grant No.~EUI2009-04167; and the Spanish Grant AYA2010-15685.


\bibliographystyle{apsrev}
\bibliography{References}

\begin{thebibliography}{70}
\expandafter\ifx\csname natexlab\endcsname\relax\def\natexlab#1{#1}\fi
\expandafter\ifx\csname bibnamefont\endcsname\relax
  \def\bibnamefont#1{#1}\fi
\expandafter\ifx\csname bibfnamefont\endcsname\relax
  \def\bibfnamefont#1{#1}\fi
\expandafter\ifx\csname citenamefont\endcsname\relax
  \def\citenamefont#1{#1}\fi
\expandafter\ifx\csname url\endcsname\relax
  \def\url#1{\texttt{#1}}\fi
\expandafter\ifx\csname urlprefix\endcsname\relax\def\urlprefix{URL }\fi
\providecommand{\bibinfo}[2]{#2}
\providecommand{\eprint}[2][]{\url{#2}}

\bibitem[{\citenamefont{Iliadis et~al.}(2002)\citenamefont{Iliadis, Champagne,
  Jos$\acute{\mbox{e}}$, Starrfield, and Tupper}}]{Iliadis:2002}
\bibinfo{author}{\bibfnamefont{C.}~\bibnamefont{Iliadis}},
  \bibinfo{author}{\bibfnamefont{A.~E.} \bibnamefont{Champagne}},
  \bibinfo{author}{\bibfnamefont{J.}~\bibnamefont{Jos$\acute{\mbox{e}}$}},
  \bibinfo{author}{\bibfnamefont{S.}~\bibnamefont{Starrfield}},
  \bibnamefont{and} \bibinfo{author}{\bibfnamefont{P.}~\bibnamefont{Tupper}},
  \bibinfo{journal}{Astrophys.~J.~Suppl.~Ser.} \textbf{\bibinfo{volume}{142}},
  \bibinfo{pages}{105} (\bibinfo{year}{2002}).

\bibitem[{\citenamefont{Jos$\acute{\mbox{e}}$
  et~al.}(2006)\citenamefont{Jos$\acute{\mbox{e}}$, Hernanz, and
  Iliadis}}]{Jose:2006}
\bibinfo{author}{\bibfnamefont{J.}~\bibnamefont{Jos$\acute{\mbox{e}}$}},
  \bibinfo{author}{\bibfnamefont{M.}~\bibnamefont{Hernanz}}, \bibnamefont{and}
  \bibinfo{author}{\bibfnamefont{C.}~\bibnamefont{Iliadis}},
  \bibinfo{journal}{Nucl.~Phys.~A} \textbf{\bibinfo{volume}{777}},
  \bibinfo{pages}{550} (\bibinfo{year}{2006}).

\bibitem[{\citenamefont{Gehrz et~al.}(1998)\citenamefont{Gehrz, Truran,
  Williams, and Starrfield}}]{Gehrz:1998}
\bibinfo{author}{\bibfnamefont{R.~D.} \bibnamefont{Gehrz}},
  \bibinfo{author}{\bibfnamefont{J.~W.} \bibnamefont{Truran}},
  \bibinfo{author}{\bibfnamefont{R.~E.} \bibnamefont{Williams}},
  \bibnamefont{and}
  \bibinfo{author}{\bibfnamefont{S.}~\bibnamefont{Starrfield}},
  \bibinfo{journal}{Publ.~Astron.~Soc.~Pac.} \textbf{\bibinfo{volume}{110}},
  \bibinfo{pages}{3} (\bibinfo{year}{1998}).

\bibitem[{\citenamefont{Starrfield et~al.}(2007)\citenamefont{Starrfield,
  Iliadis, Hix, Timmes, and Sparks}}]{Starrfield:2007}
\bibinfo{author}{\bibfnamefont{S.}~\bibnamefont{Starrfield}},
  \bibinfo{author}{\bibfnamefont{C.}~\bibnamefont{Iliadis}},
  \bibinfo{author}{\bibfnamefont{W.~R.} \bibnamefont{Hix}},
  \bibinfo{author}{\bibfnamefont{F.~X.} \bibnamefont{Timmes}},
  \bibnamefont{and} \bibinfo{author}{\bibfnamefont{W.~M.}
  \bibnamefont{Sparks}}, \bibinfo{journal}{AIP Conf.~Proc.}
  \textbf{\bibinfo{volume}{891}}, \bibinfo{pages}{364} (\bibinfo{year}{2007}).

\bibitem[{\citenamefont{Amari et~al.}(2001)\citenamefont{Amari, Gao, Nittler,
  Zinner, Jos$\acute{\mbox{e}}$, Hernanz, and Lewis}}]{Amari:2001}
\bibinfo{author}{\bibfnamefont{S.}~\bibnamefont{Amari}},
  \bibinfo{author}{\bibfnamefont{X.}~\bibnamefont{Gao}},
  \bibinfo{author}{\bibfnamefont{L.~R.} \bibnamefont{Nittler}},
  \bibinfo{author}{\bibfnamefont{E.}~\bibnamefont{Zinner}},
  \bibinfo{author}{\bibfnamefont{J.}~\bibnamefont{Jos$\acute{\mbox{e}}$}},
  \bibinfo{author}{\bibfnamefont{M.}~\bibnamefont{Hernanz}}, \bibnamefont{and}
  \bibinfo{author}{\bibfnamefont{R.~S.} \bibnamefont{Lewis}},
  \bibinfo{journal}{Astrophys.~J.} \textbf{\bibinfo{volume}{551}},
  \bibinfo{pages}{1065} (\bibinfo{year}{2001}).

\bibitem[{\citenamefont{Amari}(2002)}]{Amari:2002}
\bibinfo{author}{\bibfnamefont{S.}~\bibnamefont{Amari}}, \bibinfo{journal}{New
  Astron.~Rev.} \textbf{\bibinfo{volume}{46}}, \bibinfo{pages}{519}
  (\bibinfo{year}{2002}).

\bibitem[{\citenamefont{Jos$\acute{\mbox{e}}$
  et~al.}(2004)\citenamefont{Jos$\acute{\mbox{e}}$, Hernanz, Amari, Lodders,
  and Zinner}}]{Jose:2004}
\bibinfo{author}{\bibfnamefont{J.}~\bibnamefont{Jos$\acute{\mbox{e}}$}},
  \bibinfo{author}{\bibfnamefont{M.}~\bibnamefont{Hernanz}},
  \bibinfo{author}{\bibfnamefont{S.}~\bibnamefont{Amari}},
  \bibinfo{author}{\bibfnamefont{K.}~\bibnamefont{Lodders}}, \bibnamefont{and}
  \bibinfo{author}{\bibfnamefont{E.}~\bibnamefont{Zinner}},
  \bibinfo{journal}{Astrophys.~J.} \textbf{\bibinfo{volume}{612}},
  \bibinfo{pages}{414} (\bibinfo{year}{2004}).

\bibitem[{\citenamefont{Jos$\acute{\mbox{e}}$
  et~al.}(2001)\citenamefont{Jos$\acute{\mbox{e}}$, Coc, and
  Hernanz}}]{Jose:2001}
\bibinfo{author}{\bibfnamefont{J.}~\bibnamefont{Jos$\acute{\mbox{e}}$}},
  \bibinfo{author}{\bibfnamefont{A.}~\bibnamefont{Coc}}, \bibnamefont{and}
  \bibinfo{author}{\bibfnamefont{M.}~\bibnamefont{Hernanz}},
  \bibinfo{journal}{Astrophys.~J.} \textbf{\bibinfo{volume}{560}},
  \bibinfo{pages}{897} (\bibinfo{year}{2001}).

\bibitem[{\citenamefont{Iliadis et~al.}(2001)\citenamefont{Iliadis, D'Auria,
  Starrfield, Thompson, and Wiescher}}]{Iliadis:2001}
\bibinfo{author}{\bibfnamefont{C.}~\bibnamefont{Iliadis}},
  \bibinfo{author}{\bibfnamefont{J.~M.} \bibnamefont{D'Auria}},
  \bibinfo{author}{\bibfnamefont{S.}~\bibnamefont{Starrfield}},
  \bibinfo{author}{\bibfnamefont{W.~J.} \bibnamefont{Thompson}},
  \bibnamefont{and} \bibinfo{author}{\bibfnamefont{M.}~\bibnamefont{Wiescher}},
  \bibinfo{journal}{Astrophys.~J.~Suppl.~Ser.} \textbf{\bibinfo{volume}{134}},
  \bibinfo{pages}{151} (\bibinfo{year}{2001}).

\bibitem[{\citenamefont{Souin et~al.}(2011)\citenamefont{Souin, Ascher,
  Audirac, \"{A}yst\"{o}, Blank, Elomaa, Eronen, Giovinazzo, Hakala, Jokinen
  et~al.}}]{Souin:2011}
\bibinfo{author}{\bibfnamefont{J.}~\bibnamefont{Souin}},
  \bibinfo{author}{\bibfnamefont{P.}~\bibnamefont{Ascher}},
  \bibinfo{author}{\bibfnamefont{L.}~\bibnamefont{Audirac}},
  \bibinfo{author}{\bibfnamefont{J.}~\bibnamefont{\"{A}yst\"{o}}},
  \bibinfo{author}{\bibfnamefont{B.}~\bibnamefont{Blank}},
  \bibinfo{author}{\bibfnamefont{V.-V.} \bibnamefont{Elomaa}},
  \bibinfo{author}{\bibfnamefont{T.}~\bibnamefont{Eronen}},
  \bibinfo{author}{\bibfnamefont{J.}~\bibnamefont{Giovinazzo}},
  \bibinfo{author}{\bibfnamefont{J.}~\bibnamefont{Hakala}},
  \bibinfo{author}{\bibfnamefont{A.}~\bibnamefont{Jokinen}},
  \bibnamefont{et~al.}, \bibinfo{journal}{Eur.~J.~Phys.~A}
  \textbf{\bibinfo{volume}{47}}, \bibinfo{pages}{40} (\bibinfo{year}{2011}).

\bibitem[{\citenamefont{Wiescher and
  G$\ddot{\mbox{o}}$rres}(1988)}]{Wiescher:1988}
\bibinfo{author}{\bibfnamefont{M.}~\bibnamefont{Wiescher}} \bibnamefont{and}
  \bibinfo{author}{\bibfnamefont{J.}~\bibnamefont{G$\ddot{\mbox{o}}$rres}},
  \bibinfo{journal}{Z.~Phys.~A} \textbf{\bibinfo{volume}{329}},
  \bibinfo{pages}{121} (\bibinfo{year}{1988}).

\bibitem[{\citenamefont{Iliadis
  et~al.}(2010{\natexlab{a}})\citenamefont{Iliadis, Longland, Champagne, Coc,
  and Fitzgerald}}]{Iliadis:2010a}
\bibinfo{author}{\bibfnamefont{C.}~\bibnamefont{Iliadis}},
  \bibinfo{author}{\bibfnamefont{R.}~\bibnamefont{Longland}},
  \bibinfo{author}{\bibfnamefont{A.~E.} \bibnamefont{Champagne}},
  \bibinfo{author}{\bibfnamefont{A.}~\bibnamefont{Coc}}, \bibnamefont{and}
  \bibinfo{author}{\bibfnamefont{R.}~\bibnamefont{Fitzgerald}},
  \bibinfo{journal}{Nucl.~Phys.~A} \textbf{\bibinfo{volume}{841}},
  \bibinfo{pages}{31} (\bibinfo{year}{2010}{\natexlab{a}}).

\bibitem[{\citenamefont{Bardayan et~al.}(2007)\citenamefont{Bardayan, Blackmon,
  Fitzgerald, Hix, Jones, Kozub, Liang, Livesay, Ma, Roberts
  et~al.}}]{Bardayan:2007}
\bibinfo{author}{\bibfnamefont{D.~W.} \bibnamefont{Bardayan}},
  \bibinfo{author}{\bibfnamefont{J.~C.} \bibnamefont{Blackmon}},
  \bibinfo{author}{\bibfnamefont{R.~P.} \bibnamefont{Fitzgerald}},
  \bibinfo{author}{\bibfnamefont{W.~R.} \bibnamefont{Hix}},
  \bibinfo{author}{\bibfnamefont{K.~L.} \bibnamefont{Jones}},
  \bibinfo{author}{\bibfnamefont{R.~L.} \bibnamefont{Kozub}},
  \bibinfo{author}{\bibfnamefont{J.~F.} \bibnamefont{Liang}},
  \bibinfo{author}{\bibfnamefont{R.~J.} \bibnamefont{Livesay}},
  \bibinfo{author}{\bibfnamefont{Z.}~\bibnamefont{Ma}},
  \bibinfo{author}{\bibfnamefont{L.~F.} \bibnamefont{Roberts}},
  \bibnamefont{et~al.}, \bibinfo{journal}{Phys.~Rev.~C}
  \textbf{\bibinfo{volume}{76}}, \bibinfo{pages}{045803}
  (\bibinfo{year}{2007}).

\bibitem[{\citenamefont{Paddock}(1972)}]{Paddock:1972}
\bibinfo{author}{\bibfnamefont{R.~A.} \bibnamefont{Paddock}},
  \bibinfo{journal}{Phys.~Rev.~C} \textbf{\bibinfo{volume}{5}},
  \bibinfo{pages}{485} (\bibinfo{year}{1972}).

\bibitem[{\citenamefont{Cara\c{c}a et~al.}(1972)\citenamefont{Cara\c{c}a, Gill,
  Cox, and Rose}}]{Caraca:1972}
\bibinfo{author}{\bibfnamefont{J.~M.~G.} \bibnamefont{Cara\c{c}a}},
  \bibinfo{author}{\bibfnamefont{R.~D.} \bibnamefont{Gill}},
  \bibinfo{author}{\bibfnamefont{A.~J.} \bibnamefont{Cox}}, \bibnamefont{and}
  \bibinfo{author}{\bibfnamefont{H.~J.} \bibnamefont{Rose}},
  \bibinfo{journal}{Nucl.~Phys.~A} \textbf{\bibinfo{volume}{193}},
  \bibinfo{pages}{1} (\bibinfo{year}{1972}).

\bibitem[{\citenamefont{Kuhlmann et~al.}(1973)\citenamefont{Kuhlmann, Albrecht,
  and Hoffmann}}]{Kuhlmann:1973}
\bibinfo{author}{\bibfnamefont{E.}~\bibnamefont{Kuhlmann}},
  \bibinfo{author}{\bibfnamefont{W.}~\bibnamefont{Albrecht}}, \bibnamefont{and}
  \bibinfo{author}{\bibfnamefont{A.}~\bibnamefont{Hoffmann}},
  \bibinfo{journal}{Nucl.~Phys.~A} \textbf{\bibinfo{volume}{213}},
  \bibinfo{pages}{82} (\bibinfo{year}{1973}).

\bibitem[{\citenamefont{Yokota et~al.}(1982)\citenamefont{Yokota, Fujioka,
  Ichimaru, Mihara, and Chiba}}]{Yokota:1982}
\bibinfo{author}{\bibfnamefont{H.}~\bibnamefont{Yokota}},
  \bibinfo{author}{\bibfnamefont{K.}~\bibnamefont{Fujioka}},
  \bibinfo{author}{\bibfnamefont{K.}~\bibnamefont{Ichimaru}},
  \bibinfo{author}{\bibfnamefont{Y.}~\bibnamefont{Mihara}}, \bibnamefont{and}
  \bibinfo{author}{\bibfnamefont{R.}~\bibnamefont{Chiba}},
  \bibinfo{journal}{Nucl.~Phys.~A} \textbf{\bibinfo{volume}{383}},
  \bibinfo{pages}{298} (\bibinfo{year}{1982}).

\bibitem[{\citenamefont{Fynbo et~al.}(2000)\citenamefont{Fynbo, Borge,
  Axelsson, \"{A}yst\"{o}, Bergmann, Fraile, Honkanen, Hornsh${\o}$j, Jading,
  Jokinen et~al.}}]{Fynbo:2000}
\bibinfo{author}{\bibfnamefont{H.~O.~U.} \bibnamefont{Fynbo}},
  \bibinfo{author}{\bibfnamefont{M.~J.~G.} \bibnamefont{Borge}},
  \bibinfo{author}{\bibfnamefont{L.}~\bibnamefont{Axelsson}},
  \bibinfo{author}{\bibfnamefont{J.}~\bibnamefont{\"{A}yst\"{o}}},
  \bibinfo{author}{\bibfnamefont{U.~C.} \bibnamefont{Bergmann}},
  \bibinfo{author}{\bibfnamefont{L.~M.} \bibnamefont{Fraile}},
  \bibinfo{author}{\bibfnamefont{A.}~\bibnamefont{Honkanen}},
  \bibinfo{author}{\bibfnamefont{P.}~\bibnamefont{Hornsh${\o}$j}},
  \bibinfo{author}{\bibfnamefont{Y.}~\bibnamefont{Jading}},
  \bibinfo{author}{\bibfnamefont{A.}~\bibnamefont{Jokinen}},
  \bibnamefont{et~al.}, \bibinfo{journal}{Nucl.~Phys.~A}
  \textbf{\bibinfo{volume}{677}}, \bibinfo{pages}{38} (\bibinfo{year}{2000}).

\bibitem[{\citenamefont{Galaviz et~al.}(Geneva, 2006)\citenamefont{Galaviz,
  Amthor, Bazin, Brown, Cole, Elliot, Estrade, F\"{u}l\"{o}p, Gade, Glasmacher
  et~al.}}]{Galaviz:2006}
\bibinfo{author}{\bibfnamefont{D.}~\bibnamefont{Galaviz}},
  \bibinfo{author}{\bibfnamefont{A.~M.} \bibnamefont{Amthor}},
  \bibinfo{author}{\bibfnamefont{D.}~\bibnamefont{Bazin}},
  \bibinfo{author}{\bibfnamefont{B.~A.} \bibnamefont{Brown}},
  \bibinfo{author}{\bibfnamefont{A.}~\bibnamefont{Cole}},
  \bibinfo{author}{\bibfnamefont{T.}~\bibnamefont{Elliot}},
  \bibinfo{author}{\bibfnamefont{A.}~\bibnamefont{Estrade}},
  \bibinfo{author}{\bibfnamefont{Z.}~\bibnamefont{F\"{u}l\"{o}p}},
  \bibinfo{author}{\bibfnamefont{A.}~\bibnamefont{Gade}},
  \bibinfo{author}{\bibfnamefont{T.}~\bibnamefont{Glasmacher}},
  \bibnamefont{et~al.}, in \emph{\bibinfo{booktitle}{Proceedings of Science:
  International Symposium on Nuclear Astrophysics -- Nuclei in the Cosmos --
  IX}} (\bibinfo{year}{Geneva, 2006}), \bibinfo{note}{{P}o{S}(NIC-IX)099}.

\bibitem[{Fig()}]{Figueira:2008}
\bibinfo{howpublished}{J.~M.~Figueira {\em et al.}, Argonne National Laboratory
  ATLAS Proposal No.~1242 (2008), (unpublished)}.

\bibitem[{\citenamefont{O'Brien et~al.}(2009)\citenamefont{O'Brien, Adachi,
  Berg, Couder, Dozono, Fujita, Fujita, G$\ddot{\mbox{o}}$rres, Hatanaka,
  Ishikawa et~al.}}]{O'Brien:2009}
\bibinfo{author}{\bibfnamefont{S.}~\bibnamefont{O'Brien}},
  \bibinfo{author}{\bibfnamefont{T.}~\bibnamefont{Adachi}},
  \bibinfo{author}{\bibfnamefont{G.~P.~A.} \bibnamefont{Berg}},
  \bibinfo{author}{\bibfnamefont{M.}~\bibnamefont{Couder}},
  \bibinfo{author}{\bibfnamefont{M.}~\bibnamefont{Dozono}},
  \bibinfo{author}{\bibfnamefont{H.}~\bibnamefont{Fujita}},
  \bibinfo{author}{\bibfnamefont{Y.}~\bibnamefont{Fujita}},
  \bibinfo{author}{\bibfnamefont{J.}~\bibnamefont{G$\ddot{\mbox{o}}$rres}},
  \bibinfo{author}{\bibfnamefont{K.}~\bibnamefont{Hatanaka}},
  \bibinfo{author}{\bibfnamefont{D.}~\bibnamefont{Ishikawa}},
  \bibnamefont{et~al.}, \bibinfo{journal}{AIP Conf.~Proc.}
  \textbf{\bibinfo{volume}{1090}}, \bibinfo{pages}{288} (\bibinfo{year}{2009}).

\bibitem[{\citenamefont{Galaviz et~al.}(2010)\citenamefont{Galaviz, Amthor,
  Bazin, Becerril, Brown, Chen, Cole, Cook, Elliot, Estrade
  et~al.}}]{Galaviz:2010}
\bibinfo{author}{\bibfnamefont{D.}~\bibnamefont{Galaviz}},
  \bibinfo{author}{\bibfnamefont{A.~M.} \bibnamefont{Amthor}},
  \bibinfo{author}{\bibfnamefont{D.}~\bibnamefont{Bazin}},
  \bibinfo{author}{\bibfnamefont{A.~D.} \bibnamefont{Becerril}},
  \bibinfo{author}{\bibfnamefont{B.~A.} \bibnamefont{Brown}},
  \bibinfo{author}{\bibfnamefont{A.~A.} \bibnamefont{Chen}},
  \bibinfo{author}{\bibfnamefont{A.}~\bibnamefont{Cole}},
  \bibinfo{author}{\bibfnamefont{J.~M.} \bibnamefont{Cook}},
  \bibinfo{author}{\bibfnamefont{T.}~\bibnamefont{Elliot}},
  \bibinfo{author}{\bibfnamefont{A.}~\bibnamefont{Estrade}},
  \bibnamefont{et~al.}, \bibinfo{journal}{J.~Phys.: Conf.~Ser.}
  \textbf{\bibinfo{volume}{202}}, \bibinfo{pages}{012009}
  (\bibinfo{year}{2010}).

\bibitem[{\citenamefont{Setoodehnia et~al.}(2010)\citenamefont{Setoodehnia,
  Chen, Chen, Clark, Deibel, Geraedts, Kahl, Parker, Seiler, and
  Wrede}}]{Setoodehnia:2010}
\bibinfo{author}{\bibfnamefont{K.}~\bibnamefont{Setoodehnia}},
  \bibinfo{author}{\bibfnamefont{A.~A.} \bibnamefont{Chen}},
  \bibinfo{author}{\bibfnamefont{J.}~\bibnamefont{Chen}},
  \bibinfo{author}{\bibfnamefont{J.~A.} \bibnamefont{Clark}},
  \bibinfo{author}{\bibfnamefont{C.~M.} \bibnamefont{Deibel}},
  \bibinfo{author}{\bibfnamefont{S.~D.} \bibnamefont{Geraedts}},
  \bibinfo{author}{\bibfnamefont{D.}~\bibnamefont{Kahl}},
  \bibinfo{author}{\bibfnamefont{P.~D.} \bibnamefont{Parker}},
  \bibinfo{author}{\bibfnamefont{D.}~\bibnamefont{Seiler}}, \bibnamefont{and}
  \bibinfo{author}{\bibfnamefont{C.}~\bibnamefont{Wrede}},
  \bibinfo{journal}{Phys.~Rev.~C} \textbf{\bibinfo{volume}{82}},
  \bibinfo{pages}{022801(R)} (\bibinfo{year}{2010}).

\bibitem[{\citenamefont{Setoodehnia et~al.}(2011)\citenamefont{Setoodehnia,
  Chen, Komatsubara, Kubono, Binh, Carpino, Chen, Hashimoto, Hayakawa,
  Ishibashi et~al.}}]{Setoodehnia:2011a}
\bibinfo{author}{\bibfnamefont{K.}~\bibnamefont{Setoodehnia}},
  \bibinfo{author}{\bibfnamefont{A.~A.} \bibnamefont{Chen}},
  \bibinfo{author}{\bibfnamefont{T.}~\bibnamefont{Komatsubara}},
  \bibinfo{author}{\bibfnamefont{S.}~\bibnamefont{Kubono}},
  \bibinfo{author}{\bibfnamefont{D.~N.} \bibnamefont{Binh}},
  \bibinfo{author}{\bibfnamefont{J.~F.} \bibnamefont{Carpino}},
  \bibinfo{author}{\bibfnamefont{J.}~\bibnamefont{Chen}},
  \bibinfo{author}{\bibfnamefont{T.}~\bibnamefont{Hashimoto}},
  \bibinfo{author}{\bibfnamefont{T.}~\bibnamefont{Hayakawa}},
  \bibinfo{author}{\bibfnamefont{Y.}~\bibnamefont{Ishibashi}},
  \bibnamefont{et~al.}, \bibinfo{journal}{Phys.~Rev.~C}
  \textbf{\bibinfo{volume}{83}}, \bibinfo{pages}{018803}
  (\bibinfo{year}{2011}).

\bibitem[{\citenamefont{Setoodehnia}(2011)}]{Setoodehnia:2011b}
\bibinfo{author}{\bibfnamefont{K.}~\bibnamefont{Setoodehnia}}, Ph.D. thesis,
  \bibinfo{school}{McMaster University} (\bibinfo{year}{2011}),
  \bibinfo{note}{\url{http://digitalcommons.mcmaster.ca/opendissertations/6313%
/}}.

\bibitem[{\citenamefont{Lennard et~al.}(2011)\citenamefont{Lennard,
  Setoodehnia, Chen, and Hendriks}}]{Lennard:2011}
\bibinfo{author}{\bibfnamefont{W.~N.} \bibnamefont{Lennard}},
  \bibinfo{author}{\bibfnamefont{K.}~\bibnamefont{Setoodehnia}},
  \bibinfo{author}{\bibfnamefont{A.~A.} \bibnamefont{Chen}}, \bibnamefont{and}
  \bibinfo{author}{\bibfnamefont{J.}~\bibnamefont{Hendriks}},
  \bibinfo{journal}{Nucl.~Instr.~Meth.~Phys.~Res.~B}
  \textbf{\bibinfo{volume}{269}}, \bibinfo{pages}{2726} (\bibinfo{year}{2011}).

\bibitem[{\citenamefont{Bell et~al.}(1969)\citenamefont{Bell, L'Ecuyer, Gill,
  Robertson, Towner, and Rose}}]{Bell:1969}
\bibinfo{author}{\bibfnamefont{R.~A.~I.} \bibnamefont{Bell}},
  \bibinfo{author}{\bibfnamefont{J.}~\bibnamefont{L'Ecuyer}},
  \bibinfo{author}{\bibfnamefont{R.~D.} \bibnamefont{Gill}},
  \bibinfo{author}{\bibfnamefont{B.~C.} \bibnamefont{Robertson}},
  \bibinfo{author}{\bibfnamefont{I.~S.} \bibnamefont{Towner}},
  \bibnamefont{and} \bibinfo{author}{\bibfnamefont{H.~J.} \bibnamefont{Rose}},
  \bibinfo{journal}{Nucl.~Phys.~A} \textbf{\bibinfo{volume}{133}},
  \bibinfo{pages}{337} (\bibinfo{year}{1969}).

\bibitem[{\citenamefont{Bohne et~al.}(1982)\citenamefont{Bohne,
  B$\ddot{\mbox{u}}$chs, Fuchs, Grabisch, Hilscher, Jahnke, Kluge, Masterson,
  and Morgenstern}}]{Bohne:1982}
\bibinfo{author}{\bibfnamefont{W.}~\bibnamefont{Bohne}},
  \bibinfo{author}{\bibfnamefont{K.~D.} \bibnamefont{B$\ddot{\mbox{u}}$chs}},
  \bibinfo{author}{\bibfnamefont{H.}~\bibnamefont{Fuchs}},
  \bibinfo{author}{\bibfnamefont{K.}~\bibnamefont{Grabisch}},
  \bibinfo{author}{\bibfnamefont{D.}~\bibnamefont{Hilscher}},
  \bibinfo{author}{\bibfnamefont{U.}~\bibnamefont{Jahnke}},
  \bibinfo{author}{\bibfnamefont{H.}~\bibnamefont{Kluge}},
  \bibinfo{author}{\bibfnamefont{T.~G.} \bibnamefont{Masterson}},
  \bibnamefont{and}
  \bibinfo{author}{\bibfnamefont{H.}~\bibnamefont{Morgenstern}},
  \bibinfo{journal}{Nucl.~Phys.~A} \textbf{\bibinfo{volume}{378}},
  \bibinfo{pages}{525} (\bibinfo{year}{1982}).

\bibitem[{\citenamefont{Caggiano et~al.}(2002)\citenamefont{Caggiano,
  Bradfield-Smith, Lewis, Parker, Visser, Greene, Rehm, Bardayan, and
  Champagne}}]{Caggiano:2002}
\bibinfo{author}{\bibfnamefont{J.~A.} \bibnamefont{Caggiano}},
  \bibinfo{author}{\bibfnamefont{W.}~\bibnamefont{Bradfield-Smith}},
  \bibinfo{author}{\bibfnamefont{R.}~\bibnamefont{Lewis}},
  \bibinfo{author}{\bibfnamefont{P.~D.} \bibnamefont{Parker}},
  \bibinfo{author}{\bibfnamefont{D.~W.} \bibnamefont{Visser}},
  \bibinfo{author}{\bibfnamefont{J.~P.} \bibnamefont{Greene}},
  \bibinfo{author}{\bibfnamefont{K.~E.} \bibnamefont{Rehm}},
  \bibinfo{author}{\bibfnamefont{D.~W.} \bibnamefont{Bardayan}},
  \bibnamefont{and} \bibinfo{author}{\bibfnamefont{A.~E.}
  \bibnamefont{Champagne}}, \bibinfo{journal}{Phys.~Rev.~C}
  \textbf{\bibinfo{volume}{65}}, \bibinfo{pages}{055801}
  (\bibinfo{year}{2002}).

\bibitem[{\citenamefont{Bardayan et~al.}(2002)\citenamefont{Bardayan, Blackmon,
  Champagne, Dummer, Davinson, Greife, Hill, Iliadis, Johnson, Kozub
  et~al.}}]{Bardayan:2002}
\bibinfo{author}{\bibfnamefont{D.~W.} \bibnamefont{Bardayan}},
  \bibinfo{author}{\bibfnamefont{J.~C.} \bibnamefont{Blackmon}},
  \bibinfo{author}{\bibfnamefont{A.~E.} \bibnamefont{Champagne}},
  \bibinfo{author}{\bibfnamefont{A.~K.} \bibnamefont{Dummer}},
  \bibinfo{author}{\bibfnamefont{T.}~\bibnamefont{Davinson}},
  \bibinfo{author}{\bibfnamefont{U.}~\bibnamefont{Greife}},
  \bibinfo{author}{\bibfnamefont{D.}~\bibnamefont{Hill}},
  \bibinfo{author}{\bibfnamefont{C.}~\bibnamefont{Iliadis}},
  \bibinfo{author}{\bibfnamefont{B.~A.} \bibnamefont{Johnson}},
  \bibinfo{author}{\bibfnamefont{R.~L.} \bibnamefont{Kozub}},
  \bibnamefont{et~al.}, \bibinfo{journal}{Phys.~Rev.~C}
  \textbf{\bibinfo{volume}{65}}, \bibinfo{pages}{032801(R)}
  (\bibinfo{year}{2002}).

\bibitem[{\citenamefont{Parpottas et~al.}(2004)\citenamefont{Parpottas, Grimes,
  Quraishi, Brune, Massey, Oldendick, Salas, and Wheeler}}]{Parpottas:2004}
\bibinfo{author}{\bibfnamefont{Y.}~\bibnamefont{Parpottas}},
  \bibinfo{author}{\bibfnamefont{S.~M.} \bibnamefont{Grimes}},
  \bibinfo{author}{\bibfnamefont{S.~A.} \bibnamefont{Quraishi}},
  \bibinfo{author}{\bibfnamefont{C.~R.} \bibnamefont{Brune}},
  \bibinfo{author}{\bibfnamefont{T.~N.} \bibnamefont{Massey}},
  \bibinfo{author}{\bibfnamefont{J.~E.} \bibnamefont{Oldendick}},
  \bibinfo{author}{\bibfnamefont{A.}~\bibnamefont{Salas}}, \bibnamefont{and}
  \bibinfo{author}{\bibfnamefont{R.~T.} \bibnamefont{Wheeler}},
  \bibinfo{journal}{Phys.~Rev.~C} \textbf{\bibinfo{volume}{70}},
  \bibinfo{pages}{065805} (\bibinfo{year}{2004}).

\bibitem[{\citenamefont{Seweryniak et~al.}(2007)\citenamefont{Seweryniak,
  Woods, Carpenter, Davinson, Janssens, Jenkins, Lauritsen, Lister, Shergur,
  Sinha et~al.}}]{Seweryniak:2007}
\bibinfo{author}{\bibfnamefont{D.}~\bibnamefont{Seweryniak}},
  \bibinfo{author}{\bibfnamefont{P.~J.} \bibnamefont{Woods}},
  \bibinfo{author}{\bibfnamefont{M.~P.} \bibnamefont{Carpenter}},
  \bibinfo{author}{\bibfnamefont{T.}~\bibnamefont{Davinson}},
  \bibinfo{author}{\bibfnamefont{R.~V.~F.} \bibnamefont{Janssens}},
  \bibinfo{author}{\bibfnamefont{D.~G.} \bibnamefont{Jenkins}},
  \bibinfo{author}{\bibfnamefont{T.}~\bibnamefont{Lauritsen}},
  \bibinfo{author}{\bibfnamefont{C.~J.} \bibnamefont{Lister}},
  \bibinfo{author}{\bibfnamefont{J.}~\bibnamefont{Shergur}},
  \bibinfo{author}{\bibfnamefont{S.}~\bibnamefont{Sinha}},
  \bibnamefont{et~al.}, \bibinfo{journal}{Phys.~Rev.~C}
  \textbf{\bibinfo{volume}{75}}, \bibinfo{pages}{062801}
  (\bibinfo{year}{2007}).

\bibitem[{\citenamefont{Matic et~al.}(2010)\citenamefont{Matic, Berg, Harakeh,
  W$\ddot{\mbox{o}}$rtche, Berg, Couder, G$\ddot{\mbox{o}}$rres, LeBlanc,
  O'Brien, Wiescher et~al.}}]{Matic:2010}
\bibinfo{author}{\bibfnamefont{A.}~\bibnamefont{Matic}},
  \bibinfo{author}{\bibfnamefont{A.~M.~V.~D.} \bibnamefont{Berg}},
  \bibinfo{author}{\bibfnamefont{M.~N.} \bibnamefont{Harakeh}},
  \bibinfo{author}{\bibfnamefont{H.~J.} \bibnamefont{W$\ddot{\mbox{o}}$rtche}},
  \bibinfo{author}{\bibfnamefont{G.~P.~A.} \bibnamefont{Berg}},
  \bibinfo{author}{\bibfnamefont{M.}~\bibnamefont{Couder}},
  \bibinfo{author}{\bibfnamefont{J.}~\bibnamefont{G$\ddot{\mbox{o}}$rres}},
  \bibinfo{author}{\bibfnamefont{P.}~\bibnamefont{LeBlanc}},
  \bibinfo{author}{\bibfnamefont{S.}~\bibnamefont{O'Brien}},
  \bibinfo{author}{\bibfnamefont{M.}~\bibnamefont{Wiescher}},
  \bibnamefont{et~al.}, \bibinfo{journal}{Phys.~Rev.~C}
  \textbf{\bibinfo{volume}{82}}, \bibinfo{pages}{025807}
  (\bibinfo{year}{2010}).

\bibitem[{\citenamefont{Kwiatkowski et~al.}(2010)\citenamefont{Kwiatkowski,
  Barquest, Bollen, Campbell, Ferrer, Gehring, Lincoln, Morrissey, Pang, Savory
  et~al.}}]{Kwiatkowski:2010}
\bibinfo{author}{\bibfnamefont{A.~A.} \bibnamefont{Kwiatkowski}},
  \bibinfo{author}{\bibfnamefont{B.~R.} \bibnamefont{Barquest}},
  \bibinfo{author}{\bibfnamefont{G.}~\bibnamefont{Bollen}},
  \bibinfo{author}{\bibfnamefont{C.~M.} \bibnamefont{Campbell}},
  \bibinfo{author}{\bibfnamefont{R.}~\bibnamefont{Ferrer}},
  \bibinfo{author}{\bibfnamefont{A.~E.} \bibnamefont{Gehring}},
  \bibinfo{author}{\bibfnamefont{D.~L.} \bibnamefont{Lincoln}},
  \bibinfo{author}{\bibfnamefont{D.~J.} \bibnamefont{Morrissey}},
  \bibinfo{author}{\bibfnamefont{G.~K.} \bibnamefont{Pang}},
  \bibinfo{author}{\bibfnamefont{J.}~\bibnamefont{Savory}},
  \bibnamefont{et~al.}, \bibinfo{journal}{Phys.~Rev.~C}
  \textbf{\bibinfo{volume}{81}}, \bibinfo{pages}{058501}
  (\bibinfo{year}{2010}).

\bibitem[{\citenamefont{Wrede et~al.}(2010)\citenamefont{Wrede, Clark, Deibel,
  Faestermann, Hertenberger, Parikh, Wirth, Bishop, Chen, Eppinger
  et~al.}}]{Wrede:2010}
\bibinfo{author}{\bibfnamefont{C.}~\bibnamefont{Wrede}},
  \bibinfo{author}{\bibfnamefont{J.~A.} \bibnamefont{Clark}},
  \bibinfo{author}{\bibfnamefont{C.~M.} \bibnamefont{Deibel}},
  \bibinfo{author}{\bibfnamefont{T.}~\bibnamefont{Faestermann}},
  \bibinfo{author}{\bibfnamefont{R.}~\bibnamefont{Hertenberger}},
  \bibinfo{author}{\bibfnamefont{A.}~\bibnamefont{Parikh}},
  \bibinfo{author}{\bibfnamefont{H.-F.} \bibnamefont{Wirth}},
  \bibinfo{author}{\bibfnamefont{S.}~\bibnamefont{Bishop}},
  \bibinfo{author}{\bibfnamefont{A.~A.} \bibnamefont{Chen}},
  \bibinfo{author}{\bibfnamefont{K.}~\bibnamefont{Eppinger}},
  \bibnamefont{et~al.}, \bibinfo{journal}{Phys.~Rev.~C}
  \textbf{\bibinfo{volume}{81}}, \bibinfo{pages}{055503}
  (\bibinfo{year}{2010}).

\bibitem[{\citenamefont{{P.~D.~Kunz}}()}]{DWUCK}
\bibinfo{author}{\bibnamefont{{P.~D.~Kunz}}}, \emph{\bibinfo{title}{{DWUCK}5}},
  \bibinfo{note}{{w}ebsite: \url{http://spot.colorado.edu/~kunz/DWBA.html}}.

\bibitem[{\citenamefont{Thompson}(1988)}]{Thompson:1988}
\bibinfo{author}{\bibfnamefont{I.~J.} \bibnamefont{Thompson}},
  \bibinfo{journal}{Comput.~Phys.~Rep.} \textbf{\bibinfo{volume}{7}},
  \bibinfo{pages}{167} (\bibinfo{year}{1988}).

\bibitem[{\citenamefont{Tribble and Kubo}(1977)}]{Tribble:1977}
\bibinfo{author}{\bibfnamefont{R.~E.} \bibnamefont{Tribble}} \bibnamefont{and}
  \bibinfo{author}{\bibfnamefont{K.-I.} \bibnamefont{Kubo}},
  \bibinfo{journal}{Nucl.~Phys.~A} \textbf{\bibinfo{volume}{282}},
  \bibinfo{pages}{269} (\bibinfo{year}{1977}).

\bibitem[{\citenamefont{R.~V.~Reid}(1968)}]{Reid:1968}
\bibinfo{author}{\bibfnamefont{J.}~\bibnamefont{R.~V.~Reid}},
  \bibinfo{journal}{Ann.~Phys.~(N.~Y.)} \textbf{\bibinfo{volume}{50}},
  \bibinfo{pages}{411} (\bibinfo{year}{1968}).

\bibitem[{\citenamefont{Basunia}(2010)}]{Basunia:2010}
\bibinfo{author}{\bibfnamefont{M.~S.} \bibnamefont{Basunia}},
  \bibinfo{journal}{Nucl.~Data Sheets} \textbf{\bibinfo{volume}{111}},
  \bibinfo{pages}{2331} (\bibinfo{year}{2010}).

\bibitem[{\citenamefont{Groenevbld et~al.}(1970)\citenamefont{Groenevbld,
  Hubert, Bass, and Nann}}]{Groenevbld:1970}
\bibinfo{author}{\bibfnamefont{K.~O.} \bibnamefont{Groenevbld}},
  \bibinfo{author}{\bibfnamefont{B.}~\bibnamefont{Hubert}},
  \bibinfo{author}{\bibfnamefont{R.}~\bibnamefont{Bass}}, \bibnamefont{and}
  \bibinfo{author}{\bibfnamefont{H.}~\bibnamefont{Nann}},
  \bibinfo{journal}{Nucl.~Phys.~A} \textbf{\bibinfo{volume}{151}},
  \bibinfo{pages}{198} (\bibinfo{year}{1970}).

\bibitem[{\citenamefont{Bass et~al.}(1972)\citenamefont{Bass, Friedland,
  Hubert, Nann, and Reiter}}]{Bass:1972}
\bibinfo{author}{\bibfnamefont{R.}~\bibnamefont{Bass}},
  \bibinfo{author}{\bibfnamefont{U.}~\bibnamefont{Friedland}},
  \bibinfo{author}{\bibfnamefont{B.}~\bibnamefont{Hubert}},
  \bibinfo{author}{\bibfnamefont{H.}~\bibnamefont{Nann}}, \bibnamefont{and}
  \bibinfo{author}{\bibfnamefont{A.}~\bibnamefont{Reiter}},
  \bibinfo{journal}{Nucl.~Phys.~A} \textbf{\bibinfo{volume}{198}},
  \bibinfo{pages}{449} (\bibinfo{year}{1972}).

\bibitem[{\citenamefont{Alburger and Goosman}(1974)}]{Alburger:1974}
\bibinfo{author}{\bibfnamefont{D.~E.} \bibnamefont{Alburger}} \bibnamefont{and}
  \bibinfo{author}{\bibfnamefont{D.~R.} \bibnamefont{Goosman}},
  \bibinfo{journal}{Phys.~Rev.~C} \textbf{\bibinfo{volume}{9}},
  \bibinfo{pages}{2236} (\bibinfo{year}{1974}).

\bibitem[{Sin()}]{Singh}
\bibinfo{howpublished}{B.~Singh (private communication)}.

\bibitem[{\citenamefont{Litherland and Ferguson}(1961)}]{Litherland:1961}
\bibinfo{author}{\bibfnamefont{A.~E.} \bibnamefont{Litherland}}
  \bibnamefont{and} \bibinfo{author}{\bibfnamefont{A.~J.}
  \bibnamefont{Ferguson}}, \bibinfo{journal}{Can.~J.~Phys.}
  \textbf{\bibinfo{volume}{39}}, \bibinfo{pages}{788} (\bibinfo{year}{1961}).

\bibitem[{\citenamefont{Morinaga and Yamazaki}(1976)}]{Morinaga:1976}
\bibinfo{author}{\bibfnamefont{H.}~\bibnamefont{Morinaga}} \bibnamefont{and}
  \bibinfo{author}{\bibfnamefont{T.}~\bibnamefont{Yamazaki}},
  \emph{\bibinfo{title}{In-Beam Gamma-Ray Spectroscopy}}
  (\bibinfo{publisher}{North Holland Publishing Co.}, \bibinfo{year}{1976}).

\bibitem[{\citenamefont{Yamazaki}(1967)}]{Yamazaki:1967}
\bibinfo{author}{\bibfnamefont{T.}~\bibnamefont{Yamazaki}},
  \bibinfo{journal}{Nucl.~Data A} \textbf{\bibinfo{volume}{3}},
  \bibinfo{pages}{1} (\bibinfo{year}{1967}).

\bibitem[{\citenamefont{Iliadis}(2007)}]{Iliadis:2007}
\bibinfo{author}{\bibfnamefont{C.}~\bibnamefont{Iliadis}},
  \emph{\bibinfo{title}{Nuclear Physics of Stars}}
  (\bibinfo{publisher}{Wiley-VCH, Weinheim}, \bibinfo{year}{2007}).

\bibitem[{\citenamefont{Rasmussen and Sugihara}(1966)}]{Rasmussen:1966}
\bibinfo{author}{\bibfnamefont{J.~O.} \bibnamefont{Rasmussen}}
  \bibnamefont{and} \bibinfo{author}{\bibfnamefont{T.~T.}
  \bibnamefont{Sugihara}}, \bibinfo{journal}{Phys.~Rev.}
  \textbf{\bibinfo{volume}{151}}, \bibinfo{pages}{992} (\bibinfo{year}{1966}).

\bibitem[{\citenamefont{der Mateosian and
  Sunyar}(1974{\natexlab{a}})}]{Mateosian:1974a}
\bibinfo{author}{\bibfnamefont{E.}~\bibnamefont{der Mateosian}}
  \bibnamefont{and} \bibinfo{author}{\bibfnamefont{A.~W.}
  \bibnamefont{Sunyar}}, \bibinfo{journal}{At.~Data Nucl.~Data Tables}
  \textbf{\bibinfo{volume}{13}}, \bibinfo{pages}{391}
  (\bibinfo{year}{1974}{\natexlab{a}}).

\bibitem[{\citenamefont{Krane and Steffen}(1970)}]{Krane:1970}
\bibinfo{author}{\bibfnamefont{K.~S.} \bibnamefont{Krane}} \bibnamefont{and}
  \bibinfo{author}{\bibfnamefont{R.~M.} \bibnamefont{Steffen}},
  \bibinfo{journal}{Phys.~Rev.~C} \textbf{\bibinfo{volume}{2}},
  \bibinfo{pages}{724} (\bibinfo{year}{1970}).

\bibitem[{\citenamefont{Rose and Brink}(1967)}]{Rose:1967}
\bibinfo{author}{\bibfnamefont{H.~J.} \bibnamefont{Rose}} \bibnamefont{and}
  \bibinfo{author}{\bibfnamefont{D.~M.} \bibnamefont{Brink}},
  \bibinfo{journal}{Rev.~Mod.~Phys.} \textbf{\bibinfo{volume}{39}},
  \bibinfo{pages}{306} (\bibinfo{year}{1967}).

\bibitem[{\citenamefont{der Mateosian and
  Sunyar}(1974{\natexlab{b}})}]{Mateosian:1974b}
\bibinfo{author}{\bibfnamefont{E.}~\bibnamefont{der Mateosian}}
  \bibnamefont{and} \bibinfo{author}{\bibfnamefont{A.~W.}
  \bibnamefont{Sunyar}}, \bibinfo{journal}{At.~Data Nucl.~Data Tables}
  \textbf{\bibinfo{volume}{13}}, \bibinfo{pages}{407}
  (\bibinfo{year}{1974}{\natexlab{b}}).

\bibitem[{Nuc(2010)}]{NuclearDataSheets}
\emph{\bibinfo{title}{{S}ummary of {B}ases for {S}pin and {P}arity
  {A}ssignments}}, \bibinfo{howpublished}{Nucl.~Data Sheets, \textbf{111}, iv}
  (\bibinfo{year}{2010}).

\bibitem[{\citenamefont{Kr$\ddot{\mbox{a}}$mer-Flecken
  et~al.}(1989)\citenamefont{Kr$\ddot{\mbox{a}}$mer-Flecken, Morek, Lieder,
  Gast, Hebbinghaus, J$\ddot{\mbox{a}}$ger, and Urban}}]{Kramer-Flecken:1989}
\bibinfo{author}{\bibfnamefont{A.}~\bibnamefont{Kr$\ddot{\mbox{a}}$mer-Flecken%
}}, \bibinfo{author}{\bibfnamefont{T.}~\bibnamefont{Morek}},
  \bibinfo{author}{\bibfnamefont{R.~M.} \bibnamefont{Lieder}},
  \bibinfo{author}{\bibfnamefont{W.}~\bibnamefont{Gast}},
  \bibinfo{author}{\bibfnamefont{G.}~\bibnamefont{Hebbinghaus}},
  \bibinfo{author}{\bibfnamefont{H.~M.} \bibnamefont{J$\ddot{\mbox{a}}$ger}},
  \bibnamefont{and} \bibinfo{author}{\bibfnamefont{W.}~\bibnamefont{Urban}},
  \bibinfo{journal}{Nucl.~Instr.~Meth.~Phys.~Res.~A}
  \textbf{\bibinfo{volume}{275}}, \bibinfo{pages}{333} (\bibinfo{year}{1989}).

\bibitem[{Bro()}]{Brown}
\bibinfo{howpublished}{B.~A.~Brown (private communication)}.

\bibitem[{Ric()}]{Richter}
\bibinfo{howpublished}{W.~A.~Richter (private communication)}.

\bibitem[{\citenamefont{Mackh et~al.}(1973)\citenamefont{Mackh, Oeschler,
  Wagner, Dehnhard, and Ohnuma}}]{Mackh:1973}
\bibinfo{author}{\bibfnamefont{H.}~\bibnamefont{Mackh}},
  \bibinfo{author}{\bibfnamefont{H.}~\bibnamefont{Oeschler}},
  \bibinfo{author}{\bibfnamefont{G.~J.} \bibnamefont{Wagner}},
  \bibinfo{author}{\bibfnamefont{D.}~\bibnamefont{Dehnhard}}, \bibnamefont{and}
  \bibinfo{author}{\bibfnamefont{H.}~\bibnamefont{Ohnuma}},
  \bibinfo{journal}{Nucl.~Phys.~A} \textbf{\bibinfo{volume}{202}},
  \bibinfo{pages}{497} (\bibinfo{year}{1973}).

\bibitem[{\citenamefont{Iliadis
  et~al.}(2010{\natexlab{b}})\citenamefont{Iliadis, Longland, Champagne, and
  Coc}}]{Iliadis:2010b}
\bibinfo{author}{\bibfnamefont{C.}~\bibnamefont{Iliadis}},
  \bibinfo{author}{\bibfnamefont{R.}~\bibnamefont{Longland}},
  \bibinfo{author}{\bibfnamefont{A.~E.} \bibnamefont{Champagne}},
  \bibnamefont{and} \bibinfo{author}{\bibfnamefont{A.}~\bibnamefont{Coc}},
  \bibinfo{journal}{Nucl.~Phys.~A} \textbf{\bibinfo{volume}{841}},
  \bibinfo{pages}{251} (\bibinfo{year}{2010}{\natexlab{b}}).

\bibitem[{\citenamefont{Iliadis}(1997)}]{Iliadis:1997}
\bibinfo{author}{\bibfnamefont{C.}~\bibnamefont{Iliadis}},
  \bibinfo{journal}{Nucl.~Phys.~A} \textbf{\bibinfo{volume}{618}},
  \bibinfo{pages}{166} (\bibinfo{year}{1997}).

\bibitem[{\citenamefont{Longland et~al.}(2010)\citenamefont{Longland, Iliadis,
  Champagne, Newton, Ugalde, Coc, and Fitzgerald}}]{Longland:2010}
\bibinfo{author}{\bibfnamefont{R.}~\bibnamefont{Longland}},
  \bibinfo{author}{\bibfnamefont{C.}~\bibnamefont{Iliadis}},
  \bibinfo{author}{\bibfnamefont{A.~E.} \bibnamefont{Champagne}},
  \bibinfo{author}{\bibfnamefont{J.~R.} \bibnamefont{Newton}},
  \bibinfo{author}{\bibfnamefont{C.}~\bibnamefont{Ugalde}},
  \bibinfo{author}{\bibfnamefont{A.}~\bibnamefont{Coc}}, \bibnamefont{and}
  \bibinfo{author}{\bibfnamefont{R.}~\bibnamefont{Fitzgerald}},
  \bibinfo{journal}{Nucl.~Phys.~A} \textbf{\bibinfo{volume}{841}},
  \bibinfo{pages}{1} (\bibinfo{year}{2010}).

\bibitem[{\citenamefont{Newton et~al.}(2008)\citenamefont{Newton, Longland, and
  Iliadis}}]{Newton:2008}
\bibinfo{author}{\bibfnamefont{J.~R.} \bibnamefont{Newton}},
  \bibinfo{author}{\bibfnamefont{R.}~\bibnamefont{Longland}}, \bibnamefont{and}
  \bibinfo{author}{\bibfnamefont{C.}~\bibnamefont{Iliadis}},
  \bibinfo{journal}{Phys.~Rev.~C} \textbf{\bibinfo{volume}{78}},
  \bibinfo{pages}{025805} (\bibinfo{year}{2008}).

\bibitem[{\citenamefont{Rauscher and Thielemann}(2000)}]{NONSMOKER}
\bibinfo{author}{\bibfnamefont{T.}~\bibnamefont{Rauscher}} \bibnamefont{and}
  \bibinfo{author}{\bibfnamefont{F.-K.} \bibnamefont{Thielemann}},
  \bibinfo{journal}{At.~Data Nucl.~Data Tables} \textbf{\bibinfo{volume}{75}},
  \bibinfo{pages}{1} (\bibinfo{year}{2000}).

\bibitem[{\citenamefont{Jos$\acute{\mbox{e}}$ and Hernanz}(1998)}]{Jose:1998}
\bibinfo{author}{\bibfnamefont{J.}~\bibnamefont{Jos$\acute{\mbox{e}}$}}
  \bibnamefont{and} \bibinfo{author}{\bibfnamefont{M.}~\bibnamefont{Hernanz}},
  \bibinfo{journal}{Astrophys.~J.} \textbf{\bibinfo{volume}{494}},
  \bibinfo{pages}{680} (\bibinfo{year}{1998}).

\bibitem[{\citenamefont{Jos$\acute{\mbox{e}}$
  et~al.}(1999)\citenamefont{Jos$\acute{\mbox{e}}$, Coc, and
  Hernanz}}]{Jose:1999}
\bibinfo{author}{\bibfnamefont{J.}~\bibnamefont{Jos$\acute{\mbox{e}}$}},
  \bibinfo{author}{\bibfnamefont{A.}~\bibnamefont{Coc}}, \bibnamefont{and}
  \bibinfo{author}{\bibfnamefont{M.}~\bibnamefont{Hernanz}},
  \bibinfo{journal}{Astrophys.~J.} \textbf{\bibinfo{volume}{520}},
  \bibinfo{pages}{347} (\bibinfo{year}{1999}).

\bibitem[{\citenamefont{Grevesse et~al.}(2007)\citenamefont{Grevesse, Asplund,
  and Sauval}}]{Grevesse:2007}
\bibinfo{author}{\bibfnamefont{N.}~\bibnamefont{Grevesse}},
  \bibinfo{author}{\bibfnamefont{M.}~\bibnamefont{Asplund}}, \bibnamefont{and}
  \bibinfo{author}{\bibfnamefont{A.~J.} \bibnamefont{Sauval}},
  \bibinfo{journal}{Space Sci.~Rev.} \textbf{\bibinfo{volume}{130}},
  \bibinfo{pages}{105} (\bibinfo{year}{2007}).

\bibitem[{Eva()}]{Evans:2008}
\bibinfo{howpublished}{S.~Starrfield, C.~Iliadis and W.~R.~Hix, in
  \textit{Classical Novae}, edited by M.~F.~Bode and A.~Evans (Cambridge
  University Press, 2008 -- 2$^{nd}$ edn.)}.

\bibitem[{\citenamefont{Wrede et~al.}(2009)\citenamefont{Wrede, Caggiano,
  Clark, Deibel, Parikh, and Parker}}]{Wrede:2009}
\bibinfo{author}{\bibfnamefont{C.}~\bibnamefont{Wrede}},
  \bibinfo{author}{\bibfnamefont{J.~A.} \bibnamefont{Caggiano}},
  \bibinfo{author}{\bibfnamefont{J.~A.} \bibnamefont{Clark}},
  \bibinfo{author}{\bibfnamefont{C.~M.} \bibnamefont{Deibel}},
  \bibinfo{author}{\bibfnamefont{A.}~\bibnamefont{Parikh}}, \bibnamefont{and}
  \bibinfo{author}{\bibfnamefont{P.~D.} \bibnamefont{Parker}},
  \bibinfo{journal}{Phys.~Rev.~C} \textbf{\bibinfo{volume}{79}},
  \bibinfo{pages}{045803} (\bibinfo{year}{2009}).

\bibitem[{\citenamefont{Parikh et~al.}(2011)\citenamefont{Parikh, Wimmer,
  Faestermann, Hertenberger, Jos$\acute{\mbox{e}}$, Longland, Wirth, Bildstein,
  Bishop, Chen et~al.}}]{Parikh:2011}
\bibinfo{author}{\bibfnamefont{A.}~\bibnamefont{Parikh}},
  \bibinfo{author}{\bibfnamefont{K.}~\bibnamefont{Wimmer}},
  \bibinfo{author}{\bibfnamefont{T.}~\bibnamefont{Faestermann}},
  \bibinfo{author}{\bibfnamefont{R.}~\bibnamefont{Hertenberger}},
  \bibinfo{author}{\bibfnamefont{J.}~\bibnamefont{Jos$\acute{\mbox{e}}$}},
  \bibinfo{author}{\bibfnamefont{R.}~\bibnamefont{Longland}},
  \bibinfo{author}{\bibfnamefont{H.-F.} \bibnamefont{Wirth}},
  \bibinfo{author}{\bibfnamefont{V.}~\bibnamefont{Bildstein}},
  \bibinfo{author}{\bibfnamefont{S.}~\bibnamefont{Bishop}},
  \bibinfo{author}{\bibfnamefont{A.~A.} \bibnamefont{Chen}},
  \bibnamefont{et~al.}, \bibinfo{journal}{Phys.~Rev.~C}
  \textbf{\bibinfo{volume}{83}}, \bibinfo{pages}{045806}
  (\bibinfo{year}{2011}).

\bibitem[{\citenamefont{Doherty et~al.}(2012)\citenamefont{Doherty, Lotay,
  Woods, Seweryniak, Carpenter, Chiara, David, Janssens, Trache, and
  Zhu}}]{Doherty:2012}
\bibinfo{author}{\bibfnamefont{D.~T.} \bibnamefont{Doherty}},
  \bibinfo{author}{\bibfnamefont{G.}~\bibnamefont{Lotay}},
  \bibinfo{author}{\bibfnamefont{P.~J.} \bibnamefont{Woods}},
  \bibinfo{author}{\bibfnamefont{D.}~\bibnamefont{Seweryniak}},
  \bibinfo{author}{\bibfnamefont{M.~P.} \bibnamefont{Carpenter}},
  \bibinfo{author}{\bibfnamefont{C.~J.} \bibnamefont{Chiara}},
  \bibinfo{author}{\bibfnamefont{H.~M.} \bibnamefont{David}},
  \bibinfo{author}{\bibfnamefont{R.~V.~F.} \bibnamefont{Janssens}},
  \bibinfo{author}{\bibfnamefont{L.}~\bibnamefont{Trache}}, \bibnamefont{and}
  \bibinfo{author}{\bibfnamefont{S.}~\bibnamefont{Zhu}},
  \bibinfo{journal}{Phys.~Rev.~Lett.} \textbf{\bibinfo{volume}{108}},
  \bibinfo{pages}{262502} (\bibinfo{year}{2012}).

\end{thebibliography}

\end{document}